\PassOptionsToPackage{table}{xcolor} 
\RequirePackage{tikz}
\documentclass[pdflatex,sn-basic]{sn-jnl}% Basic Springer Nature Reference Style/Chemistry Reference Style
\usepackage{subcaption} 
\usepackage[range-phrase={~to~}]{siunitx}

\jyear{2022}%

\usepackage{anyfontsize} % to suppress font shape not available
\usepackage{longtable,tabularx,booktabs}
\usepackage{makecell} % for two-line column headings in tables

\theoremstyle{thmstyleone}%
\theoremstyle{thmstyletwo}%

\theoremstyle{thmstylethree}%

\makeatletter
\newcommand\thefontsize[1]{{#1 The current font size is: \f@size pt\par}}
\newcommand{\showfontsize}{\f@size{} pt}
\makeatother

\usepackage{layouts}

\begin{document}

\title[Transonic Buffet at Flight Reynolds Number]{Mach and Reynolds Number Effects on Transonic Buffet on the XRF-1 Transport Aircraft Wing at Flight Reynolds Number}

\author*[1]{\fnm{Andreas} \sur{Waldmann}}\email{waldmann@iag.uni-stuttgart.de}

\author[1]{\fnm{Maximilian C.} \sur{Ehrle}}%\email{iiauthor@gmail.com}
\equalcont{These authors contributed equally to this work.}

\author[1]{\fnm{Johannes} \sur{Kleinert}}%\email{iiiauthor@gmail.com}
\equalcont{These authors contributed equally to this work.}

\author[2]{\fnm{Daisuke} \sur{Yorita}}%\email{iiiauthor@gmail.com}
\equalcont{These authors contributed equally to this work.}

\author[1]{\fnm{Thorsten} \sur{Lutz}}%\email{iiiauthor@gmail.com}
\equalcont{These authors contributed equally to this work.}

\affil*[1]{\orgdiv{Institute of Aerodynamics and Gas Dynamics}, \orgname{University of Stuttgart}, \orgaddress{\city{Stuttgart}, \postcode{70569}, \country{Germany}}}

\affil[2]{\orgdiv{Institute of Aerodynamics and Flow Technology}, \orgname{German Aerospace Center (DLR)}, \orgaddress{\city{G\"ottingen}, \postcode{37037}, \country{Germany}}}

%\affil[3]{\orgdiv{Department}, \orgname{Organization}, \orgaddress{\street{Street}, \city{City}, \postcode{610101}, \state{State}, \country{Country}}}


\abstract{
This work provides an overview of aerodynamic data acquired in the European Transonic Windtunnel using an XRF-1 transport aircraft configuration both at cruise conditions and at the edges of the flight envelope. The goals and design of the wind tunnel test was described, highlighting the use of the cryogenic wind tunnel's capability to isolate the effects of $M_{\infty}$, $Re_{\infty}$ and the dynamic pressure $q/E$. The resulting dataset includes an aerodynamic baseline characterization of the full span model with vertical and horizontal tailplanes and without engine nacelles. The effects of different inflow conditions were studied using data from continuous polars, evaluating the changes in aeroelastic deformation which are proportional to $q/E$ and the influence of $M_{\infty}$ and $Re_{\infty}$ on the shock position. Off-design data was analyzed at the lowest and highest measured Mach numbers of $0.84$ and $0.90$, respectively. Wing lower surface flow and underside shock motion was analyzed at negative angles of attack using $c_p$ distribution and unsteady pressure transducer fluctuation data, identifying significant upstream displacement of the shock close to the leading edge. Wing upper side flow and the shock motion near buffet onset and beyond was analyzed using unsteady pressure data from point transducers and unsteady pressure sensitive paint (PSP) measurements. Buffet occurs at lower angles of attack at high Mach number, and without clearly defined lift break. Spectral contents at the acquired data points in the buffet range suggest broadband fluctuations at Strouhal numbers between 0.2 and 0.6, which is consistent with recent literature. The spanwise shock propagation velocities were determined independently via analysis of unsteady PSP and pressure transducers to be in the range between $u_s / u_{\infty} = 0.24$ and $0.32$, which is similarly in line with published datasets using other swept wing aircraft models.
}


\maketitle

\section{Introduction}\label{sec1}

The flight envelope of transonic transport aircraft is bounded at high speeds by the occurrence of unsteady phenomena. When the flight Mach number $M_{\infty}$ or the angle of attack $\alpha$ exceed the design range of a given aircraft, transonic buffet and high speed stall may occur, which are undesirable conditions associated with unsteady flow on the wing surfaces. Shock unsteadiness occurring at such conditions exhibits complex behaviour, causing oscillatory loads that can pose a safety hazard. Associated flow separation and interaction between wing wake and tailplane all contribute unsteady aerodynamic phenomena which are challenging to predict.

Experimental replication of operating conditions at cruise altitudes and speeds is difficult to achieve and requires specialized experimental facilities. High Reynolds numbers representative of flight conditions can be replicated only in cryogenic, pressurized wind tunnels. Computational studies of such conditions also require experimental validation. For this reason, the research initiative FOR 2895 was established by the German Research Foundation (DFG), the Helmholtz Association of German Research Centres (HGF) and the German Aerospace Center (DLR) in order to pool the resources of several universities and research institutions and advance the state of knowledge in this area. The aims of the research include improved understanding of buffet phenomena, wing-tailplane interaction and integration issues associated with novel Ultra High Bypass Ratio (UHBR) engine nacelles. The initiative includes several wind tunnel campaigns carried out in the European Transonic Windtunnel (ETW) using a 1/37th scale aircraft model provided by Airbus. Further details can be found in the corresponding publication by Lutz et al.~\cite{lutz:2022}.

\subsection{Flight Envelope Boundaries}

Aircraft behavior at the flight envelope limits is a notoriously difficult research topic. The operating regime is bounded by limits in terms of speed and load factor which result from a variety of physical phenomena. Operation of aircraft in the vicinity of the design point and over a large part of the flight envelope in linear lift regime is characterized by smooth and largely attached flow. Approaching the flight envelope's limits often gives rise to unacceptable loads or undesirable unsteadiness associated with flow separation. Separation on the wing and empennage is a complex phenomenon, and has been identified to be particularly challenging in the case of moderately swept wings so common on transonic airliners (\cite{mabey:1999}). Tail buffet (\cite{abdrashitov:1939}) has been known for decades to be an undesirable unsteady condition, with research into the interactions between the wing wake and the tail shedding light on the occuring flow physics. While computational fluid dynamics (CFD) have achieved a state enabling consistent prediction of performance and flow phenomena near the design range (\cite{abbasbayoumi:2011}), flow physics at and beyond the envelope boundaries remain a challenge.
Recent large scale research efforts such as the ESWI$^{\mathrm{RP}}$ project placed a focus on the investigation envelope boundaries and the fluid dynamics beyond them. Lutz et al.~\cite{lutz:2016} provided an overview of the work conducted in this project on wake phenomena and high angle of attack flight on realistic configurations and conditions relevant for full scale aircraft. 

\subsection{Buffet}
\label{chp:intro_buffet}

Transonic buffet is an aerodynamic phenomenon involving a self-sustained oscillation of a shock present at a surface when sufficient Mach number $M_{\infty}$ and angle of attack $\alpha$ are reached. This aerodynamic oscillation can precipitate oscillating structural loads as a response, causing buffeting. Two-dimensional buffet on airfoils has been researched for decades using many techniques and spawned comprehensive literature reviews such as by \cite{lee:2000}. The same author provided a popular explanation of the phenomenon in \cite{lee:1992}, in which the thickening separating boundary layer feeds the oscillation using acoustic scattering of pressure fluctuations at the trailing edge. \cite{crouch:2009} employed stability analysis of linearized RANS equations and were able to identify a global mode associated to the buffet motion. In addition to the aforementioned interaction between the shock and the downstream separation, this mechanism involves pressure disturbances traveling along the pressure side of the airfoil. This interaction    involving the pressure side was experimentally observed by \cite{jacquin:2009}. \cite{garnier:2010} observed both the suction side and the pressure side disturbance transport paths using an LES approach, fully resolving the entire field.  On the other hand, \cite{hartmann:2013} performed measurements on the DRA 2303 airfoil and concluded that, while both types of disturbance propagation are possible, the more direct interaction over the suction side dominates. They also modified Lee's model by arguing that the disturbances traveling upstream from the trailing edge are in fact sound waves with a varying sound pressure level. This variation is due to the changing boundary layer thickness behind the shock, and it is this variation that causes the shock movement.

There is a degree of agreement on the essentially two-dimensional nature of the buffet phenomenon on 2D airfoils. Wing sweep and other three-dimensional features introduce an additional degree of complexity. \cite{iovnovich:2014} and \cite{daguanno:2022} also studied the effect of wing sweep on the buffet phenomenon, in particular the switch from 2D single-frequency shock  fluctuation to 3D buffet characterized by broadband spectral structure. \cite{dandois:2016} described an experimental study of the AVERT wing-body configuration with an elastic wing, with \cite{koike:2016} and \cite{sugioka:2018} conducting buffet experiments on the Common Research Model (CRM) configuration, while Lawson et al.~\cite{lawson} describing data from experiments on the RBC12 configuration. All of these experiments employed unsteady pressure sensitive paint (PSP) measurements in order to characterize the surface pressure unsteadiness on the wing. \cite{timme:2020} argued from 3D global stability analysis that buffet onset originates with one unstable oscillatory eigenmode. \cite{paladini:2019} gathered data from four published half-model transonic aircraft experiments and analyzed 3D buffet behavior, deriving buffet Strouhal number ranges which are case-independent and exhibit a degree of universality. \cite{giannelis:2017} presented a survey detailing fluctuation characteristics in 2D and 3D. Further analysis of the experiments described by \cite{lawson} was presented by \cite{masini:2020}, focusing on the time-resolved wing surface pressure data near buffet onset. Modal analysis techniques were used to extract characteristic Strouhal number ranges, largely agreeing with the results by the other authors.

Most of the aforementioned authors suggested that three-dimensional buffet on a swept wing is associated to a spanwise movement of spatially constrained buffet cells toward the wing tip. It is also a fundamentally different process then the purely chordwise 2D buffet motion with very narrowband characteristics. The most recent of the aforementioned experimental swept wing buffet studies state that buffet motion tends to occur outboard at Strouhal number ranges between $Sr = 0.2$ and $Sr = 0.6$, based on the mean aerodynamic chord. Multiple authors attempt a classification of different regimes, such as \cite{paladini:2019} with thei pre-onset, well-established and deep buffet regimes. Similarly, \cite{sugioka:2018} described three regimes based on the slope of the pressure root mean square (RMS) curve over $\alpha$, while \cite{dandois:2016} separated the observed phenomena to onset and deep buffet conditions. \cite{sugioka:2021} associated the detection of buffet cells at $Sr = $ \SIrange{0.2}{0.5}{} with buffet onset.

Several of the above publications observed a second, lower Strouhal number range. \cite{sugioka:2018} observed it mainly on the inboard wing, and \cite{paladini:2019} detected a roughly constant phase angle at this frequency over the wing span, deducing that this is a chordwise shock motion akin to two-dimensional buffet. \cite{masini:2020}, on the other hand, used modal analysis to detect an inboard propagation of low frequency disturbances. The survey by Paladini et al. shows that there is a degree of universality in the Strouhal number range between 0.2 and 0.6, for which they obtained propagation velocities around $u_s / u_{\infty} = 0.25 .. 0.3$ for the four different geometries. At the same time, some spectral characteristics differ between the four wings, which the authors ascribe to different taper ratios and other geometric differences. In addition to this, a recent study by \cite{uchida:2021} showed that the wing twist plays an important role for the buffet and disturbance propagation characteristics by comparing the CRM to the ONERA M4 model wing.

\subsection{Aims}

The purpose of the present work is the characterization of the XRF-1 aircraft model at flight relevant Mach and Reynolds numbers, and the evaluation of effects associated with variation of $M_{\infty}$ and $Re_{\infty}$. The properties of both steady and unsteady phenomena at very low and very high incidences are analyzed using various data obtained using different measurement techniques during the wind tunnel measurement campaign. With the exception of FLIRET described by \cite{paladini:2019}, most of the published data exists for conditions at Reynolds numbers on the order of $10^6$. The experiment described in this work presents a unique opportunity to investigate flight-like conditions and evaluate the effects of Mach and Reynolds number separately.

\section{Wind Tunnel Test}

The wind tunnel data analyzed in this work is the result of the first of several wind tunnel entries of the research initiative summarized by Lutz et al.~\cite{lutz:2022}. The present experimental campaign is referred to as MK1a in that work. Its focus was on acquiring a baseline set of aerodynamic data on a preselected range of Mach and Reynolds numbers using the XRF-1 aircraft model without any engine nacelles. In addition, optical measurements using pressure sensitive paint (PSP) were conducted to acquire data at incidences of interest.

\subsection{European Transonic Windtunnel}
The experimental campaign took place in 2020 in the European Transonic Windtunnel (ETW) in Cologne, Germany. It is the first of several wind tunnel entries envisioned for the present research initiative. The ETW has long been used by industry and research organizations for testing of transonic airliner configurations at flight-like conditions. ETW's operating regime spans Mach numbers between 0.15 and 1.35 and Reynolds numbers up to $50\cdot 10^6$ using full span aircraft models. This is enabled by pressurization up to \SI{450}{\kilo\pascal} and cooling down to temperatures of \SI{110}{\kelvin} using injection of liquid nitrogen.

\begin{figure}[!htbp]
  \centering
     \includegraphics[width=0.70\linewidth]{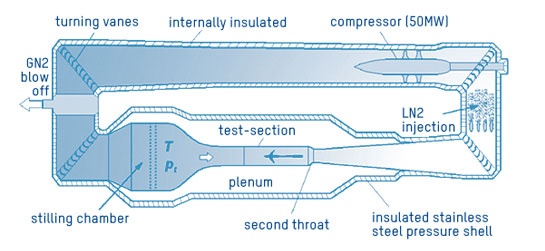}
    \caption{Aerodynamic circuit of the European Transonic Windtunnel (courtesy of ETW).}\label{f:ETW_circuit}
\end{figure}

Cryogenic testing in the ETW permits independent variation of Mach number, Reynolds number and dynamic pressure by the ability to decouple temperature and pressure changes inside the tunnel. The facility is a G\"ottingen type wind tunnel whose aerodynamic circuit (Fig.~\ref{f:ETW_circuit}) is powered by a \SI{50}{\mega\watt} compressor. State of the art features such as slotted and diverging walls and an optically accessible measurement section permit precise control of the flow condition and the application of advanced measurement techniques.

\subsection{XRF-1 Aircraft Model}

The RWF\.55\_1 wind tunnel model was provided by Airbus UK and represents the generic XRF-1 long range twin engine airliner research model from that same manufacturer. The XRF-1 research model is similar to a modern transonic commercial aircraft~\cite{goertz:2020}. The corresponding wind tunnel model was used in Airbus' FeRIT campaign conducted in 2018 in ETW~\cite{mann:2019} to create a database of aerodynamic data covering a broad range of Mach and Reynolds numbers, and to prove out the use of the model in the ETW in conjunction with pressure sensitive paint techniques. The model includes a fixed vertical tailplane (VTP) and adjustable ailerons, with the option of mounting through-flow nacelles to the wing. In contrast to FeRIT, the present work includes data acquired in the most basic configuration without the nacelles or aileron deflection.

\begin{figure}[!htbp]
  \centering
     \includegraphics[width=0.495\linewidth]{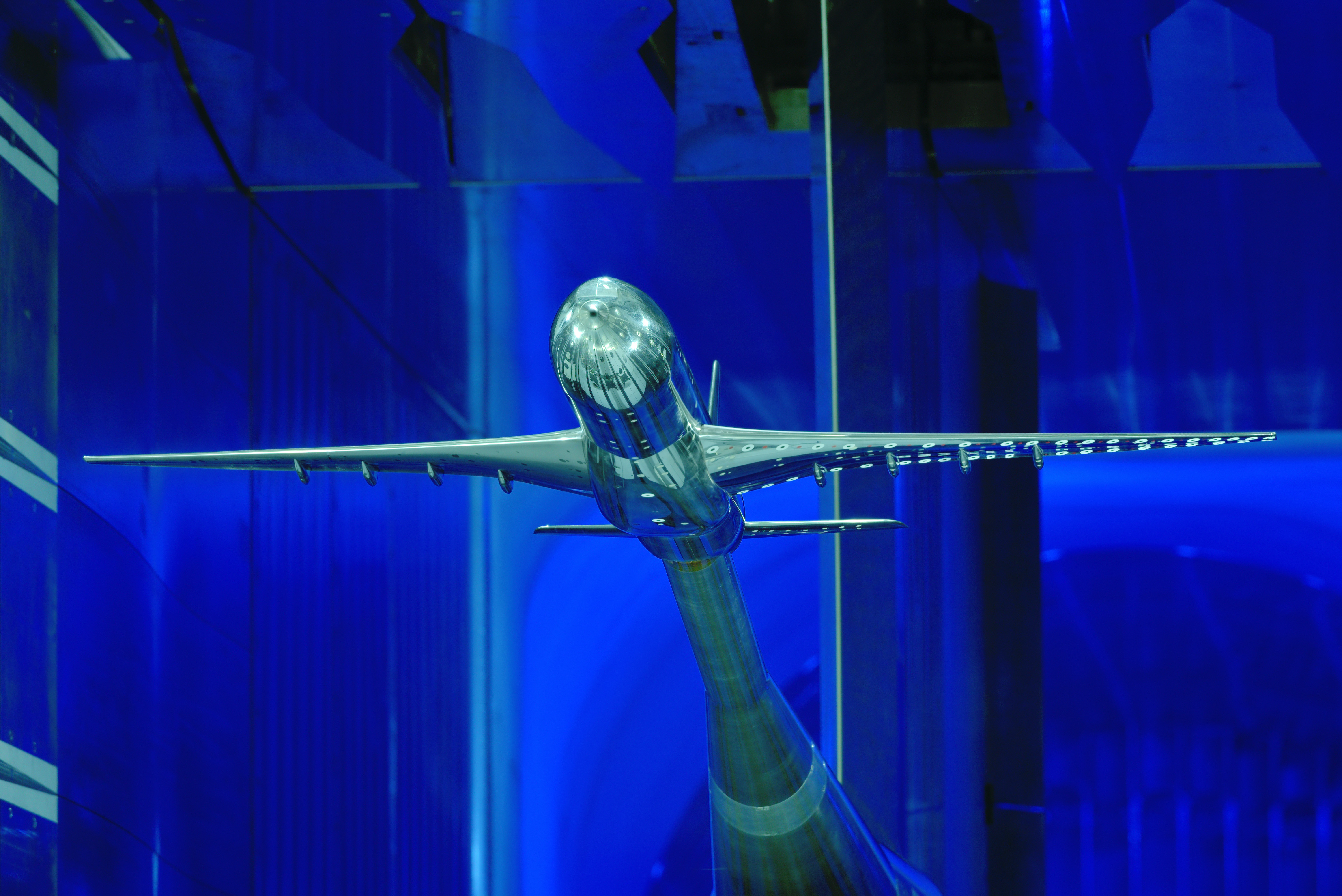}
    \caption{XRF-1 in the European Transonic Windtunnel test section (courtesy of Airbus and ETW).}
   \label{f:XRF1_photo_ETW}
\end{figure}

The model configuration used in FeRIT did not include a horizontal tailplane (HTP), which was specifically designed and manufactured prior to the present measurements. The tailplane was installed at a fixed incidence angle of \SI{-2}{\degree}, which was maintained throughout the campaign.

The main wing has a swept planform with a trailing edge crank at about $29\%$. Its aspect ratio is $AR = 9.302$ with a mean aerodynamic chord of $c_{\mathrm{ref}} = \SI{0.1965}{\meter}$ and a leading edge sweep of about $\Lambda = \SI{30}{\degree}$. The only features present on the model (shown in Fig.~\ref{f:XRF1_photo_ETW}) wing were four flap track fairings, with no engine nacelles. The model does include a wing-body fairing. No configuration changes were made during the course of the present campaign, with the exception of different types of transition fixing devices on the wing, HTP and VTP surfaces. The fuselage boundary layer was tripped near the nose using a carborundum strip which remained in place at all conditions. No tripping was applied the high Reynolds number runs at $Re_{\infty} = 25.0\cdot 10^6$ after prior studies had shown that the flow can be expected to be fully turbulent in the entire region. The lower Reynolds number settings were run with tripping dots of appropriate height depending on the Reynolds number. The tripping devices were placed at $5\%$ of the local chord on both sides of the wing, HTP and VTP.

\subsection{Instrumentation, Data Acquisition and Processing}

The full span aircraft model was placed on a straight sting entering the lower fuselage via a cavity. The total model/sting offset is \SI{5}{\degree}. The assembly included an anti-vibration system consisting of one component mounted in the sting and one at the sting/balance interface inside the model~\cite{etw_user_guide}. The instrumentation aspects relevant for aerodynamic analysis shall be described below.

The proprietary nature of the XRF-1 precludes the publication of absolute values. The main focus of this paper is on off-design conditions, with an emphasis on qualitative and quantitative comparisons of flow phenomena at different flow conditions.

\subsubsection{Forces and Moments}

Forces and moments were acquired using a six component strain gauge balance inside the fuselage. Continous polars were recorded at both fin-up ($\phi = \SI{0}{\degree}$) and fin-down ($\phi = \SI{180}{\degree}$) orientations. The fin-down polars were recorded starting at $\alpha = \SI{-4}{\degree}$ up to the maximum positive incidence deemed safe for the wind tunnel operation. The maximum incidence is therefore different at each condition. 

Each continuous polar was recorded a second time in fin-up configuration using a smaller $\alpha$ range. Apart from accuracy and repeatability considerations, the different orientations were made necessary by the fact that the cameras for pressure sensitive paint measurements were situated in the bottom wind tunnel wall. Therefore it is essential for the wind tunnel to be able to reproduce the same flow conditions in both orientations.

\subsubsection{Static Pressure Taps}

\begin{figure*}[!htb]
\centering

\begin{subfigure}{.40\textwidth}

\begin{tikzpicture}
\node [anchor=south west,inner sep=0pt] (image) at (0,0) {\includegraphics[width=\textwidth]{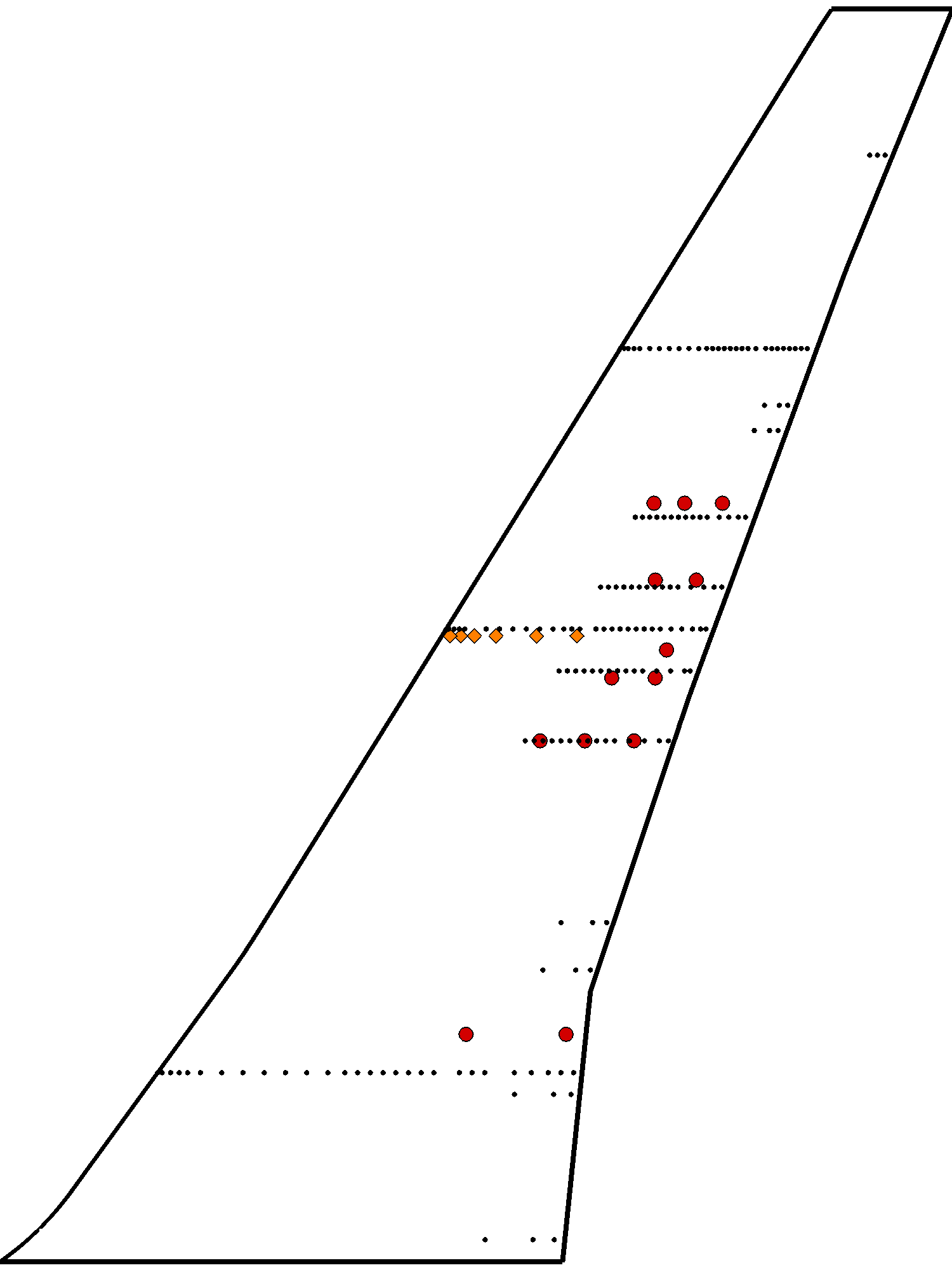}};

    \begin{scope}[x={(image.south east)},y={(image.north west)}]

\node at (0.3, 0.725) (eta075) {$\eta = 75\%$};
\draw[thick,dash dot] (eta075) -- (0.65, 0.725); 

\node at (0.2, 0.505) (eta055) {$\eta = 55\%$};
\draw[thick,dash dot] (eta055) -- (0.45, 0.505); 

\node at (0.2, 0.3) (eta023) {$\eta = 23\%$};
\draw[thick,dash dot] (eta023) -- (0.3, 0.155);

    \end{scope}
\end{tikzpicture}

  \caption{Wing upper side}
  \label{f:cp_taps_wing}
\end{subfigure}%
\begin{subfigure}{.40\textwidth}
\begin{tikzpicture}

\node [anchor=south west,inner sep=0pt] (image) at (0,0) {\includegraphics[width=\textwidth]{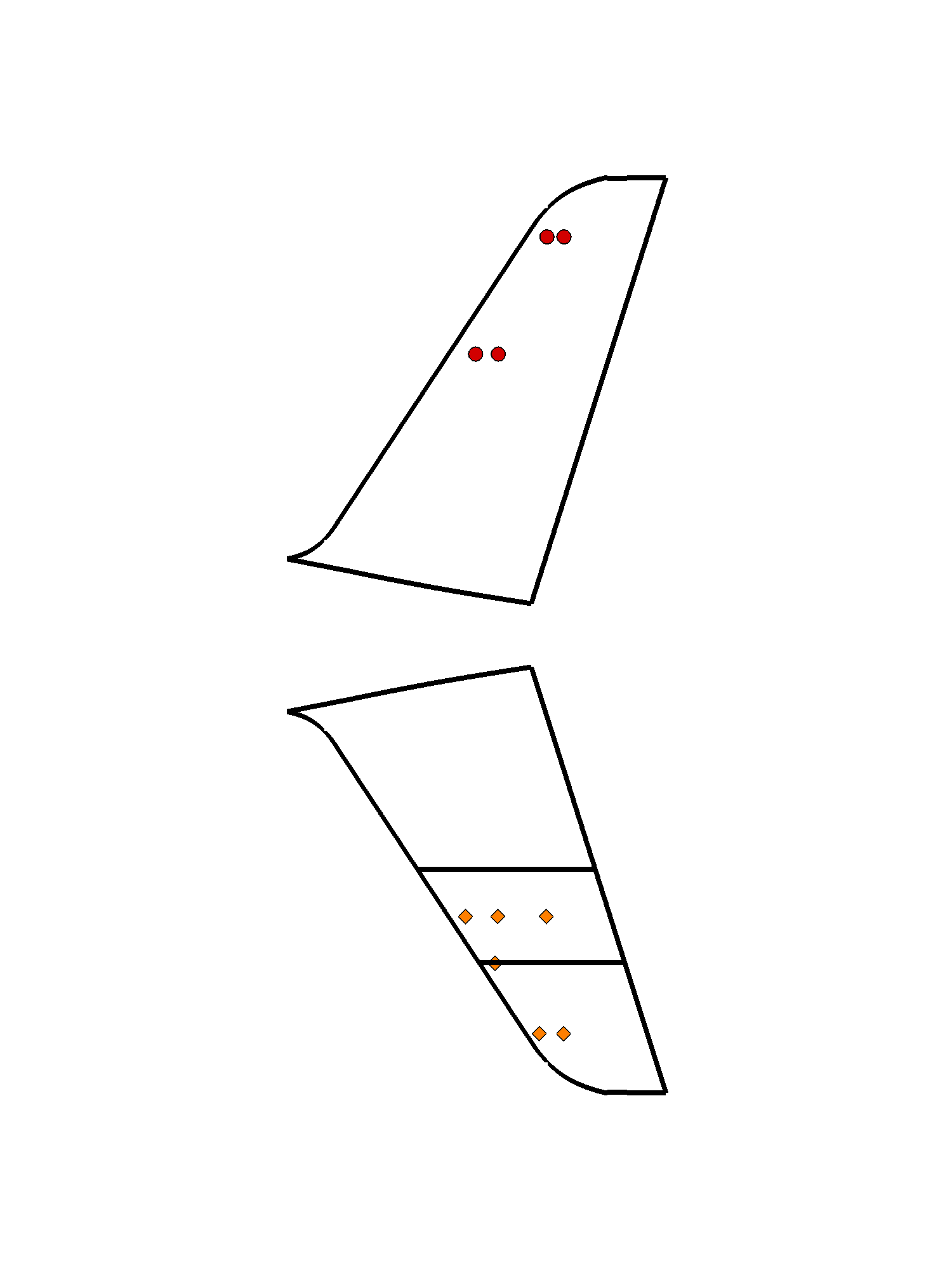}};

    \begin{scope}[x={(image.south east)},y={(image.north west)}]

\node at (0.2, 0.315) (eta07) {$\eta = 70\%$};
\draw[thick,dash dot] (eta07) -- (0.45, 0.315); 

\node at (0.2, 0.24) (eta05) {$\eta = 50\%$};
\draw[thick,dash dot] (eta05) -- (0.49, 0.24); 

    \end{scope}
\end{tikzpicture} 
  \caption{HTP upper and lower side}
  \label{f:cp_taps_HTP}
\end{subfigure}

  \caption{Static pressure tappings and dynamic pressure transducers on the wing and tailplane surfaces. Black dots: static taps, red circles: transducers on starboard wing upper surface, orange diamonds: transducers on port wing lower surface.}
  \label{f:cp_taps_wing_HTP}
\end{figure*}
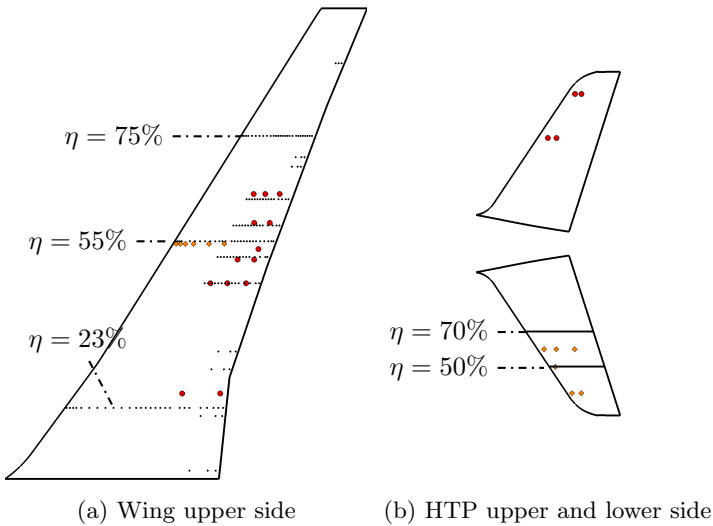

The wing and tailplane were heavily instrumented with a total of 317 pressure tappings on the wing and 50 on the horizontal tailplane, which were active during both the continuous traverse polars and the pitch/pause polars. Full-chord pressure sensor rows were at spanwise positions $\eta = 23.3\%, 55\%, 75.1\%$ on the wing, as indicated in Fig.~\ref{f:cp_taps_wing}. The figure also shows several sets of taps covering only a fraction of the chord which are located at other spanwise positions, with the rear half of the chord on the upper side covered at mid-wing at $\eta = 47\%$, $\eta = 52\%$, $\eta = 58\%$ and $\eta = 63\%$.

The horizontal tailplane shown in Fig.~\ref{f:cp_taps_HTP} is instrumented at two spanwise positions at $\eta_{HTP} = 50\%$ and $\eta_{HTP} = 70\%$. In contrast to the wing, the chordwise density of pressure tappings is higher on the lower side of the taiplane. An additional 5 tappings on the fuselage surface were also recorded.

The model and tunnel wall pressure data channels were acquired at a sampling rate of \SI{20}{\kilo\hertz}. The system recorded a data point one time per second, with the value of a data point computed from the mean over 35 samples.
\subsubsection{Dynamic Data}

Dynamic data consisted mainly of unsteady pressure transducers on the wing and tailplane surfaces. Their locations are depicted in Fig.~\ref{f:cp_taps_wing_HTP}. Most of the transducers on the wing are distributed on the wing upper side and at mid wing at the rear half of the chord. The wing upper side pressure transducers are located on the port wing at $\eta = 26\%, 47\%, 51.4\%, 54.4\%, 58.4\%, 63.8\%$. The sensors are intended to capture the separation characteristics at high angles of attack. There is one additional row of unsteady pressure sensors on the lower side near the wing leading edge at $\eta = 54.4\%$. The horizontal tailplane houses ten additional sensors, with six of them placed on the lower side. The HTP sensors are concentrated near the leading edge, which is motivated by the expectation that high angles of attack cause wing wake turbulence to impinge first near the lower side HTP tip due to the geometrical arrangement of the aircraft.

Several accelerometers were placed inside the model. Devices measuring the acceleration in the model's $z$ direction are located in the nose, the fuselage rear and the port and starboard outboard flap track fairings beneath the wings. In addition, a three-component accelerometer is located in the balance providing separate signals for each direction. Several unsteady sensors placed throughout the tunnel's aerodynamic circuit are also part of the delivered data.

The high speed data acquisition system records the above signals at a sampling rate of \SI{10000}{\hertz}, synchronized with a common time stamp. The lengths of the signals was driven by the requirements of the PSP systems, with the pressure transducers active during each run. The raw signals analyzed in this paper were checked for stationarity and cropped accordingly. Signals acquired during the continuous pitch traverse polars are mainly used to compute rolling standard deviation of signal values over $\alpha$. Sliding window sizes between 1000 and 5000 samples are used in the results discussion below, which corresponds to \SIrange{0.1}{0.5}{\second} or $\Delta \alpha = $\SIrange{0.012}{0.06}{\degree\per\second} at the pitch traversal rate of \SI{0.12}{\degree\per\second}. This approach averages much of the random variance at the cost of precision, permitting qualitative interpretation of unsteadiness magnitude over $\alpha$.

Signals acquired during fixed incidence measurements are used for spectral estimation. Welch's method~\cite{welch:1967} is used throughout this work, with the number of segments varying with the signal length. 50\% overlap and a Hann window is used in each case.

\subsubsection{Pressure Sensitive Paint Measurements}
\label{chp:psp}

Steady and unsteady PSP (pressure sensitive paint) measurements were conducted on the wing and the horizontal tailplane surfaces simultaneously. Two pairs of the steady PSP and the unsteady PSP systems for both wing and HTP view were installed on the wind tunnel bottom wall. All PSP measurements were conducted in a pitch/pause mode, i.e. at fixed incidences. The wing/HTP upper or lower surface were measured by the same PSP setup by rotating the model about the roll axis to a fin-up and a fin-down orientation. The static and unsteady pressure sensors installed on the wing and HTP were used for the direct comparison
of the steady and unsteady PSP data.

Both the wing and the HTP were coated with a thin layer of a paint which changes its emitted luminescent intensity and lifetime under ultraviolet (UV) light depending on the local pressure. This PSP coating can be used for both steady and unsteady PSP measurement (\cite{yorita:2017, klein:2022}). The thickness of the paint was below \SI{5}{\micro\meter}. The PSP coating was applied on the entire surface of the wing and HTP except on the flap track fairing and the row of pressure sensors. The steady PSP measurement was based on the lifetime method with a gated CCD sensor camera and pulsed UV light emitting diode (LED) units. A data acquisition time was about 5 seconds per one data point. Details on the PSP system used to capture steady pressure distributions can be obtained from (\cite{yorita:2018}).

The unsteady PSP measurement was based on the intensity method with a high-speed CMOS sensor camera and constant-power LED units (\cite{klein:2022}). The unsteady PSP system allows to capture a time-resolved PSP data with a sampling frequency of up to \SI{2}{\kilo\hertz} depending on the test conditions. A sampling rate of \SI{2}{\kilo\hertz} was used for tests conducted at temperatures of \SI{180}{\kelvin} and above, while \SI{1}{\kilo\hertz} sampling rate was used at \SI{115}{\kelvin} because of lower PSP light intensity at lower temperature. The camera memory permits storage of 21840 images. The number of acquired camera images per data point was set to 3640, 4368 or 5460, driven by data point prioritization and wind tunnel efficiency considerations. A camera record timing was recorded in the same data acquisition system of the pressure transducers.

The steady and unsteady PSP data were in-situ calibrated to pressure by the static and unsteady pressure sensors, respectively. The steady PSP measurement can capture the time-averaged pressure data and the unsteady PSP measurement can capture the temporal pressure variation (amplitude) data. Finally, the time-resolved absolute pressures can be reconstructed by superposition of the steady PSP data and the unsteady PSP data in data post-processing. %(CITATION, Yorita, FOR Symposium).

\subsubsection{Model Deformation}

Aeroelastic deformation of the main wing was measured using a stereo pattern tracking (SPT) system. Cameras tracking the displacement of 40 markers placed on the lower surface of the port wing were installed in the top wall. Two markers are placed at each spanwise position, near the leading edge and near the trailing edge, respectively. Comparison of recorded positions with reference positions permits deriving the twist and bend of the wing. The data was acquired during the continuous polars for all configurations. The data is transformed to bend and twist referred to the 50\% wing chord line.

\section{General Overview of Inflow Condition Effects}\label{results}

While much of the research initiative's efforts focus on envelope boundaries at incidences far from the design range, the acquired dataset permits a comprehensive characterization of the baseline XRF-1 geometry in a wing-body-tail configuration with a clean wing. The ability to vary $M$, $Re$ and $q$ independently in a cryogenic, pressurized environment enables insights into effects of these parameters isolated from each other. 

The test was conducted at Mach numbers of $0.84$, $0.87$ and $0.90$. The Reynolds number was varied between $3.3\cdot 10^6$ at ambient temperature and $12.9\cdot 10^6$ and $25.0 \cdot 10^6$ at cryogenic conditions down to $T = \SI{110}{\kelvin}$. Two levels of dynamic pressure $q/E$ were used, with the data at the intermediate Reynolds number of $12.9\cdot 10^6$ acquired at both $q/E = 0.2 \cdot 10^{-6}$ and $q/E = 0.4 \cdot 10^{-6}$. Since the unsteady PSP measurement system requires data from a steady PSP measurement for reference, each data point at which unsteady measurements were carried out included a prior steady-state PSP measurement point.

\begin{figure}[h]
\begin{center}
\begin{subfigure}{0.49\textwidth}
\centering
\includegraphics[width=1\linewidth]{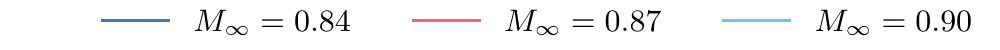}
\end{subfigure}
\end{center}
  \centering
     \includegraphics[width=0.49\linewidth]{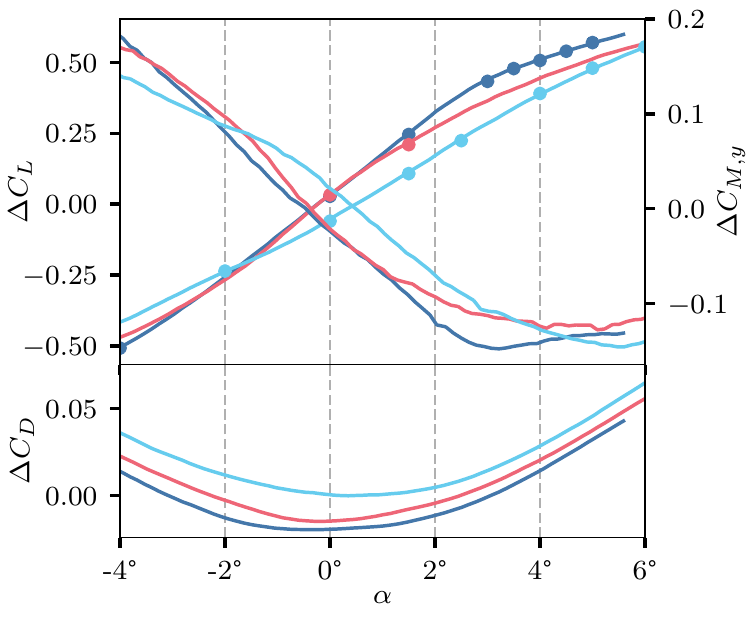}
    \caption{Force balance data at $Re_{\infty}=25.0 \cdot 10^6$. Incidences measured using PSP are highlighted. Values represent increments with respect to a reference point.}.
   \label{f:polars_Re12}
\end{figure}

Polars for all three Mach numbers are shown in Fig.~\ref{f:polars_Re12} for $Re_{\infty}=25.0 \cdot 10^6$, with the and the points acquired using PSP marked with symbols. The lift coefficient shows a distinct linear region at $M_{\infty} = 0.84$, which becomes less pronounced at higher Mach numbers. In this incidence range, the lift gradient $\partial C_L / \partial \alpha$ is highest at $M_{\infty} = 0.84$ and decreases toward higher Mach numbers. The $C_{My}$ slope magnitude is also greatest at the lowest Mach number. In both the $C_L$ and $C_{My}$ curves, there is a clear change in slope around $\alpha = \SI{3}{\degree}$ at $M_{\infty} = 0.84$, which is less pronounced at the higher Mach numbers. The drag coefficient in the lower part of Fig.~\ref{f:polars_Re12} shows a distinct broad minimum, with a significant increase toward higher incidences. High Mach numbers shift the drag minimum slightly towards higher incidences. As expected, $C_D$ increases significantly across the board at higher $M_{\infty}$.

The goal of the PSP measurements was to characterize the flow phenomena from buffet onset until incidences beyond buffet. A commonly used buffet detection criterion based on the break in the lift coefficient curve was employed to achieve this. The method described by Lawson et al.~\cite{lawson} is based on finding the point of departure of the $C_L$ polar from a linear shape. The lift break is determined using the $\Delta \alpha$ criterion, in which a line is drawn parallel to the linear portion of the $C_L$ polar with an offset of $\Delta \alpha = 0.1^{\circ}$. The intersection of this line with the lift coefficient curve is considered the incidence of buffet onset. The shape of the polars in Fig.~\ref{f:polars_Re12} shows that this is a rough estimate, due to the lack of a distinct lift break at high $M_{\infty}$. For this reason, the PSP measurement points were distributed over a broader $\alpha$ range at high Mach number, and clustered closer around the lift break where it is detectable at lower $M_{\infty}$. Other popular buffet detection methods described by \cite{lawson}, such as determination of root mean square increase of a wing root strain gauge signal, were not applicable due to a lack of such as sensor.

The following discussion focuses on the moderate incidences range and on conditions near the design lift coefficient, where possible.

\subsection{Model Wing Deformation}

The geometry of a wing with positive sweep generating positive lift causes it to bend and twist when it undergoes aeroelastic deformation. The wing deflects upward and twists in a manner increasing washout, i.e. reducing the geometric incidence of the outboard part. A consequence of this is an alteration of the spanwise lift distribution compared to the wind-off wing shape. The magnitude of deformation scales with dynamic pressure $q$, which was varied independently of $M$ and $Re$. Before the impact of Mach and Reynolds numbers can be assessed, an analysis of the isolated variation of wing shape changes

\begin{figure}[!htb]
\begin{center}
\begin{subfigure}{1\textwidth}
\centering
\includegraphics[width=1\linewidth]{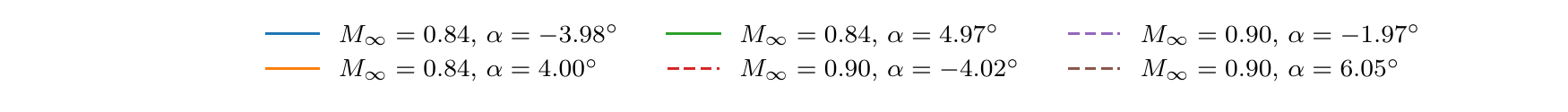}
\end{subfigure}
\end{center}
%\vspace*{-5mm}

\centering
\begin{subfigure}{.49\textwidth}
  \centering
  \includegraphics[width=1\linewidth]{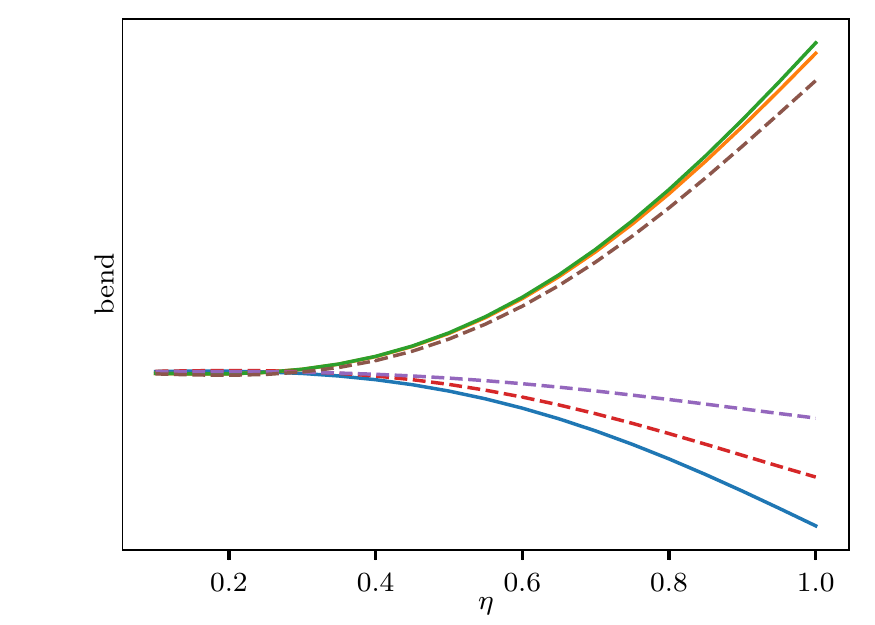}
  \caption{Bend}
  \label{f:M084_deform_bend}
\end{subfigure}
\begin{subfigure}{.49\textwidth}
  \centering
  \includegraphics[width=1\linewidth]{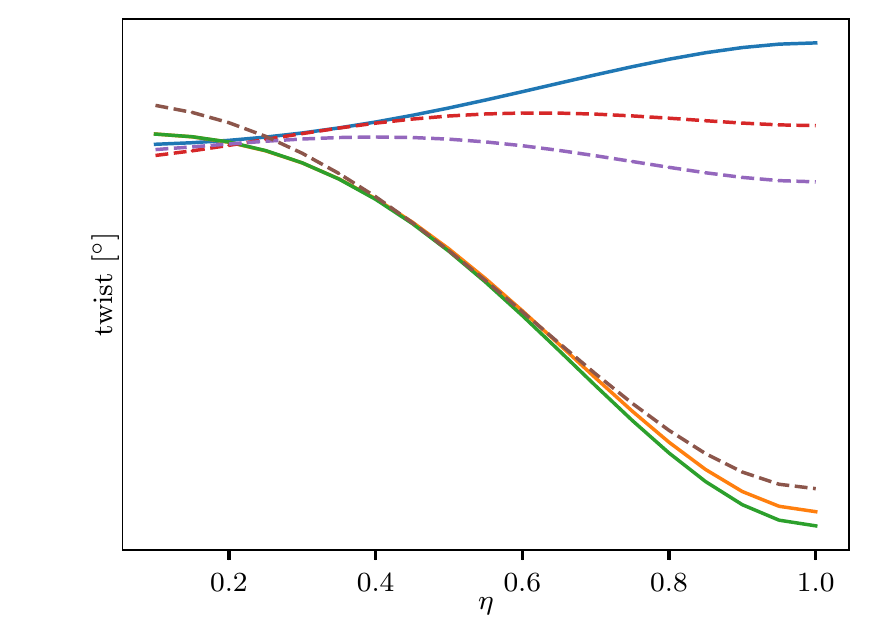}
  \caption{Twist}
  \label{f:M084_deform_twist}
\end{subfigure}
  \caption{Aeroelastic wing deformation at $q/E = 0.4 \cdot 10^{-6}$ ($Re_{\infty} = 12.9\cdot 10^6$) as spanwise distribution across the $\alpha$ range. Bend and twist are referred to the 50\% chord line, increments are depicted with respect to wind-off condition.}
  \label{f:M084_M090_deform_bend_twist}
\end{figure}

Typical shapes of bend and twist over the wing span are shown for $Re_{\infty} = 12.9 \cdot 10^6$ at $q/E = 0.4 \cdot 10^{-6}$ in Fig.~\ref{f:M084_M090_deform_bend_twist} over the $\alpha$ range. The bend and twist magnitudes at equal $\alpha$ are smaller for $M_{\infty} = 0.90$ than for $M_{\infty} = 0.84$. This is consistent with the differences in the lift polars in Fig.~\ref{f:polars_Re12}. The wing tip deflection ceases to increase at higher $\alpha$ for $M_{\infty} = 0.84$, which does not occur for $M_{\infty} = 0.90$ inside the measured $\alpha$ range.

The measured twist in Fig.~\ref{f:M084_deform_twist} is largely negative at positive $\alpha$, with the shapes again similar across the Mach numbers. The main difference is in the twist distribution at negative incidences, with $M_{\infty} = 0.84$ showing the largest magnitudes of twist at the tip, and $M_{\infty} = 0.90$ at about mid-span. The pressure distributions relating to this are shown further below in Section~\ref{chp:negative}.

\begin{figure}[!htb]
\begin{center}
\begin{subfigure}{1\textwidth}
\centering
\includegraphics[width=1\linewidth]{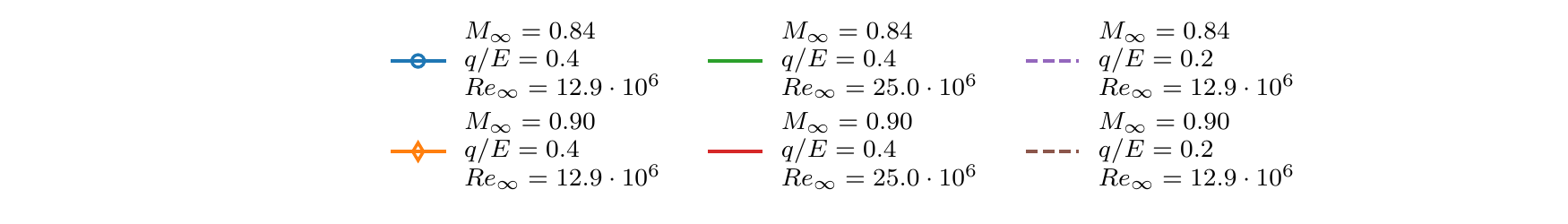}
\end{subfigure}
\end{center}
%\vspace*{-5mm}
\centering
\begin{subfigure}{.49\textwidth}
  \centering
  \includegraphics[width=1\linewidth]{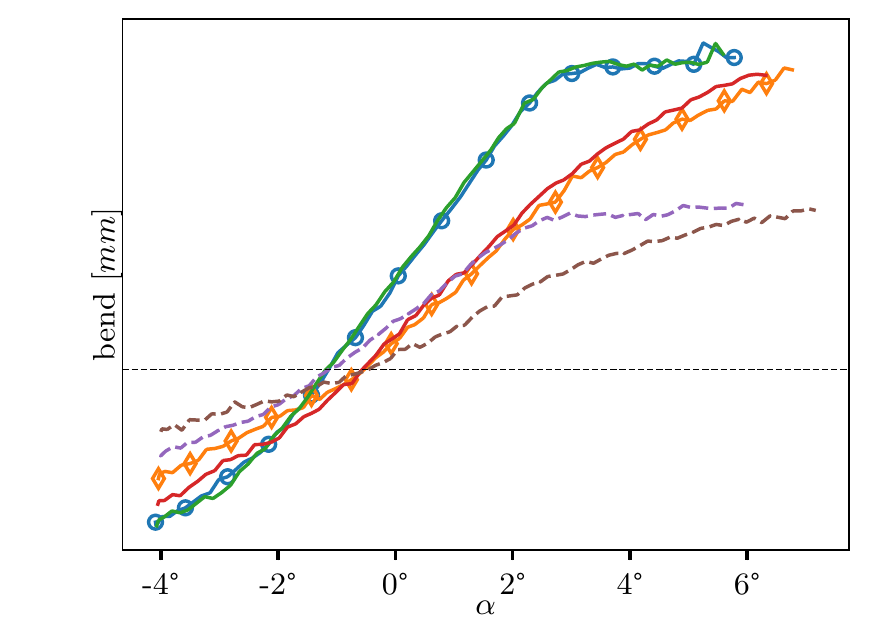}
  \caption{Bend}
  \label{f:M084_M090_deform_bend_eta075}
\end{subfigure}
\begin{subfigure}{.49\textwidth}
  \centering
  \includegraphics[width=1\linewidth]{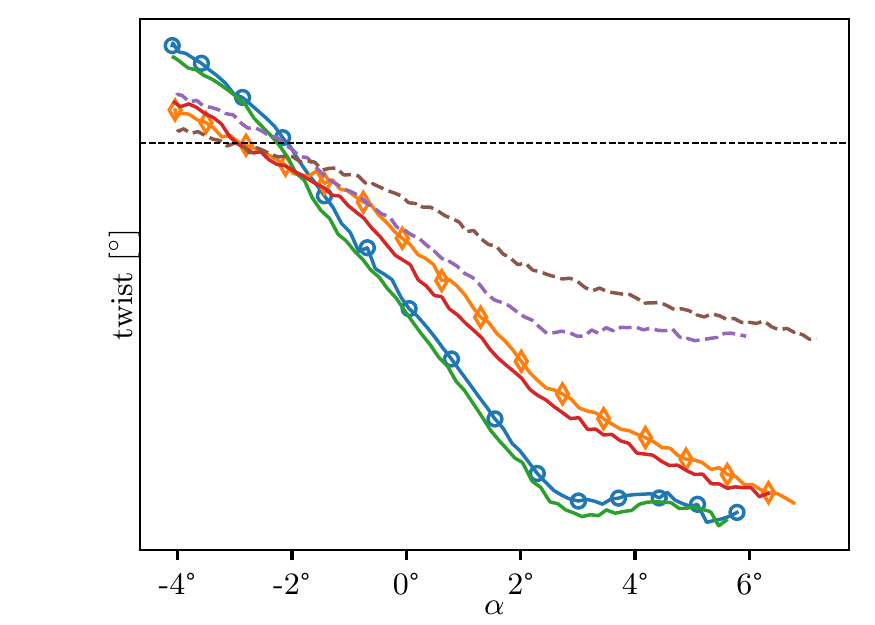}
  \caption{Twist}
  \label{f:M084_M090_deform_twist_eta075}
\end{subfigure}
  \caption{Aeroelastic wing deformation at the $\eta = 75\%$ wing station. Bend and twist are increments with respect to wind-off condition.}
  \label{f:M084_M090_bend_twist_eta075}
\end{figure}

The deformation scales roughly proportionally with lift, as evidenced by the bend and twist magnitude measured at $\eta = 75\%$ in Fig.~\ref{f:M084_M090_bend_twist_eta075}. In fact, the shapes of the bend deflections mirror those of $C_L$, albeit with a plateau at $M_{\infty} = 0.84$ indicating that maximum deflection is reached at about $\alpha = \SI{3}{\degree}$. There is no further increase beyond that $\alpha$, which is also true of the twist magnitude in Fig.~\ref{f:M084_M090_deform_twist_eta075}. The $M_{\infty} = 0.90$ deformation does not show such a clean break, and maximum deformation is attained at the highest measured incidence. Scaling of both bend and twist in terms of $q/E$ is nearly perfectly linear. The impact of $Re_{\infty}$ at constant $q/E = 0.4$ is a slight increase of twist at $Re_{\infty} = 25.0 \cdot 10^6$ compared to $Re_{\infty} = 12.9 \cdot 10^6$, suggesting higher local lift on the outboard wing at higher $Re_{\infty}$.

\subsection{Mach Number Effects}
\label{chp:mach_effect}

\begin{figure}[h]
\centering

\begin{subfigure}{.49\textwidth}
  \centering
  \includegraphics[width=1\linewidth]{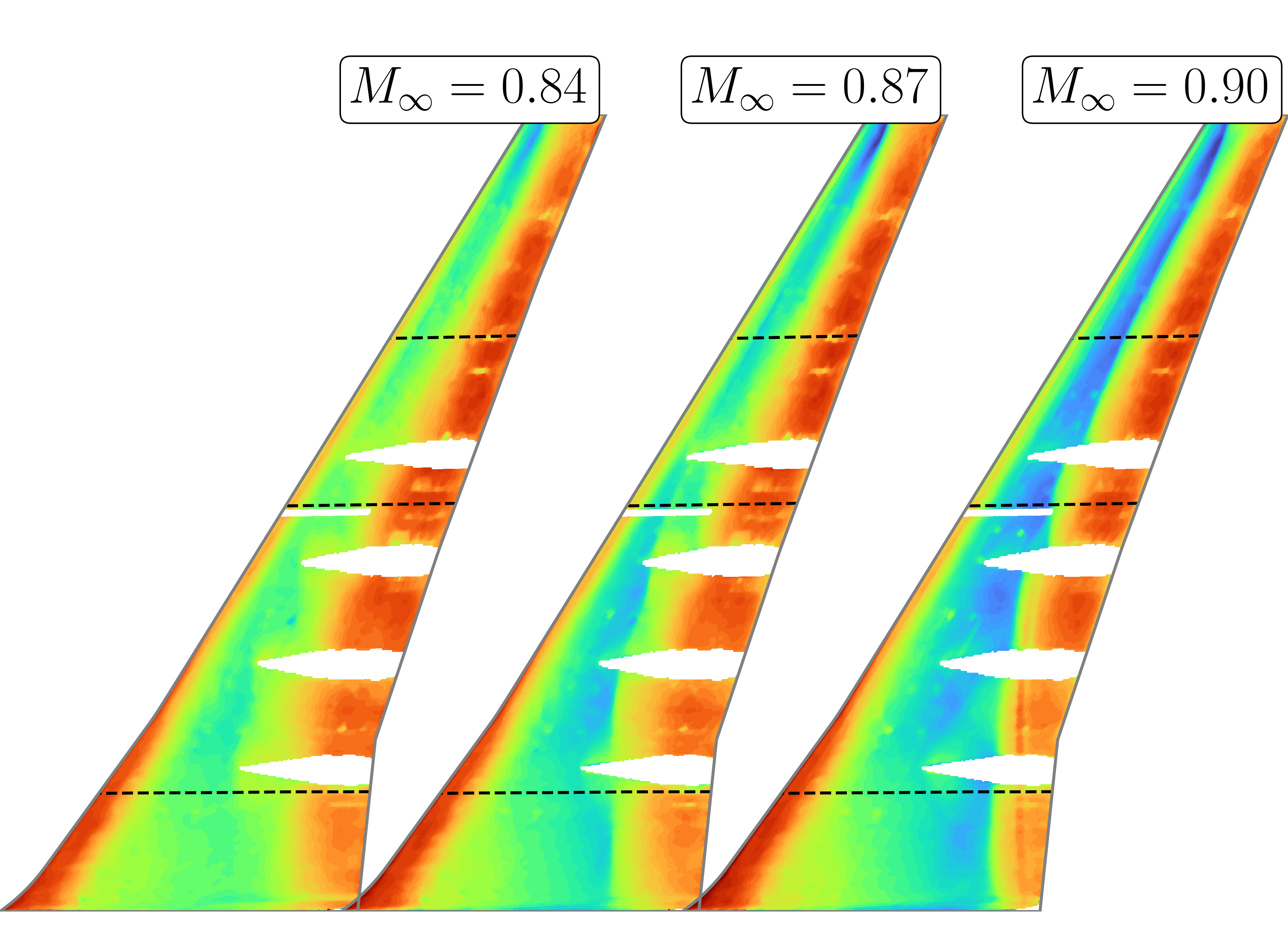}
  \caption{Wing lower side}
  \label{f:PSP_cpmean_wing_AoA15_lower}
\end{subfigure}
\begin{subfigure}{.49\textwidth}
  \centering
  \includegraphics[width=1\linewidth]{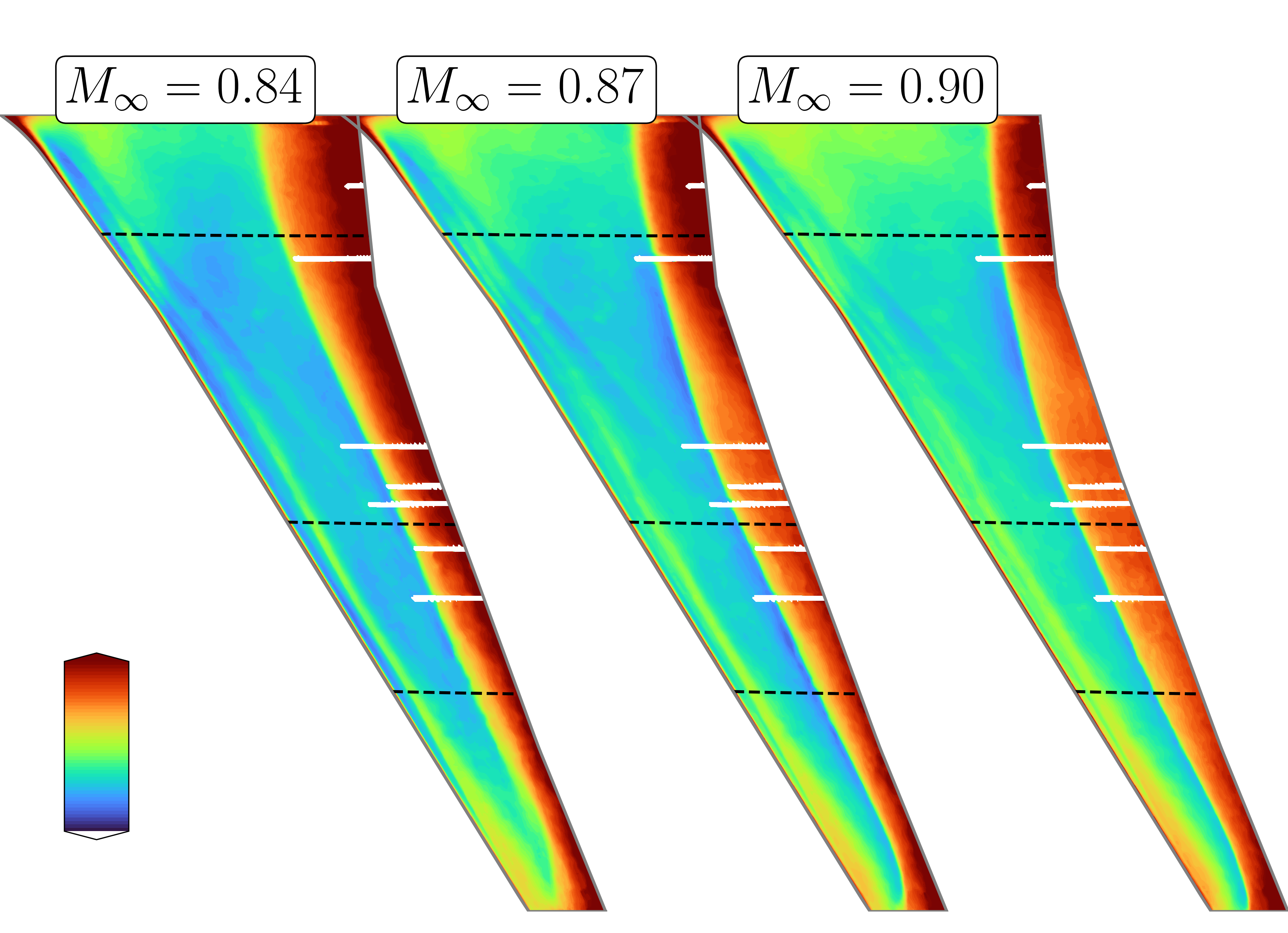}
\put(-148,10){\small{low}}
\put(-148,30){\small{high}}
\put(-160,40){\small{$c_p$}}
  \caption{Wing upper side}
  \label{f:PSP_cpmean_wing_AoA15_upper}
\end{subfigure}
\begin{subfigure}{.26\textwidth}
  \centering
    \scalebox{-1}[1]{\includegraphics[angle=-90,origin=c
,width=1\linewidth]{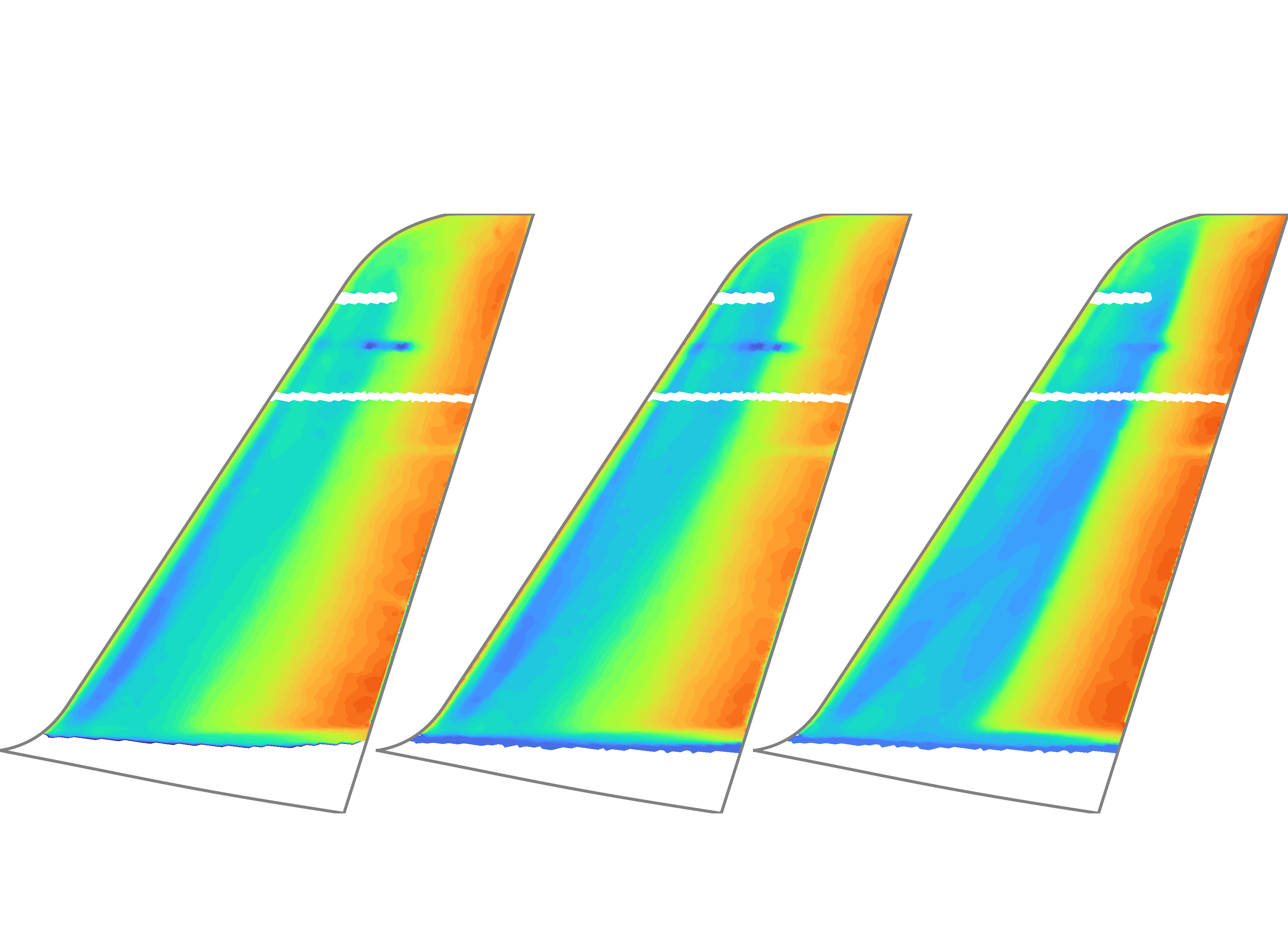}}
    \put(-10,10){\small{$M_{\infty}  = 0.90$}}
    \put(-10,50){\small{$M_{\infty}  = 0.87$}}
    \put(-10,90){\small{$M_{\infty}  = 0.84$}}
  \caption{HTP lower side}
  \label{f:PSP_cpmean_HTP_AoA15_lower}
\end{subfigure} \qquad
\begin{subfigure}{.26\textwidth}
  \centering
    \scalebox{-1}[1]{\includegraphics[angle=-90,origin=c
,width=1\linewidth]{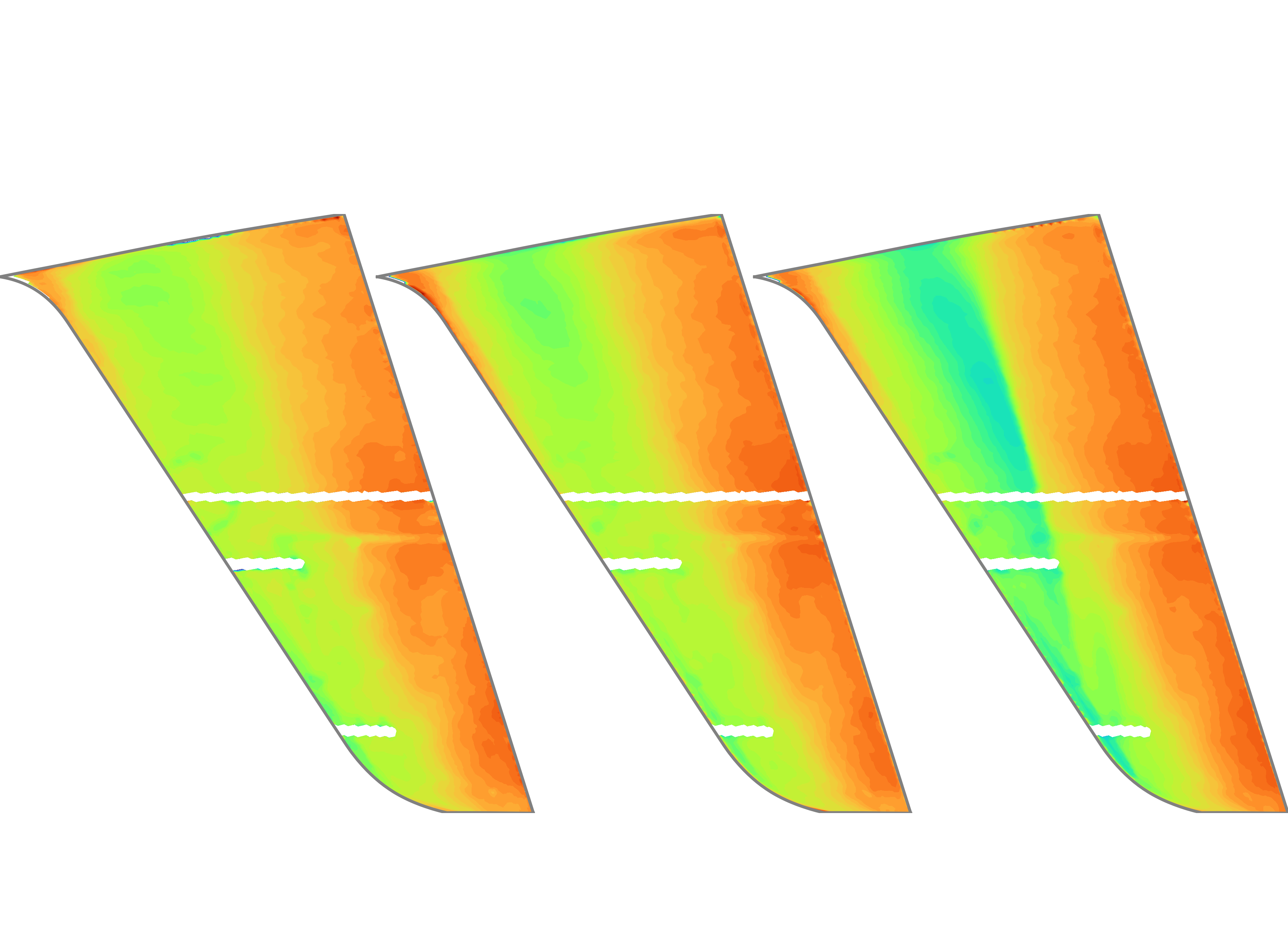}}
  \caption{HTP upper side}
  \label{f:PSP_cpmean_HTP_AoA15_upper}
\end{subfigure}
  \caption{Mean $c_p$ from PSP on the wing at $Re_{\infty} = 12.9\cdot 10^6$, $q/E = 0.4 \cdot 10^{-6}$, $\alpha = \SI{1.5}{\degree}$. The color ranges are equal across all datasets. HTP is enlarged relative to wing for clarity. Dashed lines denote the $c_p$ measurement locations $\eta = 23.3\%$, $\eta = 55\%$ and $\eta = 75.1\%$ on the wing.}
  \label{f:PSP_cpmean_wing_AoA15}
\end{figure}

Increased $M_{\infty}$ causes a successive deformation of the lift polar in Fig.~\ref{f:polars_Re12} and a departure from a clearly delineated shape with linear and non-linear regions. PSP measurements were acquired in the moderate incidence range at $\alpha = \SI{0}{\degree}$ and $\alpha = \SI{1.5}{\degree}$. The surface $c_p$ distributions in Fig.~\ref{f:PSP_cpmean_wing_AoA15} at $\alpha = \SI{1.5}{\degree}$ provide an impression of the flow at moderate positive lift coefficients. The upper side $c_p$ in Fig.~\ref{f:PSP_cpmean_wing_AoA15_upper} shows a two-shock pattern at all three conditions with a $\lambda$-like shape near the wing root. The main shock is visible over the rear part of the chord across the entire span, whereas a weaker oblique shock is present over the inboard portion. It originates near the leading edge at the root, with a sweep angle growing with Mach number. This behavior is consistent with the explanation of the forward shock by \cite{rogers:1960}. Its origin is the flow accelerating over the leading edge and it propagates according to its Mach angle, which increases with $M_{\infty}$. Therefore the triple point at the intersection between the front and aft shocks moves inboard at increasing $M_{\infty}$.

\begin{figure}[!htb]
\begin{center}
\begin{subfigure}{1\textwidth}
\centering
\includegraphics[width=1\linewidth]{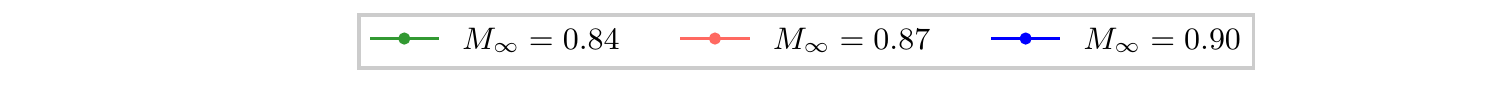}
\end{subfigure}
\end{center}
%\vspace*{-5mm}
\centering
\begin{subfigure}{.32\textwidth}
  \centering
  \includegraphics[width=1\linewidth]{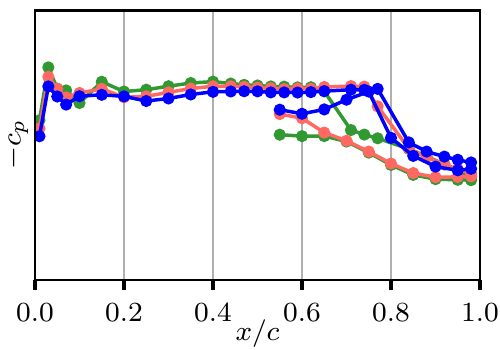}
  \caption{$\eta=23.3\%$}
  \label{f:cp_allMach_Re12_CL05_wing_23}
\end{subfigure}
\begin{subfigure}{.32\textwidth}
  \centering
  \includegraphics[width=1\linewidth]{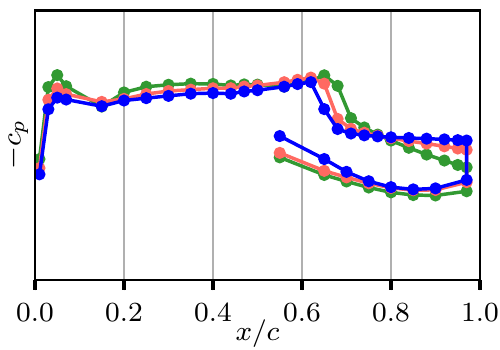}
  \caption{$\eta=55\%$}
  \label{f:cp_allMach_Re12_CL05_wing_55}
\end{subfigure}
\begin{subfigure}{.32\textwidth}
  \centering
  \includegraphics[width=1\linewidth]{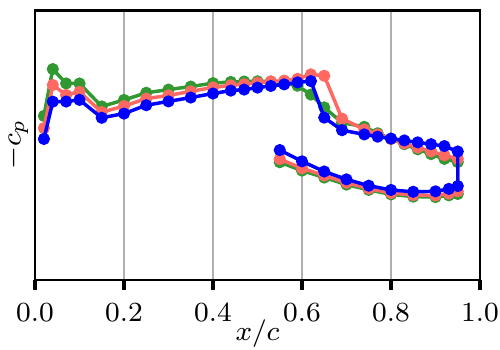}
  \caption{$\eta=75.1\%$}
  \label{f:cp_allMach_Re12_CL05_wing_75}
\end{subfigure}

  \caption{Wing static $c_p$ data at $Re_{\infty}=12.9 \cdot 10^6$ and all three Mach numbers at $\alpha = \SI{1.5}{\degree}$.}
  \label{f:cp_allMach_Re12_CL05_wing_HTP}
\end{figure}

Pressure recovery at mid-wing downstream of the shock differs between Mach, with the highest $c_p$ values achieved at trailing edge at $M_{\infty}$. Increasing Mach number causes a loss of pressure recovery and results in a more uniform $c_p$ distribution in the region between shock and trailing edge. The mid-wing pressure distributions in Fig.~\ref{f:cp_allMach_Re12_CL05_wing_HTP} underscore this. The chordwise $c_p$ gradient downstream from the shock at $\eta = 55\%$ decreases in magnitude, indicating separation in this region. Apart from that the $M_{\infty} = 0.84$ $c_p$ distribution exhibits a largely typical shape for a supercritical wing section at moderate lift coefficients, with a shock between $60\%$ and $80\%$ chord. The shock moves slightly upstream at increasing Mach. The smooth lower side $c_p$ increase seen at low Mach numbers gives way to a shock at $M_{\infty} = 0.90$. It is visible in PSP in Fig.~\ref{f:PSP_cpmean_wing_AoA15_lower} as well, where the smooth pressure gradient is replaced by a sharp change at high Mach number. The lower wing surface PSP data involves more masked regions due to the presence of the four flap track fairings. The fairings block the flow and deform the isobars, which results in disrupted shock shapes. The shock at high $M_{\infty}$ is visible over virtually the entire span, consistently with all three positions shown in Fig.~\ref{f:cp_allMach_Re12_CL05_wing_HTP}.

Considering the large differences between forces and the wing upper side pressure recovery between the Mach numbers, it is evident that comparison at constant $\alpha$ does not necessarily represent similar flow regimes. However, this data still constitutes a baseline for high $\alpha$ unsteady comparisons in the next chapter.

\begin{figure}[!htb]
\centering
  \includegraphics[width=0.32\linewidth]{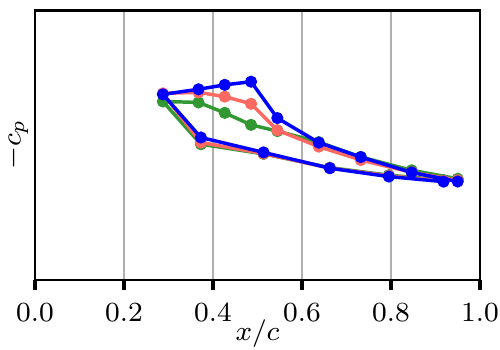}
  \caption{Tailplane static $c_p$ data at $\eta_{HTP} = 70\%$ at $Re_{\infty}=12.9 \cdot 10^6$ and all three Mach numbers at $\alpha = \SI{1.5}{\degree}$. Line legend is the same as in Fig.~\ref{f:cp_allMach_Re12_CL05_wing_HTP}.}
  \label{f:cp_allMach_Re12_CL05_HTP}
\end{figure}

The pressure data at the tailplane surfaces in Fig.~\ref{f:cp_allMach_Re12_CL05_wing_HTP} indicates low overall load on the tailplane, with the surface generating downforce at these conditions. There is little difference between on the upper side, and the pressure distributions are largely smooth. By contrast, the lower side data shows the appearance of a more significant adverse pressure gradient for $M_{\infty} = 0.90$. The lack of shocks and the overall low load on the tailplane justify the a priori choice of the tailplane incidence setting to $\SI{-2}{\degree}$.  

\subsection{Reynolds Number and Dynamic Pressure Effects}
\label{chp:re_effect}

\begin{figure}[!htb]
\begin{center}
\begin{subfigure}{1\textwidth}
\centering
\includegraphics[width=1\linewidth]{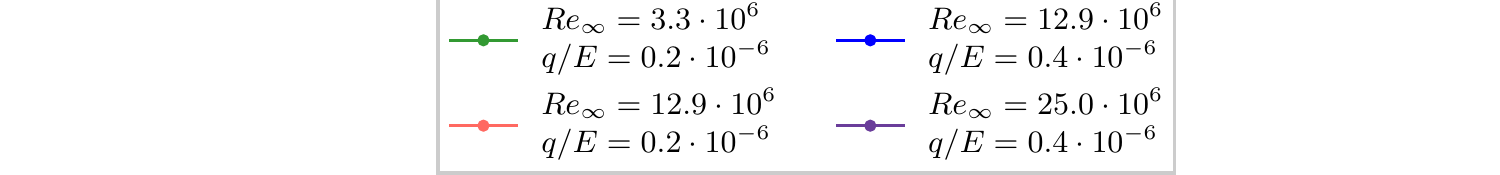}
\end{subfigure}
\end{center}
%\vspace*{-5mm}
\centering
\begin{subfigure}{.32\textwidth}
  \centering
  \includegraphics[width=1\linewidth]{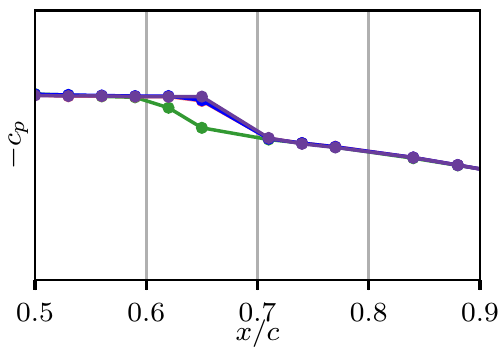}
  \caption{$\eta=23.3\%$}
  \label{f:cp_allRe_M084_CL05_wing_23}
\end{subfigure}
\begin{subfigure}{.32\textwidth}
  \centering
  \includegraphics[width=1\linewidth]{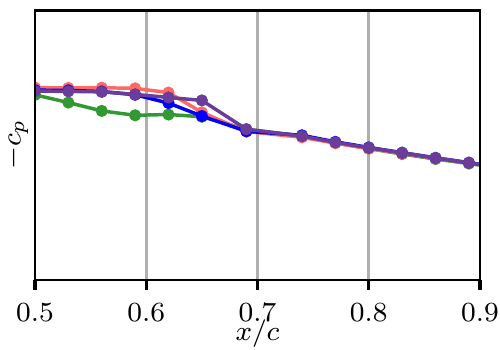}
  \caption{$\eta=75.1\%$}
  \label{f:cp_allRe_M084_CL05_wing_75}
\end{subfigure}

  \caption{Wing upper side static $c_p$ data at $M_{\infty}=0.84$ and all three Reynolds numbers at $\alpha = \SI{1.5}{\degree}$.}
  \label{f:cp_allRe_M084_CL05_wing_HTP}
\end{figure}

\begin{figure}[!htb]

\centering
\begin{subfigure}{.32\textwidth}
  \centering
  \includegraphics[width=1\linewidth]{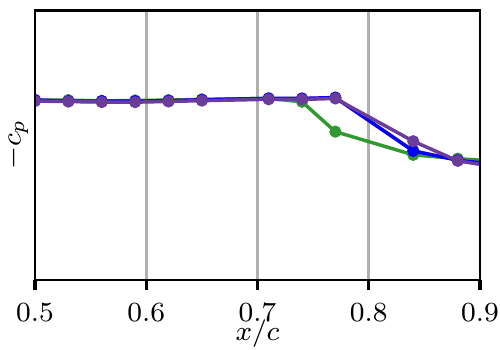}
  \caption{$\eta=23.3\%$}
  \label{f:cp_allRe_M090_CL05_wing_23}
\end{subfigure}
\begin{subfigure}{.32\textwidth}
  \centering
  \includegraphics[width=1\linewidth]{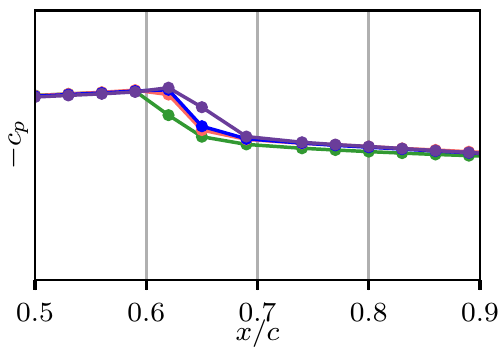}
  \caption{$\eta=75.1\%$}
  \label{f:cp_allRe_M090_CL05_wing_75}
\end{subfigure}
  \caption{Wing upper side static $c_p$ data at $M_{\infty}=0.90$ and all three Reynolds numbers at $\alpha = \SI{1.5}{\degree}$. Line legend is the same as in Fig.~\ref{f:cp_allRe_M084_CL05_wing_HTP}.}
  \label{f:cp_allRe_M090_CL05_wing_HTP}
\end{figure}

The effects of Reynolds number and $q/E$ on the forces and wing pressures are more subtle than those of the Mach number. While the pressure distributions at constant $\alpha$ differ greatly due to changing Mach number in Fig.~\ref{f:cp_allMach_Re12_CL05_wing_HTP}, the changes due to $Re_{\infty}$ and $q/E$ in Figs.~\ref{f:cp_allRe_M084_CL05_wing_HTP} and ~\ref{f:cp_allRe_M090_CL05_wing_HTP} are much less significant. $Re_{\infty} = 12.9\cdot 10^6$ was measured twice at two different dynamic pressure settings, enabling both the comparison of a $q/E$ effect at constant $Re_{\infty}$, and of a Reynolds number effect at both high and low levels of $q/E$.

The thickening boundary layer and lower effective camber at low Reynolds number conditions such as at $Re_{\infty} = 3.3 \cdot 10^6$ tends to move the shock upstream. This is most evident inboard in Figs.~\ref{f:cp_allRe_M084_CL05_wing_23} and \ref{f:cp_allRe_M090_CL05_wing_23}. There is no visible difference due to $q/E$ at that position, which is consistent with the small deformation magnitude close to the wing root shown in Fig.~\ref{f:M084_M090_deform_bend_twist}. Therefore the in shock position differences at $\eta = 23.3\%$ can be attributed to the Reynolds number, with the two runs at $Re_{\infty} = 12.9 \cdot 10^6$ yielding identical pressure distributions. There is a significant difference in shock position between $Re_{\infty} = 3.3 \cdot 10^6$ and $Re_{\infty} = 12.9 \cdot 10^6$ at both Mach numbers. The change in boundary layer thickness between $Re_{\infty} = 12.9 \cdot 10^6$ and $Re_{\infty} = 25.0 \cdot 10^6$ appears insufficient to significantly impact the shock position. The difference in effective camber resulting from Reynolds number variation scales with $Re^{-1/5}$ following \cite{agard303}, which reduces the relative sensitivity at higher values of $Re_{\infty}$.

As deformation becomes more important toward the tip due to reduction of effective incidence, the effects of $Re_{\infty}$ and $q/E$ are compounded. Low $Re$ and low $q/E$ both cause an upstream shift of the shock, the former through increased boundary layer thickness and the latter through less washout and therefore higher local wing section incidence. The chordwise shift of the shock position is slightly more pronounced in Fig.~\ref{f:cp_allRe_M084_CL05_wing_75} at $M_{\infty} = 0.84$ than at the inboard location. In addition, the Reynolds number effect itself is significantly greater than at $\eta = 23.3\%$, with differences noticeable between all values of $Re_{\infty}$. The $c_p$ differences on the tailplane are of negligilble magnitude and are therefore omitted.

\section{Wing Buffet and Separation at Different $M_{\infty}$ and $Re_{\infty}$}

Having focuced on moderate incidences and time-averaged data in the previous section, the high $\alpha$ region where buffet is expected is discussed in the following. Data acquired using time-resolved PSP at the sampling points shown in Fig.~\ref{f:polars_Re12} helps shed light onto sensitivity of the flow with respect to $\alpha$ and $M_{\infty}$ at high Reynolds numbers.

The polars in Fig.~\ref{f:polars_Re12} indicate that the departure from linear region differs significantly between the Mach numbers. The $C_L$ polar at $M_{\infty} = 0.84$ has a fairly pronounced linear region, whereas $M_{\infty} = 0.90$ does not. As the PSP incidences were determined a priori, they were spaced over a wider range of $\alpha$ values at $M_{\infty} = 0.90$ in order to ensure that some amount of unsteadiness can be captured. The PSP incidences at $M_{\infty} = 0.84$ are grouped closer together. 

Apart from the polar's curvature, $M_{\infty} = 0.90$ exhibits lower $C_L$ across most of the $\alpha$ range. Significant unsteadiness tends to occur at higher incidences. The nature of the unsteadiness is markedly different between the two Mach numbers, meriting a detailed discussion of the respective results. The following chapter is focused on the discussion of shock behavior and its sensitivity to Mach number, Reynolds number, dynamic pressure, and angle of attack.

\subsection{Surface Shock Pattern}

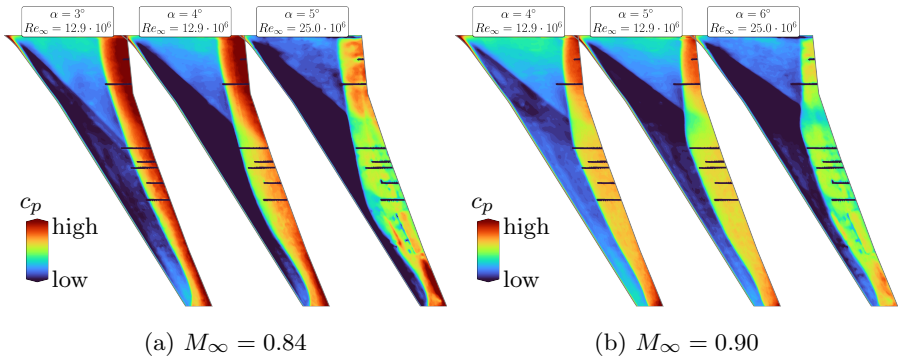
\begin{figure}[htb]
\centering
\begin{subfigure}{.49\textwidth}
  \centering
  \includegraphics[width=1\linewidth]{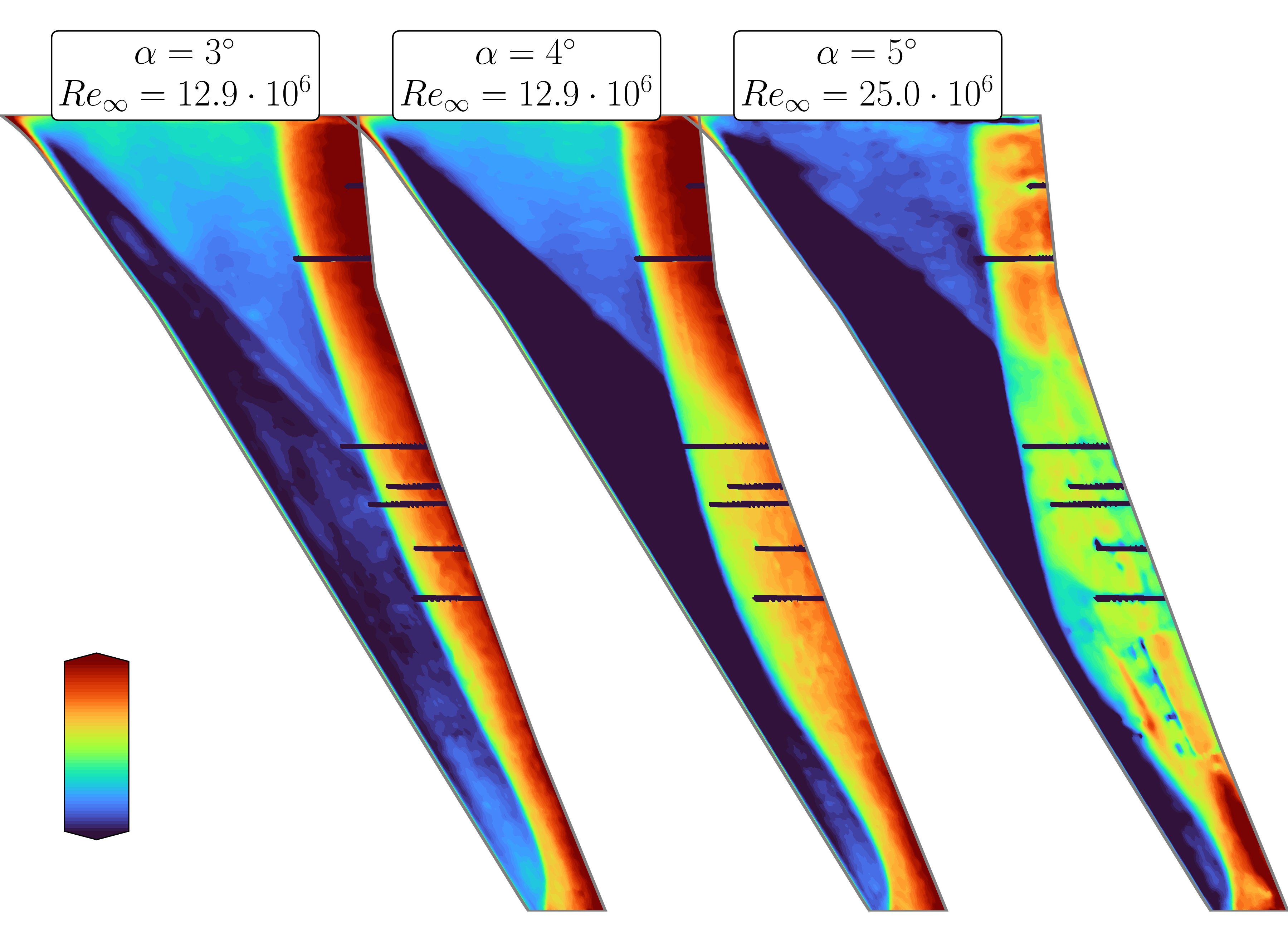}
  \put(-148,10){\small{low}}
\put(-148,30){\small{high}}
\put(-160,40){\small{$c_p$}}
  \caption{$M_{\infty} = 0.84$}
  \label{f:PSP_cpmean_wing_M084}
\end{subfigure}
\begin{subfigure}{.49\textwidth}
  \centering
\includegraphics[width=1\linewidth]{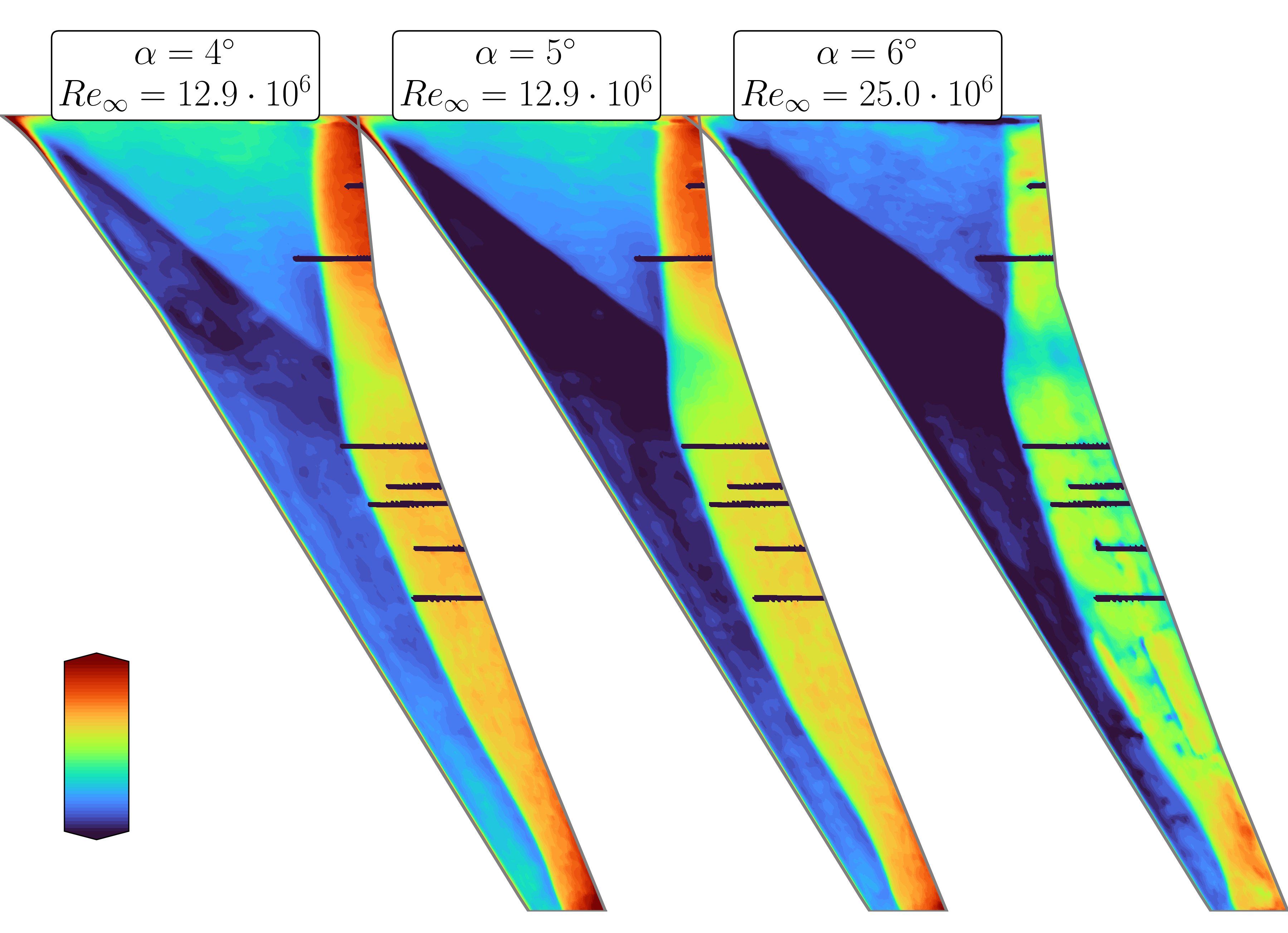}
\put(-148,10){\small{low}}
\put(-148,30){\small{high}}
\put(-160,40){\small{$c_p$}}
  \caption{$M_{\infty} = 0.90$}
  \label{f:PSP_cpmean_wing_M090}
\end{subfigure}

  \caption{Mean $c_p$ from PSP on the upper side of the wing at high $\alpha$. Data is shown at $Re_{\infty}=12.9 \cdot 10^6$ due to better image quality, with the highest angles of attack only available at $Re_{\infty}=25.0 \cdot 10^6$. The color ranges are equal across all datasets.}
  \label{f:PSP_cpmean_wing_upper_allMach_Re12_aL2_aB}
\end{figure}

Wing upper surface $c_p$ distributions in Fig.~\ref{f:PSP_cpmean_wing_upper_allMach_Re12_aL2_aB} provide an overview of how the effects discussed in Section~\ref{chp:mach_effect} are distributed. $\alpha = $ \SIlist{3;4;5}{\degree} are shown for $M_{\infty} = 0.84$, while data for $M_{\infty} = 0.90$ is available at higher incidences at $\alpha = $ \SIlist{4;5;6}{\degree}. Where possible, data is shown for $Re_{\infty}=12.9 \cdot 10^6$ due to superior image clarity at the higher test temperature during these measurements. The highest incidence was acquired only at $Re_{\infty}=25.0 \cdot 10^6$ and is shown in the rightmost panel at both Mach numbers. The higher amount of noise in these images is due to the lower wind tunnel temperature of \SI{115}{\kelvin} as opposed to the \SI{180}{\kelvin} at the lower Reynolds number. Low temperature decreases the signal to noise ratio of the PSP acquisition system, resulting in less clear data. Such data artifacts are visible on the outboard wing at the highest $\alpha$ at both Mach numbers.

The data points in Fig.~\ref{f:PSP_cpmean_wing_upper_allMach_Re12_aL2_aB} showcase the effect of increasing incidence, which is significantly larger compared to the Reynolds number effect. As a general rule, the shock position inboard of the crank is significantly farther downstream at $M_{\infty} = 0.90$ than at the lower Mach number. The faintly visible oblique shock originating at the wing root leading edge has a noticeably higher sweep angle at higher Mach number. Higher angles of attack at constant $M_{\infty}$ generally alter the shock pattern in a similar manner to an increase of the Mach number: the inboard shock position slightly shifts aft, while the mid-span and outboard shock moves upstream. The inboard oblique shock increases its sweep angle, and the intersection between the outboard normal shock and the oblique front shock from the wing root moves inboard. \cite{sugioka:2018} observed similar effects on the Common Research Model (CRM) configuration's wing using similar techniques.

\begin{figure}[htb]
\centering
\begin{subfigure}{.49\textwidth}
  \centering
  \includegraphics[width=1\linewidth]{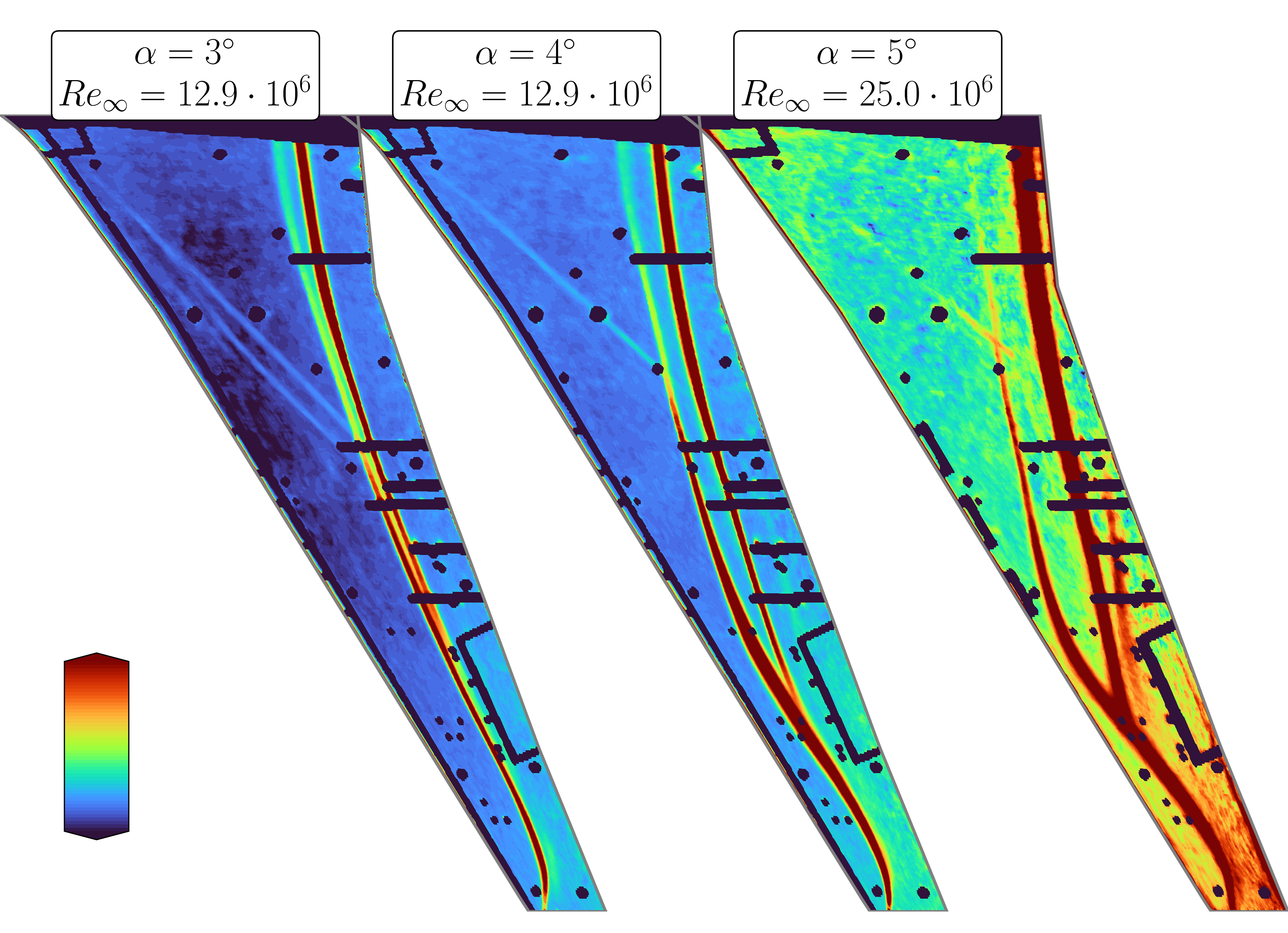}
  \put(-148,10){\small{low}}
\put(-148,30){\small{high}}
\put(-160,40){\small{$c_p$}}
  \caption{$M_{\infty} = 0.84$}
  \label{f:PSP_cpmean_wing_M084_Re12_aL2}
\end{subfigure}
\begin{subfigure}{.49\textwidth}
  \centering
\includegraphics[width=1\linewidth]{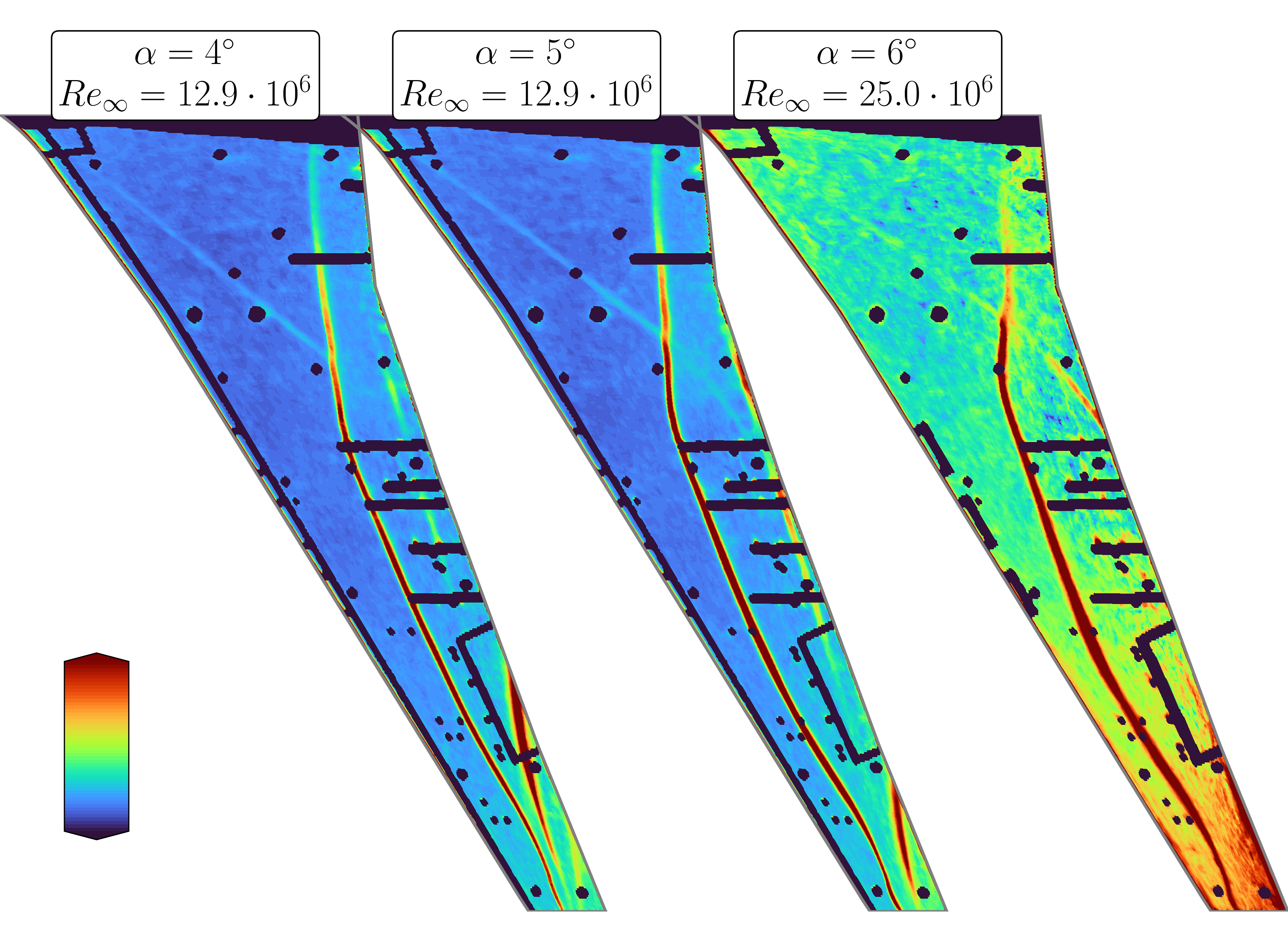}
\put(-148,10){\small{low}}
\put(-148,30){\small{high}}
\put(-160,40){\small{$c_p$}}
  \caption{$M_{\infty} = 0.90$}
  \label{f:PSP_cpmean_wing_M090_Re12_aL2}
\end{subfigure}

  \caption{Pointwise root mean square of $c_p$ via unsteady PSP on the upper side of the wing at high $\alpha$. Data is shown at $Re_{\infty}=12.9 \cdot 10^6$ due to better image quality, with the highest angles of attack only available at $Re_{\infty}=25.0 \cdot 10^6$.}
  \label{f:PSP_cprms_wing_upper_M090_allRe_aL2_aB}
\end{figure}
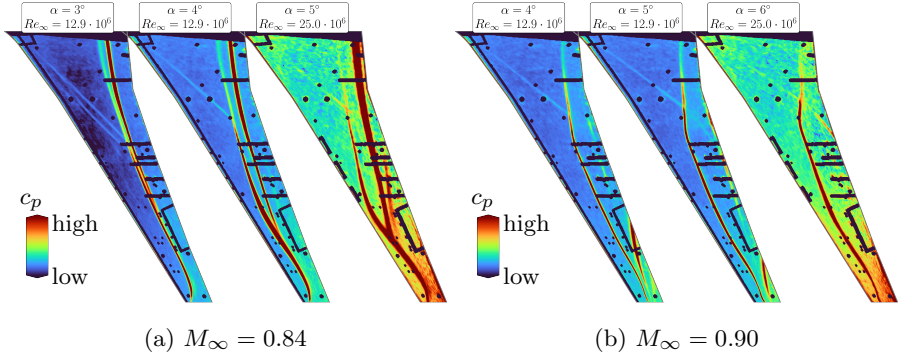

The root mean square (RMS) distributions of $c_p$ in Fig.~\ref{f:PSP_cprms_wing_upper_M090_allRe_aL2_aB} were obtained by computing the RMS value of $c_p$ for each pixel from the time series acquired by the PSP camera for unsteady measurement. The distributions provide information on the unsteadiness related to the shock. There are bands of elevated RMS values which are higher than their surroundings near the shock, consistent with the shock locations in Fig.~\ref{f:PSP_cpmean_wing_upper_allMach_Re12_aL2_aB}. These areas represent the shock and its unsteady motion over time at constant incidence. A shock which is only slightly displaced over time can cause large jumps in intensity at a given pixel of the camera sensor, leaving a large RMS footprint. A second band with higher rms values over much of the inboard wing in Fig.~\ref{f:PSP_cpmean_wing_M084_Re12_aL2} has been identified to be an optical error arising from the camera line of sight passing through the shock above the wing surface. The same phenomenon occurs in Fig.~\ref{f:PSP_cpmean_wing_M090_Re12_aL2} near the wing tip.

Most of the shock displacement due to $\alpha$ occurs on the outboard wing surface. At $M_{\infty} = 0.84$, the shock advances to $20~\%$ of the chord around $70~\%$ span at $\alpha = \SI{5}{\degree}$, with significant spanwise curvature of the shock footprint. The shock at $M_{\infty} = 0.90$ at high $\alpha$ maintains a more straight shape and moves upstream in a more uniform manner. However, it does so over a larger portion of the span. The maximum displacement magnitude at $M_{\infty}=0.90$ is smaller than at $M_{\infty}=0.84$.

\begin{figure}[!htb]
\begin{center}
\begin{subfigure}{1\textwidth}
\centering
\includegraphics[width=1\linewidth]{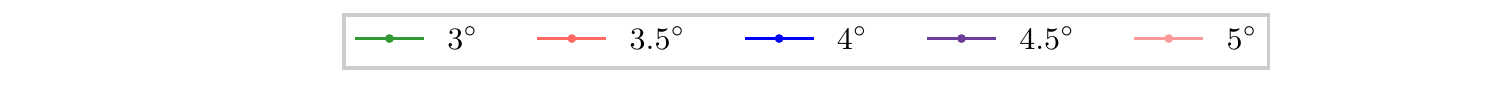}
\end{subfigure}
\end{center}
\centering
\begin{subfigure}{.32\textwidth}
  \centering
  \includegraphics[width=1\linewidth]{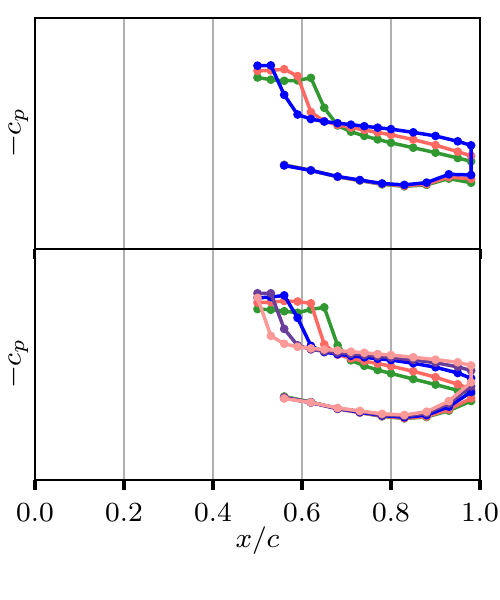}
  \caption{$\eta=47\%$}
  \label{f:cp_M084_Re12_Re25_pos_cp_aB_range_47}
\end{subfigure}
\begin{subfigure}{.32\textwidth}
  \centering
  \includegraphics[width=1\linewidth]{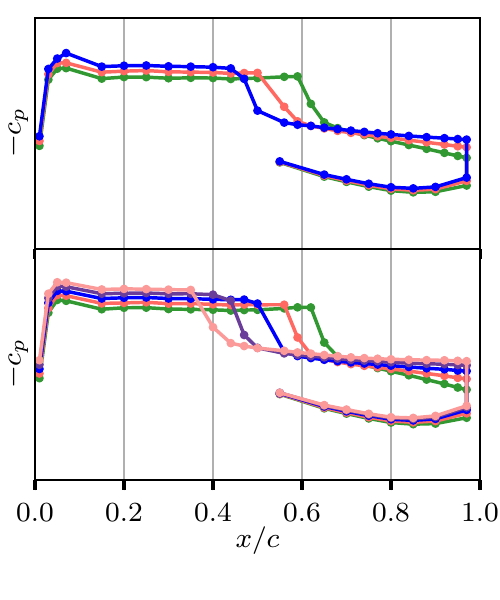}
  \caption{$\eta=55\%$}
  \label{f:cp_M084_Re12_Re25_pos_cp_aB_range_55}
\end{subfigure}
\begin{subfigure}{.32\textwidth}
  \centering
  \includegraphics[width=1\linewidth]{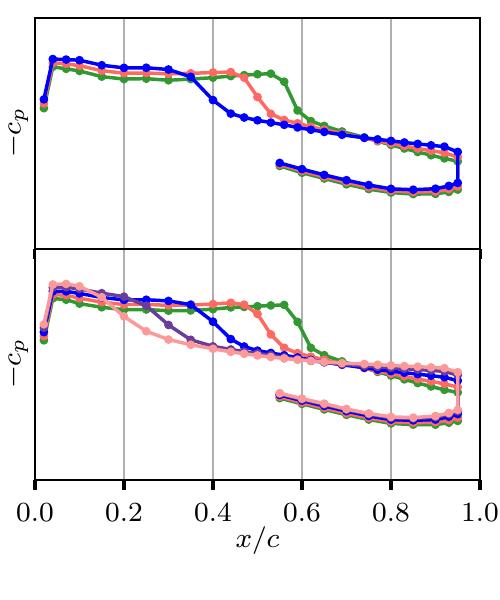}
  \caption{$\eta=75.1\%$}
  \label{f:cp_M084_Re12_Re25_pos_cp_aB_range_75}
\end{subfigure}%

  \caption{Static $c_p$ data $M_{\infty} = 0.84$ at $Re_{\infty}=12.9 \cdot 10^6$ (upper row) $Re_{\infty}=25.0 \cdot 10^6$ (lower row) at buffet onset and beyond acquired during the PSP polars.}
  \label{f:cp_M084_Re12_Re25_pos_cp_aB_range}
\end{figure}

In more quantitative terms, the static pressure coefficients in Fig.~\ref{f:cp_M084_Re12_Re25_pos_cp_aB_range} confirm that the largest displacement magnitude over $\alpha$ is encountered at $\eta=75.1\%$ at $M_{\infty} = 0.84$. The upper and lower rows in the figure show data for $Re_{\infty}=12.9 \cdot 10^6$ and $Re_{\infty}=25.0 \cdot 10^6$, respectively. The shock position is slightly further aft at the higher Reynolds number, which can be observed at all positions. This is most evident at $\eta=75.1\%$, but it can also be discerned at the other positions.

The $c_p$ data shown is not time-averaged, but it involves implicit time averaging inherent in the pneumatic measurement system. This results in a reduced chordwise pressure gradient in case of an oscillating shock. Larger chordwise motion amplitudes such as at $\eta = 75.1\%$ at high $\alpha$ cause the averaged shock to appear as a smeared chordwise pressure increase. This is consistent with the widening of the high RMS band in the PSP image in Fig.~\ref{f:PSP_cprms_wing_upper_M090_allRe_aL2_aB}. The pressure distributions downstream from the shock over the aft portion of the chord exhibit decreasing chordwise gradients at growing $\alpha$ at all positions, indicating shock-induced separation of an increasing magnitude. The aft pressure distribution becomes nearly horizontal at the highest measured incidence of $\alpha = \SI{5}{\degree}$ (at $Re_{\infty}=25.0 \cdot 10^6$). This reduced pressure recovery is a typical indication of largely separated flow.

\begin{figure}[!htb]
\begin{center}
\begin{subfigure}{1\textwidth}
\centering
\includegraphics[width=1\linewidth]{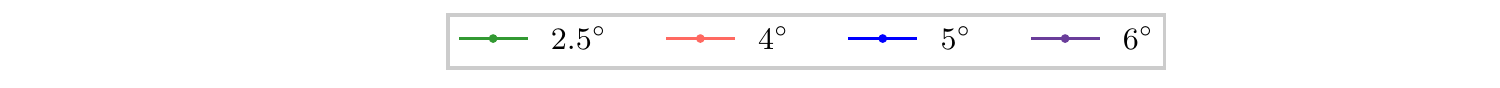}
\end{subfigure}
\end{center}
%\vspace*{-5mm}
\centering
\begin{subfigure}{.32\textwidth}
  \centering
  \includegraphics[width=1\linewidth]{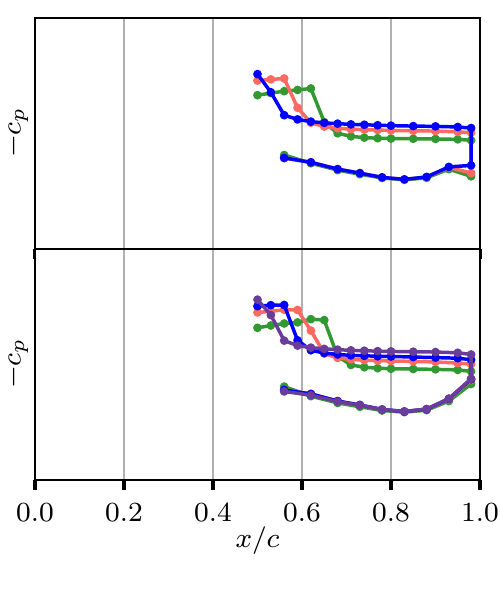}
  \caption{$\eta=47\%$}
  \label{f:cp_M090_Re12_Re25_pos_cp_aB_range_47}
\end{subfigure}
\begin{subfigure}{.32\textwidth}
  \centering
  \includegraphics[width=1\linewidth]{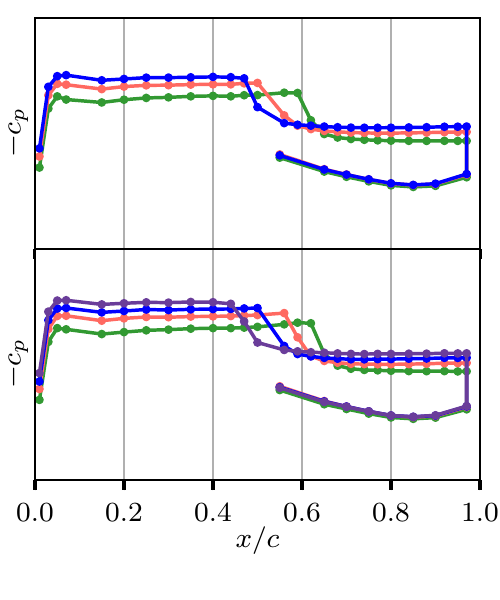}
  \caption{$\eta=55\%$}
  \label{f:cp_M090_Re12_Re25_pos_cp_aB_range_55}
\end{subfigure}
\begin{subfigure}{.32\textwidth}
  \centering
  \includegraphics[width=1\linewidth]{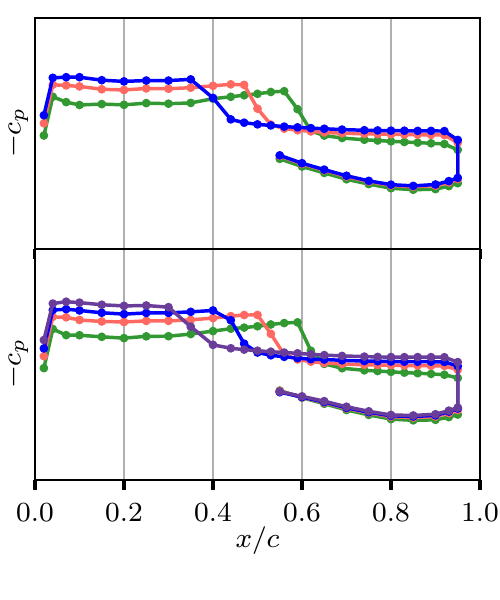}
  \caption{$\eta=75.1\%$}
  \label{f:cp_M090_Re12_Re25_pos_cp_aB_range_75}
\end{subfigure}%

  \caption{Static $c_p$ data at $M_{\infty} = 0.90$ at $Re_{\infty}=12.9 \cdot 10^6$ (upper row) $Re_{\infty}=25.0 \cdot 10^6$ (lower row) at buffet onset and beyond acquired during the PSP polars.}% Vertical black lines denote the chordwise positions of operational dynamic pressure sensors located on the port wing at the same spanwise stations.}
  \label{f:cp_M090_Re12_Re25_pos_cp_aB_range}
\end{figure}

The shock pattern at high $\alpha$ at $M_{\infty} = 0.90$ is less curved and moves more uniformly upstream. This is reflected in Fig.~\ref{f:cp_M090_Re12_Re25_pos_cp_aB_range}, where the magnitudes of shock displacement with increasing $\alpha$ are more similar than at $M_{\infty} = 0.84$. The extreme smearing of the shock pressure increase at $\eta = 75.1\%$ in Fig.~\ref{f:cp_M084_Re12_Re25_pos_cp_aB_range_75} does not occur in Fig.~\ref{f:cp_M090_Re12_Re25_pos_cp_aB_range_75}. This is consistent with the unsteady PSP RMS band in Fig.~\ref{f:PSP_cpmean_wing_M090_Re12_aL2}, which is narrower than at $M_{\infty} = 0.84$.

In addition, the differences in chordwise shock position due to Reynolds number in Fig.~\ref{f:cp_M090_Re12_Re25_pos_cp_aB_range} are slightly larger than those at $M_{\infty} = 0.84$ observed in Fig.~\ref{f:cp_M084_Re12_Re25_pos_cp_aB_range}. The upper surface pre-shock suction increases with higher incidences, with the difference due to $\alpha$ significantly larger than at $M_{\infty} = 0.84$. In addition, the wing lower surface shock at mid-chord observed in Section~\ref{chp:mach_effect} is still well visible at $\alpha = \SI{2.5}{\degree}$ and reduces its strength only at high angles of attack.

\begin{figure}[!h]
  \centering

\begin{center}
\begin{subfigure}{0.99\textwidth}
\centering
\includegraphics[width=1\linewidth]{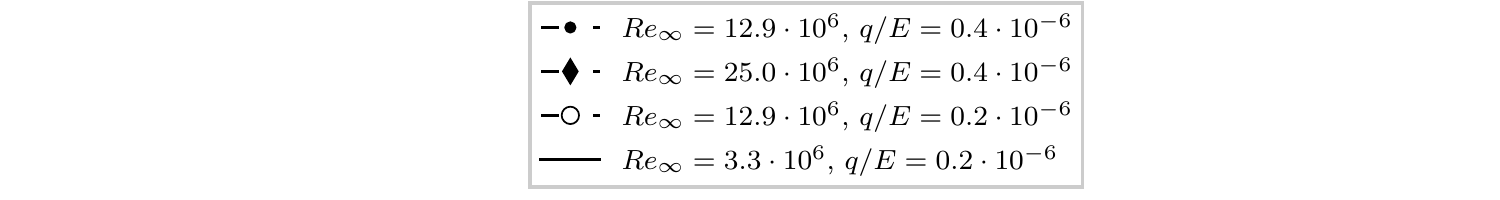}
\end{subfigure}
\end{center}
%\vspace*{-5mm}  
    \includegraphics[width=0.49\textwidth]{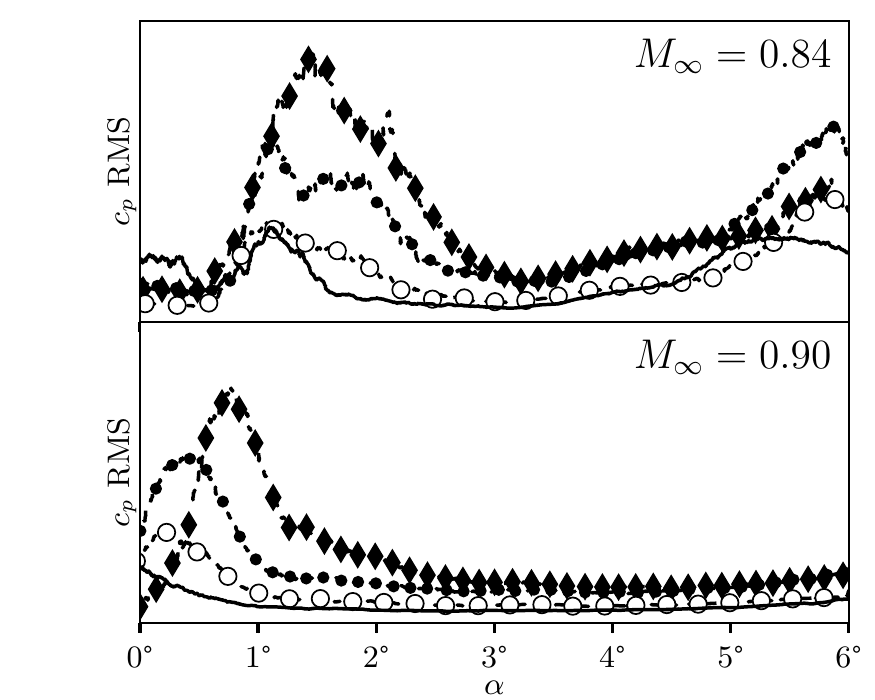}
    \caption{$c_{p,RMS}$ over $\alpha$ at $\eta = 63\%$ at different $M_{\infty}$ and $Re_{\infty}$ recorded at pressure transducers at $x/c = 55\%$. Rolling RMS computed using a sliding \SI{0.5}{\second} window.}
      \label{f:kulite_rms_M084_M090_Re12}

\end{figure}

The large shock displacement ranges on the outboard surface visible in the $c_p$ distributions in Figs.~\ref{f:cp_M084_Re12_Re25_pos_cp_aB_range} and Fig.~\ref{f:cp_M090_Re12_Re25_pos_cp_aB_range} can be also detected in the root mean square data of the dynamic pressure sensors. The chordwise passage of the shock at mid-wing is shown in Fig.~\ref{f:kulite_rms_M084_M090_Re12} via RMS values at the same four conditions as discussed in Section~\ref{chp:re_effect}.  
The narrow RMS peaks associated with the passage of the shocks are due to the large pressure gradients occurring at a strong shock, similar to the PSP RMS images. Generally, the shock moves aft at increasing incidences before reaching its most downstream position and beginning to travel upstream with further increasing incidences, which is typical for transonic airfoils and wings. The two peaks of elevated $c_{p,RMS}$ present at both Reynolds numbers at $x/c = 0.55$ at $M_{\infty} = 0.84$ in Fig.~\ref{f:kulite_rms_M084_M090_Re12} represent this process, with each passage causing a sharp peak. Across the board, the RMS values are higher downstream from the shock than upstream from it, which is due to the increased turbulence and thicker boundary layers downstream. The turbulent structures inside the boundary layer grow in size downstream from the shock, causing increased amplutides of wall pressure fluctuations. This is reflected by higher RMS values at high $\alpha$ after the shock passes upstream from the sensor. 

Fig.~\ref{f:kulite_rms_M084_M090_Re12} also demonstrates the differences in terms of shock position related to $Re_{\infty}$ and dynamic pressure. For $M_{\infty} = 0.90$ and $q/E = 0.4 \cdot 10^{-6}$, the peak positions indicate that the shock passes the sensor at $x/c = 55\%$ at $\alpha = \SI{4}{\degree}$ ($Re_{\infty}=25.0 \cdot 10^6$) and $\alpha = \SI{3.2}{\degree}$ ($Re_{\infty}=12.9 \cdot 10^6$). This is consistent with the generally downstream shock location associated with higher $Re_{\infty}$ at constant $\alpha$ in Fig.~\ref{f:cp_M090_Re12_Re25_pos_cp_aB_range}. The shift of $\Delta \alpha = \SI{0.8}{\degree}$ due to $Re$ is also apparent at $M_{\infty}  = 0.84$, albeit it is smaller in magnitude. Similar differences are apparent between $Re_{\infty} = 3.3 \cdot 10^6$ and $Re_{\infty}=12.9 \cdot 10^6$ at $q/E = 0.2 \cdot 10^{-6}$. Generally, the shock passes upstream over the sensor at lower $\alpha$ at low $Re_{\infty}$, owing to the thicker boundary layers which are less able to overcome adverse pressure gradients. In contrast, the influence of $q/E$ is smaller in comparison. When comparing the two runs at $Re_{\infty} = 12.9 \cdot 10^6$, the slightly earlier upstream passage at lower $q/E$  is consistent with the smaller wing deformation magnitude, which causes higher lift due to lower amount of negative aeroelastic twist.

\subsection{Buffet Onset and Dynamics}
While buffet onset had been estimated a priori using the lift curve break, that method may be inaccurate according to \cite{lawson}, as separation occurring on one part of the wing may be offset by a lift increase elsewere. Furthermore, the gradual nature of the lift curve break renders the approach imprecise. The recorded data can be used to more precisely characterize the occurrence of buffet.% The lift curve break method

All spectra in the following sections are shown using the Strouhal number $Sr = f c_{\mathrm{ref}}/u_{\infty}$ on the horizontal axis, based on the mean aerodynamic chord $c_{\mathrm{ref}}$. The reference velocity $u_{\infty}$ changes with $M_{\infty}$ and $Re_{\infty}$, the Strouhal number ensures comparability of aerodynamic phenomena in these cases.

\paragraph*{Structural and Aerodynamic Oscillations}

Detection of buffet onset is complicated by the lack of a wing root strain gauge and by the fact that the experiment was conducted using a full aircraft model mounted on a sting. Structural vibrations need to be taken into consideration when interpreting the dynamic data obtained from pressure transducers and accelerometers. The observed pressure fluctuations may result from pure aerodynamic phenomena, but could also be a consequence of the model vibrating at a certain natural frequency. 

\begin{figure}[!htb]
\centering
  \includegraphics[width=.49\linewidth]{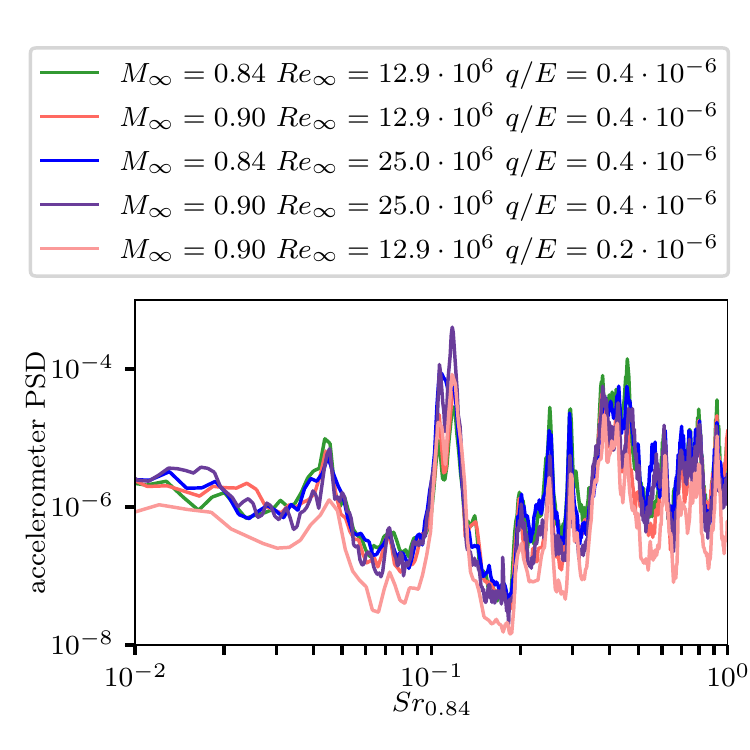}
  \caption{Fuselage accelerometer $z$ component at different flow conditions and $\alpha = \SI{4}{\degree}$. $Sr$ for all spectra calculated using $u_{\infty}$ at $M_{\infty} = 0.84$ and $Re_{\infty} = 12.9 \cdot 10^6$. Data at $q/E = 0.2 \cdot 10^{-6}$ is shown at $\alpha = \SI{5}{\degree}$.}
  \label{f:ACC_Port_Re12_Re25_PSD_AccZ}
\end{figure}
  
These phenomena are difficult to separate from each other. Fig.~\ref{f:ACC_Port_Re12_Re25_PSD_AccZ} shows the spectra of the fuselage accelerometer's $z$ component at $\alpha = \SI{4}{\degree}$ at different Mach and Reynolds numbers, indicating similar spectra at very different flow conditions. The horizontal axis in this figure is proportional to a frequency, as the Strouhal number in this case is computed from a common reference velocity of $u_{\infty,M=0.84}$, resulting in $Sr_{0.84} = f \cdot u_{\infty,M=0.84} / c_{\mathrm{ref}}$. As the horizontal axis in Fig.~\ref{f:ACC_Port_Re12_Re25_PSD_AccZ} is based on the same reference values $u_{\infty,M=0.84}$ and $c_{\mathrm{ref}}$, the shown peaks occur at the same absolute frequency. The fact that these peaks occur at the same frequencies in each case is an indication that they result from non-aerodynamic eigenmodes. They may be attributable to the natural frequencies of the wing tunnel model, the sting assembly or other wind tunnel influences. The frequencies of the peaks below $Sr_{0.84} = 0.2$ are consistent with eigenfrequencies of the sting and the wing observed during a wind-off vibration test. All peaks in Fig.~\ref{f:ACC_Port_Re12_Re25_PSD_AccZ} unaffected by temperature or dynamic pressure, and exhibit sharp and prominent peaks. Therefore, the wind tunnel fan frequency does not appear to be an important driver of these characteristics either. The lower dynamic pressure $q/E$ causes lower amplitudes across the board, but does not alter the shape of the spectra.

 \begin{figure}[!htb]
\centering 
  \centering
  \includegraphics[width=.49\linewidth]{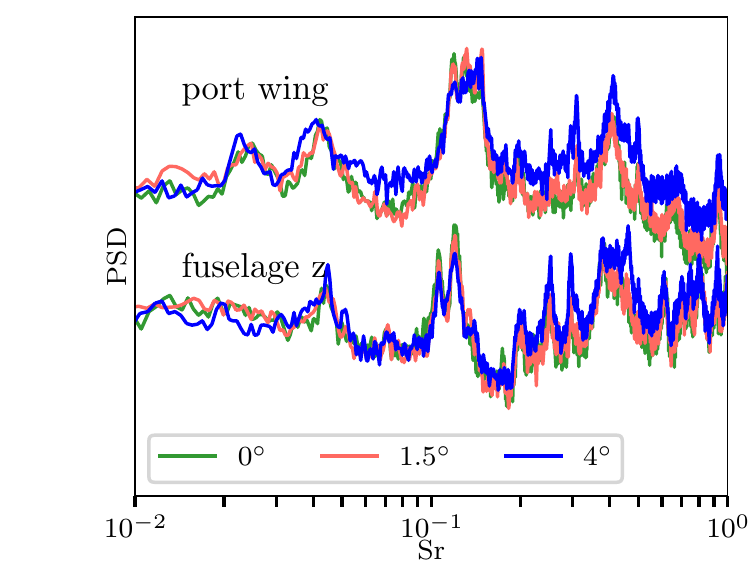}
  \caption{Spectra of port wing and fuselage accelerometers at different angles of attack at $M_{\infty} = 0.84$, $Re_{\infty} = 12.9 \cdot 10^6$.}
  \label{f:ACC_Port_Re12_Re25_PSD}
\end{figure}

The same is valid for changing aerodynamic conditions due to $\alpha$ at constant Mach and Reynolds number, shown in Fig.~\ref{f:ACC_Port_Re12_Re25_PSD} using the spectra of the fuselage accelerometer and the port wing accelerometer for a series of incidences at $M_{\infty} = 0.84$ and $Re_{\infty} = 12.9 \cdot 10^6$. While there is an increase in overall amplitudes at high angles of attack at high frequencies, the spectral characteristics again do not change. It is notable in particular that even $\alpha = \SI{0}{\degree}$ involves the same significant peaks as $\alpha = \SI{4}{\degree}$. Since the flow at $\alpha = \SI{0}{\degree}$ can be assumed to not involve any buffet motion, this behavior indicates that these model accelerations are caused by broadband excitation from the turbulent flow in the test section resulting in the structure vibrating at its natural frequencies.

The dominant oscillation occurs as two sharp peaks at $Sr = 0.105$ and $Sr = 0.12$ measured at the fuselage center. The corresponding high amplitude bump at the wing accelerometer extends up to a frequency of $Sr = 0.145$. The latter is consistent with a bending mode of the wing measured in the ground vibration test at wind-off conditions, albeit with some indications of a possible acoustic phenomenon occurring in the wind tunnel. This observation is under research at the time of submission~\cite{you:2022}. Higher frequency peaks around $Sr \approx 0.4$ occur as well. While this is in the widely accepted frequency range for swept wing buffet oscillation, it occurs at all incidences and is therefore unlikely to be caused by shock motion. One explanation is that it may constitute a harmonic of the $Sr = 0.145$ oscillation.

\paragraph*{Buffet Onset}

Detection of buffet onset proved difficult using the popular method involving lift polar nonlinearity in the present dataset, especially at high Mach numbers. ~\cite{lawson} gave a comprehensive overview of different buffet detection methods. In the following, the onset of oscillations will be analyzed in terms of fluctuation level increase of both accelerometer and pressure transducer signals, as well as via trailing edge pressure divergence.

\begin{figure}[!htb]
\centering
  \includegraphics[width=0.49\linewidth]{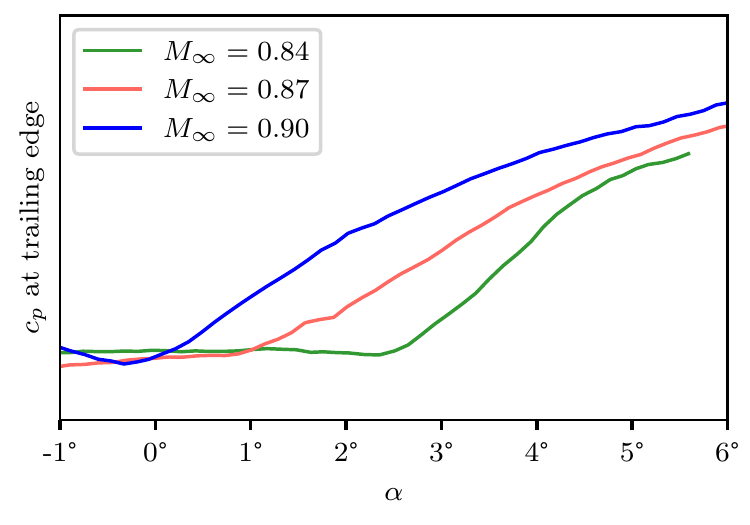}
  \caption{$c_p$ decrease at the aft-most static port on the wing upper side at $\eta = 75.1\%$ at $Re_{\infty} = 12.9\cdot 10^6$.}
  \label{f:cp_TE_alpha}
\end{figure}

As in the discussion by \cite{dandois:2016}, both RMS values of unsteady sensors and pressure divergence of static taps near the trailing edge can serve as additional indicators of buffet onset. The trailing edge pressure divergence, as described by \cite{lawson}, can be evaluated by determining the point at which the trailing edge $c_p$ deviates by $\Delta c_p = 0.05$ from a linear trend. The linear trend at low incidence is clearly identifiable at low Mach number in Fig.~\ref{f:cp_TE_alpha}, and disappears at high Mach number. The maximum $c_p$ value measured during that run is then taken as the baseline for $\Delta c_p$ instead.

Nevertheless there is a clear in Fig.~\ref{f:cp_TE_alpha} trend over $M_{\infty}$, indicating occurrence of $c_p$ divergence at decreasing $\alpha$ at growing Mach number. Pressure data at $\eta = 75.1\%$ is used here, as it is the spanwise location closest to the most significant upstream shock motion in Fig.~\ref{f:PSP_cpmean_wing_upper_allMach_Re12_aL2_aB}. The $c_p$ divergence criterion is fulfilled at $\alpha = \SI{3}{\degree}$, $\alpha = \SI{1.6}{\degree}$, and $\alpha = \SI{0.6}{\degree}$ at the three Mach numbers, respectively. Note that the sensors are not at the trailing edge, but on the upper surface at $x/c = 0.98$.

\begin{figure}[!htb]
\centering
  \includegraphics[width=0.49\linewidth]{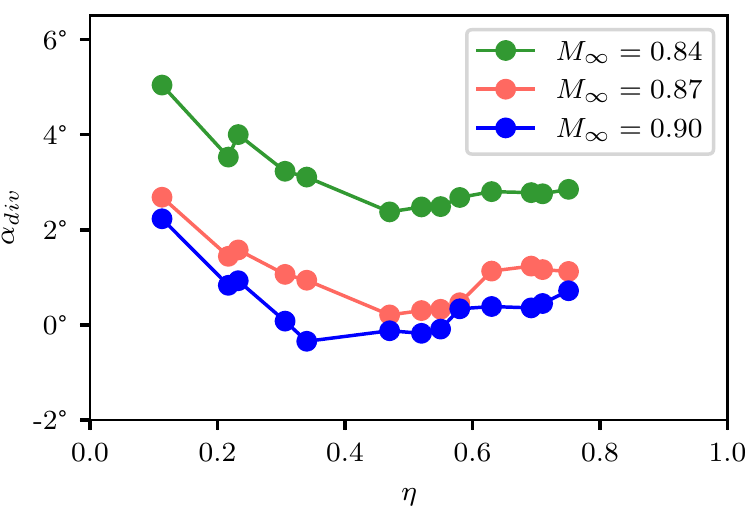}
  \caption{Incidence of $c_p$ divergence occurrence at the aft-most static ports on the wing upper side at $Re_{\infty} = 12.9\cdot 10^6$.}
  \label{f:cp_TE_alpha_all}
\end{figure}

Moving towards the tip, the rearmost sensors are at $x/c = 0.95$. Fig.~\ref{f:cp_TE_alpha_all} shows the angles of attack at which the $c_p$ divergence criterion is first fulfilled. The initial onset of shock forward motion and trailing edge separation is confirmed here, with the spanwise location of initial $c_p$ divergence visible outboard of about $\eta = 0.4$ in all cases. The Mach number trend is very clear, with $c_p$ divergence occurring at or even below $\alpha = \SI{0}{\degree}$ at $M_{\infty} = 0.90$. 

\begin{figure}[!htb]
  \centering
\begin{center}
\begin{subfigure}{0.49\textwidth}
\centering
\includegraphics[width=1\linewidth]{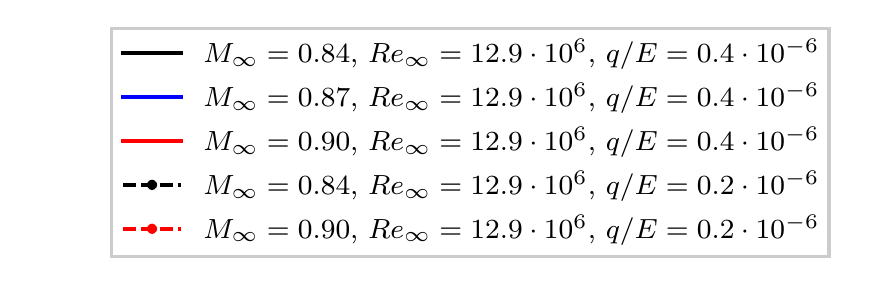}
\end{subfigure}
\end{center}  
      \includegraphics[width=0.49\textwidth]{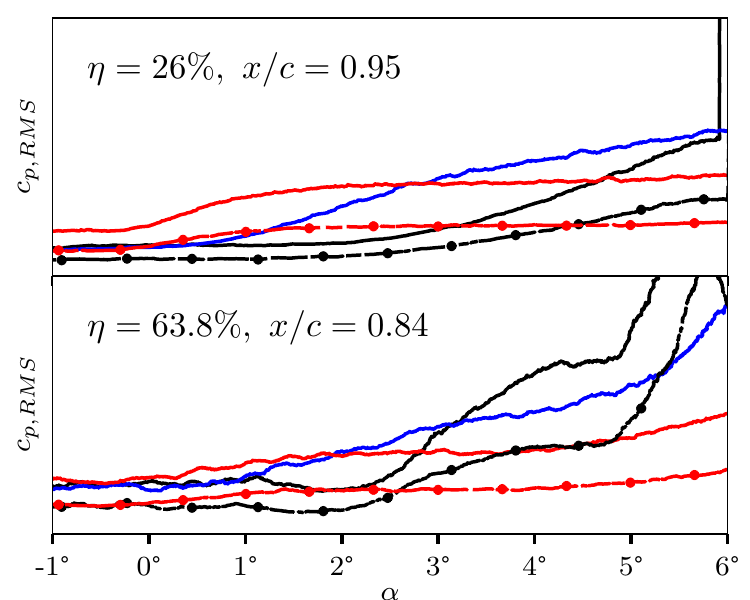}
    \caption{$c_{p,RMS}$ near the trailing edge on the inboard (upper) and outboard (lower) wing surface at $Re_{\infty}=12.9 \cdot 10^6$ at different $M_{\infty}$ and $q/E$.}
      \label{f:kulite_rms_M084_M090_Re25_wing063_TE}    
  \end{figure}

A similar picture of the Mach number trend is painted by the RMS values of unsteady pressure transducers nearest to the trailing edge. Fig.~\ref{f:kulite_rms_M084_M090_Re25_wing063_TE} shows inboard and outboard trailing edge RMS data for all Mach numbers. The outboard unsteady sensors are located further inboard than the pressure taps shown in Fig.\ref{f:cp_TE_alpha}, but they nevertheless indicate similar incidences at which significant RMS increase occurs. At $M_{\infty} = 0.84$ and $Re_{\infty} = 12.9 \cdot 10^6$, both the initial $c_p$ increase (Fig.~\ref{f:cp_TE_alpha}) and the pressure RMS increase outboard (Fig.~\ref{f:kulite_rms_M084_M090_Re25_wing063_TE}) occur at $\alpha \approx \SI{2.5}{\degree}$. The RMS increase begins at $\alpha \approx \SI{1}{\degree}$ and $\alpha \approx \SI{0}{\degree}$ at $M_{\infty} = 0.87$ and $M_{\infty} = 0.90$, respectively. The trend of significant $c_{p, \mathrm{RMS}}$ growth occurring at lower $\alpha$ for lower $M_{\infty}$ is recognizable and a conseqeuence of earlier upstream shock motion at lower Mach number.

The changes are much more gradual than on the outboard wing, which is mostly due to significantly smaller amount of chordwise shock motion over $\alpha$. High Mach number causes separation downstream from the shock at very low angles of attack, which is consistent with the flat $c_p$ distributions downstream from the shock in Fig.~\ref{f:cp_M090_Re12_Re25_pos_cp_aB_range} and with the increase of the pressure transducer RMS level around $\alpha = \SI{0}{\degree}$ at $M_{\infty} = 0.90$. At the same time, the maximum RMS magnitudes at the higher Mach numbers are lower than those at lower Mach number. The changes at $\eta = 26\%$ are much more gradual than on the outboard wing, which is mostly due to significantly less amount of chordwise shock motion over $\alpha$. 

Overall, the loss of lift due to upstream shock motion is most abrupt at low Mach and much more gradual at high Mach numbers. The gradient of RMS increase in Fig.~\ref{f:kulite_rms_M084_M090_Re25_wing063_TE} reflects this. The local forward bulge occurring in the shock shape on the outboard wing at $M_{\infty} = 0.84$ is reflected by early and strong growth of trailing edge pressure RMS at the $\eta = 63\%$ position. The dynamic pressure does not alter the incidences at which the increases occur, it only decreases the RMS level. The Reynolds number does not alter these characteristics fundamentally, therefore the results at $Re_{\infty} = 25.0 \cdot 10^6$ are omitted for clarity.

Pressure divergence and transducer RMS were termed aerodynamic criteria by \cite{masini:2020}, as opposed to strain gauge or accelerometer measurements. One drawback of these aerodynamic criteria is that they are local by definition and yield results valid only in the immediate vicinity. Nevertheless, these two criteria are consistent and the general trends with respect to Mach number are largely independent of the spanwise position. They help tackle augment the results of the $\Delta \alpha$ criterion and are mostly consistent with it. As the $\Delta \alpha$ criterion fails at $M_{\infty} = 0.90$, these aerodynamic criteria are useful for the understanding of trends.

\subsection{Buffet Spectra}

The growth of pressure fluctuation amplitudes at increasing $\alpha$ discussed above will be discussed in the following. Shock-induced separation and buffet motion induce characteristic spectral features measured by the unsteady pressure transducers. Recent swept wing experimental studies mentioned in Sec.~\ref{chp:intro_buffet} showed two distinct spectral regions. Typically, there is one low frequency range below $Sr = 0.1$ which occurs over a large part of the wing and which has been associated with inboard propagation of disturbances by \cite{masini:2020} or with with 2D-like behavior by \cite{paladini:2019}. The classical 3D buffet shock unsteadiness with outboard propagation tends to occur outboard and at higher Strouhal numbers around $Sr = $ \SIrange{0.2}{0.6}{}. \cite{dandois:2016} also observed a Kelvin-Helmholtz (K-H) instability at very high frequencies at $Sr = $ \SIrange{1}{4}{}. While Nyquist frequency of the pressure sensors in the present experiment also reached $Sr = 4$, no indication of spectral features attibutable to K-H instabilities were found.

\begin{figure}[!h]

\centering
\begin{subfigure}{.3\textwidth}

\begin{tikzpicture}
\node [anchor=south west,inner sep=0pt] (image) at (0,0) {\includegraphics[width=\textwidth]{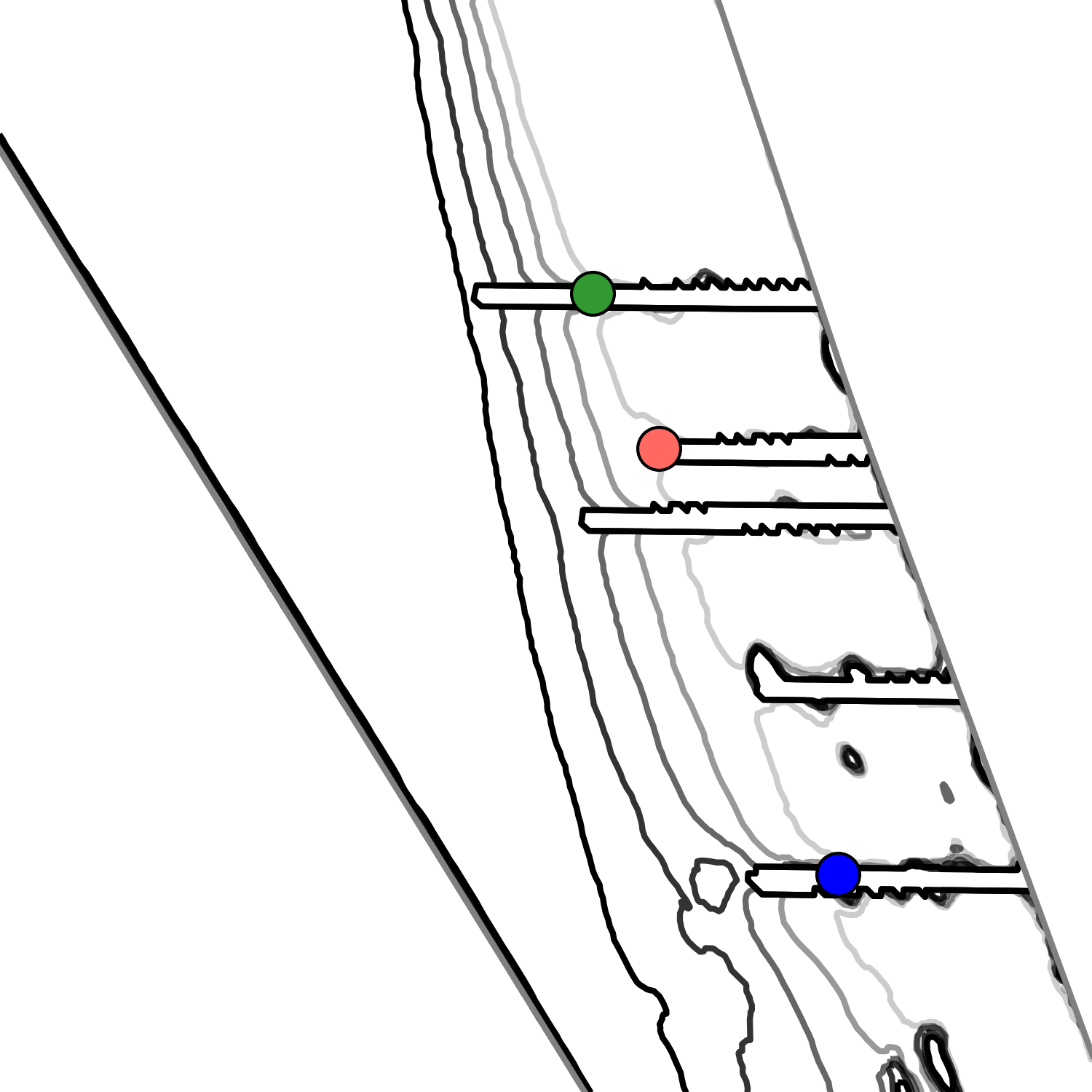}};

    \begin{scope}[x={(image.south east)},y={(image.north west)}]

\node at (0.95, 0.8) (eta04) {$\eta = 47\%$};
\draw[thick] (eta04) -- (0.55, 0.75); 

\node at (0.95, 0.7) (eta07) {$\eta = 51\%$};
\draw[thick] (eta07) -- (0.63, 0.6); 

\node at (0.95, 0.5) (eta05) {$\eta = 63\%$};
\draw[thick] (eta05) -- (0.75, 0.2);

    \end{scope}

\end{tikzpicture} 
  \caption{Shock and sensor positions.}
  \label{f:M084_Re25_PSP_mean_contours_cp05_aoarange_kulite_pos_midwing_zoom}
\end{subfigure}%

\begin{subfigure}{.49\textwidth}
  \centering
  \includegraphics[width=1\linewidth]{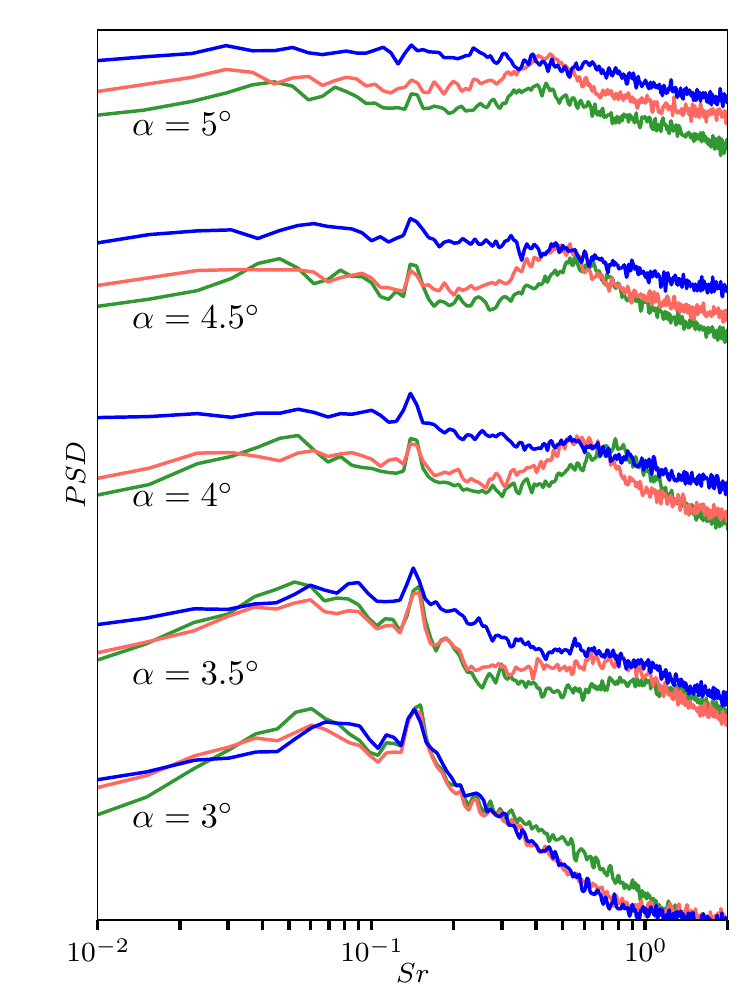}
  \caption{Power spectral densities at three pressure transducers. Each $\alpha$ is separated from the previous by multiplying the amplitudes by a factor of 100 for clarity.}
  \label{f:M084_Re25_multiAoA_midrow_spectra}
\end{subfigure}%

  \caption{$M_{\infty} = 0.84$, $Re_{\infty} = 25.0 \cdot 10^6$ pressure power spectral density at different incidences. The shock positions are indicated using isolines of $c_p$ for five incidences. The line colors in the spectra correspond to the symbols showing the sensor positions. $c_p$ isolines indicate shock positions at $\alpha = $ \SIlist{3;3.5;4;4.5;5}{\degree} from light to dark grey. Spurious $c_p$ lines over the rear part are artifacts due to masking.}
  \label{f:M084_Re25_kulite_iPSP_per_aoa}
\end{figure}
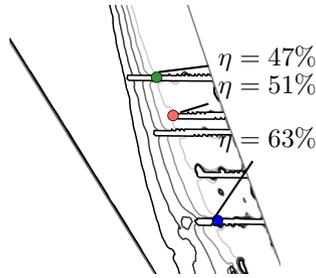
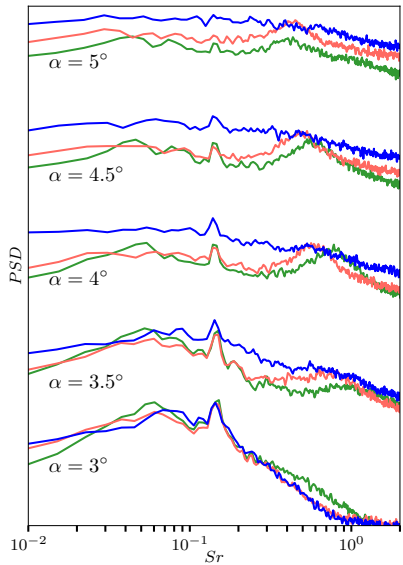

The unsteady pressure transducers are located in the inboard two thirds of the wing, with between two and three sensors active during the experiment at different spanwise positions. The positions and spectra of three such devices near 70\% chord and at three spanwise locations at $\eta = 47\%, 51\%$ and $63\%$ are shown in Fig.~\ref{f:M084_Re25_kulite_iPSP_per_aoa}. The sensors are consistently downstream from the shock at all shown conditions. The mean shock position is shown using isolines at a constant $c_p$ value, giving a qualitative indication of the shape and relative motion at increasing $\alpha$. The isoline of constant $c_p$ shows how the shock moves upstream relative to the sensors with increasing $\alpha$.

The spectra in Fig.~\ref{f:M084_Re25_kulite_iPSP_per_aoa} show the wide bump of high amplitude corresponding to the transonic buffet motion at $M_{\infty} = 0.84$. This buffet bump's center frequency shifts with the inflow conditions. The image shows the spectra at five different angles of attack for three unsteady pressure transducers, with the sensors' location and the position shock on the wing surface indicated in the sketch above the plots. The narrow peak associated with structural oscillation at $Sr \approx 0.14$ is visible in the spectra at all angles of attack. 

The lowest incidence of \SI{3}{\degree} is beyond the incidence at which initial separation was detected above via RMS and $c_p$ divergence. The shock is just upstream from the transducers at $x/c = 0.7$ at this incidence. The corresponding unsteady PSP RMS image in Fig.~\ref{f:PSP_cpmean_wing_M084_Re12_aL2} shows an increase of pressure fluctuations downstream from the shock, but no wide region of high RMS indicating a large scale motion of the shock itself. The high amplitude measured by all three transducers at $Sr = 0.07$ is consistent with the observation of \cite{masini:2020} and \cite{sugioka:2018}, who discussed that low frequency shock oscillation occurs even before buffet onset.

With further upstream motion of the shock at higher angles of attack, more activity becomes apparent in the frequency range between $Sr = 0.1$ and $Sr = 1$. Wide amplitude bumps appear at the two inboard sensors at $\eta = 47\%$ and $\eta = 51\%$. This is consistent with swept wing buffet studies described above. The buffet bump frequency decreases in outboard spanwise direction. This spanwise frequency shift disappears at $\alpha = \SI{5}{\degree}$, with the two inner sensors showing the bump at the same frequency. \cite{paladini:2019} described the observation that the frequency ceases to shift over wing span in a regime they termed deep buffet, as opposed to well-established buffet at $\alpha < \SI{5}{\degree}$.

The buffet bump is much less distrinct at $\eta = 63\%$ than at the two inboard sensors. It is only somewhat distinguishable at $\alpha = \SI{3.5}{\degree}$ and $\alpha = \SI{4}{\degree}$. However, the fluctuation level is much higher overall at this location than at the other two, which may mask a bump by increased amplitudes across the entire frequency range. The shock is much farther upstream at $\eta = 63\%$ than further inboard, causing a larger scale separation. The unsteadiness resulting from this causes broadband fluctuation with high amplitudes across the entire frequency range.

The frequency shift toward lower frequencies associated with lower $\alpha$ is significant at the two inner sensors. The buffet bump center frequency at $\eta =47\%$ starts at $Sr = 0.8$ for $\alpha = \SI{3.5}{\degree}$ and shifts to about $Sr = 0.4$ at $\alpha = \SI{5}{\degree}$.

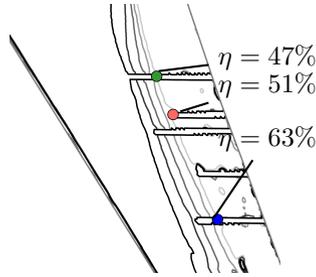
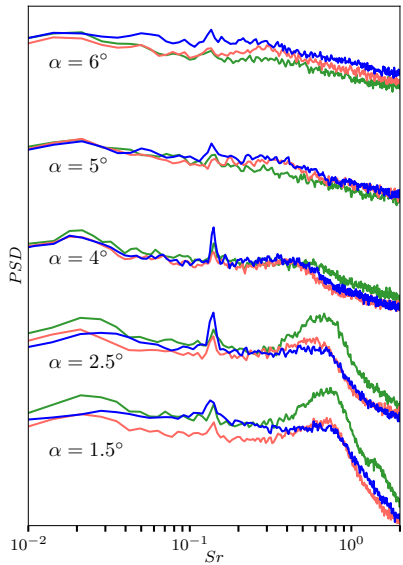
\begin{figure}[!h]
\centering

\begin{subfigure}{.3\textwidth}
\begin{tikzpicture}
\node [anchor=south west,inner sep=0pt] (image) at (0,0) {\includegraphics[width=\textwidth]{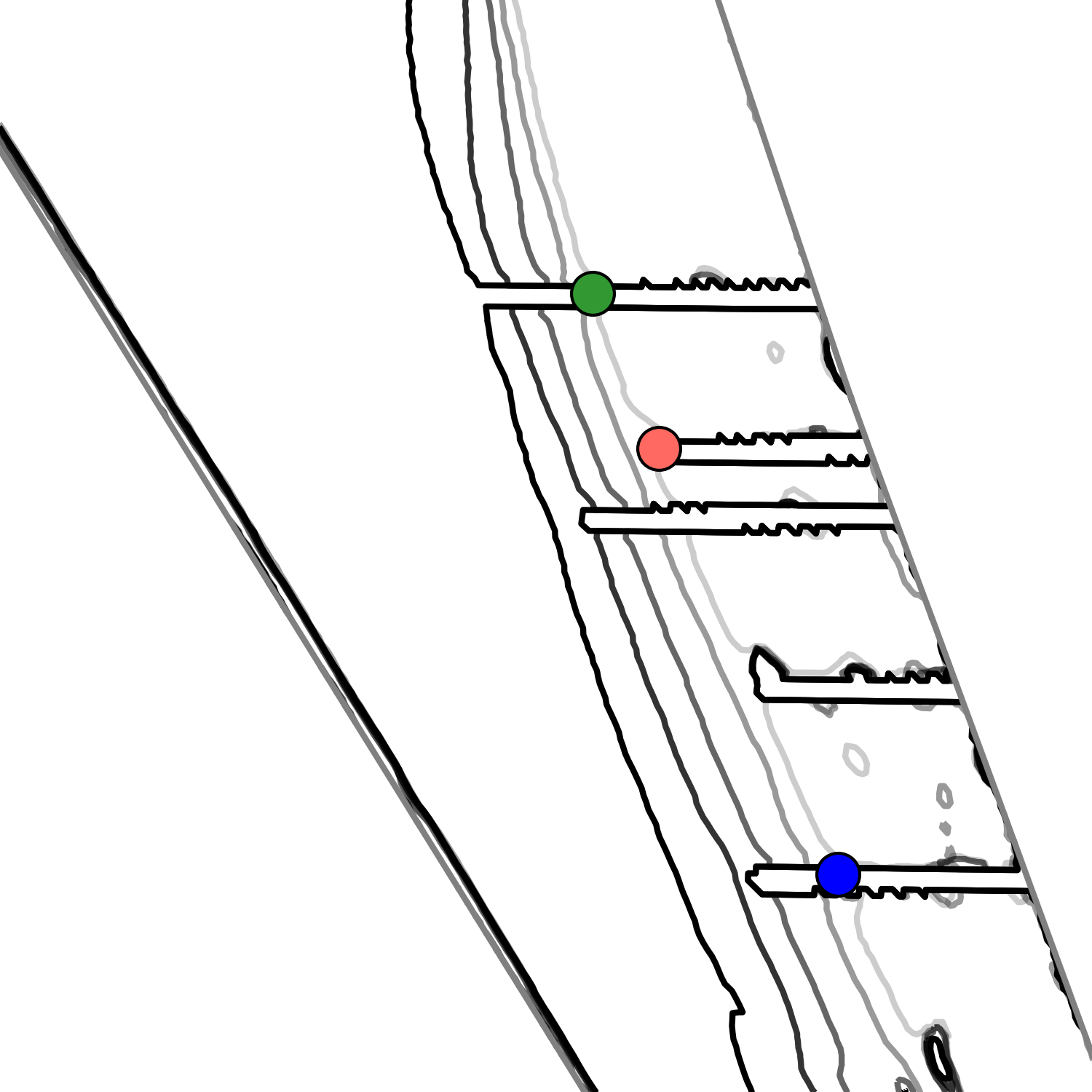}};

    \begin{scope}[x={(image.south east)},y={(image.north west)}]

\node at (0.95, 0.8) (eta04) {$\eta = 47\%$};
\draw[thick] (eta04) -- (0.55, 0.75); 

\node at (0.95, 0.7) (eta07) {$\eta = 51\%$};
\draw[thick] (eta07) -- (0.63, 0.6); 

\node at (0.95, 0.5) (eta05) {$\eta = 63\%$};
\draw[thick] (eta05) -- (0.75, 0.2);

    \end{scope}

\end{tikzpicture} 

  \caption{Shock and sensor positions.}
  \label{f:M090_Re25_PSP_mean_contours_cp05_aoarange_kulite_pos_midwing_zoom}
\end{subfigure}%

\begin{subfigure}{.49\textwidth}
  \centering
  \includegraphics[width=1\linewidth]{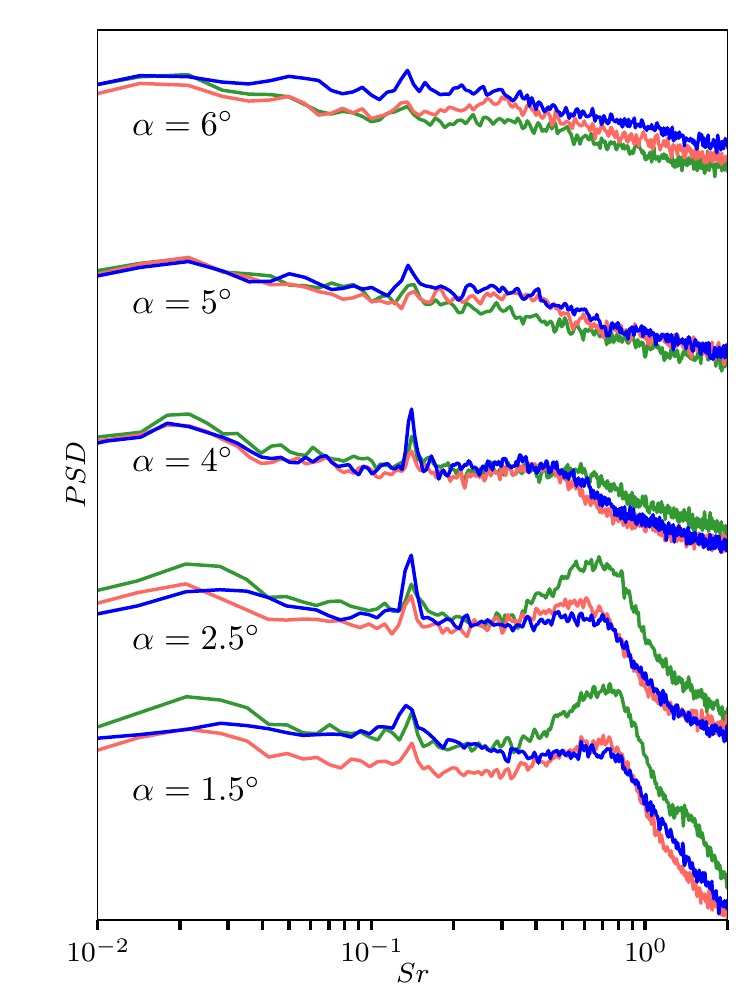}
  \caption{Power spectral densities at the pressure transducers at $\alpha$ from \SIrange{1.5}{6}{\degree}. Each $\alpha$ is shifted by 100 in $y$ axis direction for clarity.}
  \label{f:M090_Re25_multiAoA_midrow_spectra}
\end{subfigure}%

  \caption{$M_{\infty} = 0.90$, $Re_{\infty} = 25.0 \cdot 10^6$ pressure power spectral density at different incidences. The shock positions are indicated using isolines of $c_p$ for five incidences. The line colors in the spectra correspond to the symbols showing the sensor positions. $c_p$ isolines indicate shock positions at $\alpha = $ \SIlist{1.5;2.5;4;5;6}{\degree} from light to dark grey. Spurious $c_p$ lines over the rear part are artifacts due to masking.}
  \label{f:M090_Re25_kulite_iPSP_per_aoa}
\end{figure}

Earlier buffet onset due to higher Mach number becomes apparent in the spectra at $M_{\infty} = 0.90$ in Fig.~\ref{f:M090_Re25_kulite_iPSP_per_aoa}. A buffet bump around $Sr = 0.8$ is visible at very low angles of attack at the two inboard sensors $\eta = 47\%$ and $\eta = 51\%$, beginning at $\alpha = \SI{1.5}{\degree}$. The buffet bump is most pronounced at the two inboard sensors at low $\alpha$, while at higher incidences the differences disappear. There is no spanwise shift of buffet bump center frequency at $\alpha = \SI{1.5}{\degree}$, which confirms the observation of \cite{paladini:2019} that there is little shift near buffet onset. In more well-established buffet regimes there is a spanwise variation of frequency, which is visible in Fig.~\ref{f:M090_Re25_kulite_iPSP_per_aoa} at $\alpha = \SI{2.5}{\degree}$ and $\alpha = \SI{4}{\degree}$.

There is some indication of similar broadband fluctuation phenomena downstream from the shock, which is especially evident at $\alpha = \SI{2.5}{\degree}$ at the positions shown in Fig.~\ref{f:M090_Re25_kulite_iPSP_per_aoa}. The comparatively high amplitude at that incidence is the consequence of the close proximity of the sensors to the shock.

The frequency shift due to $\alpha$ occurs here as well. While the buffet bump is most clearly recognizable at $\eta = 51.4\%$, is can be discerned at all three sensors and moves to lower frequencies up to $\alpha = \SI{4}{\degree}$. The very early buffet onset at this Mach number means that the highest measured incidences are in deep buffet. There is barely any visible buffet bump at $\alpha = \SI{5}{\degree}$ and $\alpha = \SI{6}{\degree}$, indicating that there is little coherent motion and high levels of broadband unsteadiness downstream from the shock.%The narrow peak at $Sr = 0.13$ persists at all positions and angles of attack, which is independent from Mach number. 

The sharp peak at $Sr \approx 0.14$ is observed at this Mach number as well, again disappearing at high $\alpha$. Although this peak occurs at a constant absolute frequency across the different flow conditions, the difference between $u_{\infty,M=0.84}$ and $u_{\infty,M=0.90}$ used for the computation of Strouhal numbers in the plots is sufficiently small for it to remain at the same approximate $Sr$.

\begin{figure}[!h]
\centering
  \includegraphics[width=0.49\linewidth]{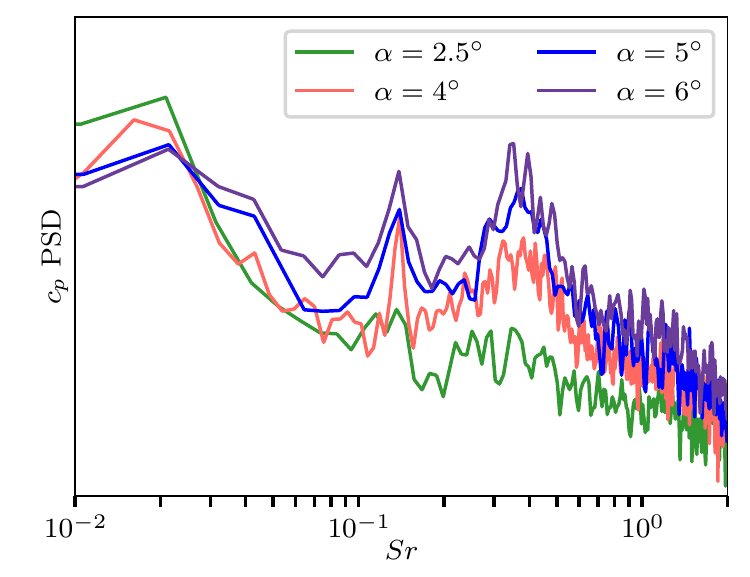}
  \caption{Inboard unsteady pressure transducer PSD at $x/c = 0.95$, $\eta = 26\%$ at different $\alpha$ for $M_{\infty} = 0.90$, $Re_{\infty}=25.0\cdot 10^6$.}
  \label{f:spectra_buffet_M090_Re25PSD_per_alpha_KUP10401}
\end{figure}

As opposed to $M_{\infty} = 0.84$, there is significant inboard expansion of buffet-related unsteadiness toward the inboard wing. The unsteady PSP images in Fig.~\ref{f:PSP_cpmean_wing_M090_Re12_aL2} show increasing RMS values toward the wing root, and the inboard pressure transducer at $\eta = 26\%$ and $x/c = 0.95$ detects a growing spectral bump at $Sr = 0.4$ with increasing angle of attack. This is shown in Fig.~\ref{f:spectra_buffet_M090_Re25PSD_per_alpha_KUP10401}. Since that sensor is near the trailing edge it follows that the oscillating shock motion expands in spanwise direction and causes increasing pressure oscillation at the typical buffet frequency even inboard.

\begin{figure}[!htb]
\centering
  \includegraphics[width=0.49\linewidth]{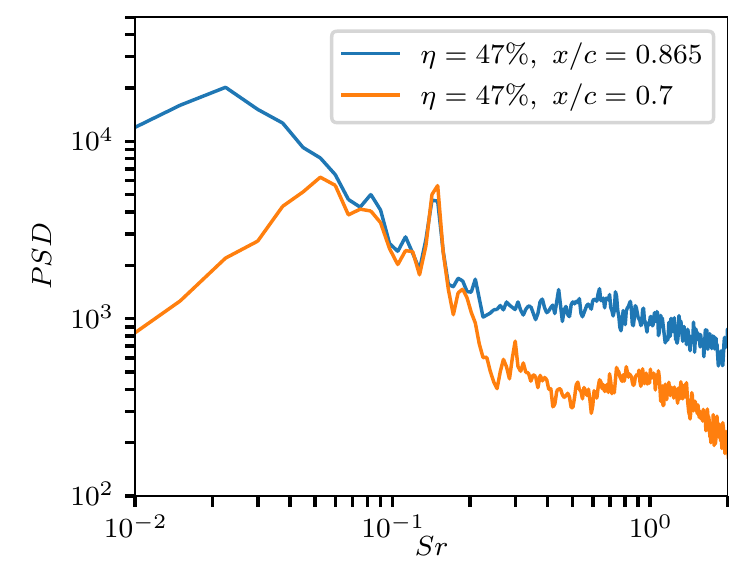}
  \caption{Chordwise spectral differences at unsteady pressure transducers at $\eta = 47\%$ in chordwise direction at $\alpha = \SI{3.5}{\degree}$ for $M_{\infty} = 0.84$, $Re_{\infty}=25.0\cdot 10^6$.}
  \label{f:M084_Re25_AoA35_eta047_chordwise_sensors_spectra}
\end{figure}

Both \cite{dandois:2016} and \cite{paladini:2019} describe a slight chordwise reduction of buffet bump frequency at low to moderate angles of attack. There is no such shift at most of the conditions encountered in the present dataset. One reason may be that most of the measured incidences are in significant buffet conditions and beyond onset. Also, there is only a low number of unsteady sensors downstream from a shock installed in the XRF-1 model, i.e. there is no sufficient chordwise resolution. In most cases the buffet bump is at the same frequency for successive sensors in the same spanwise position. The measurement point just beyond buffet onset at $\alpha = \SI{3.5}{\degree}$ at $M_{\infty} = 0.84$ shown in Fig.~\ref{f:M084_Re25_AoA35_eta047_chordwise_sensors_spectra} offers some indication of a slight reduction of measured buffet frequency in chordwise direction. There is no shift at any other Mach number or incidence.

Incidentally, one of the four geometries analyzed by \cite{paladini:2019} displayed a similar behavior. That experiment, FLIRET, used a configuration with a wing with a high taper ratio which is similar to that of the presently investigated XRF-1. While no direct causal relation between chordwise buffet frequency shift and swept wing taper ratio is known to the authors, this is an observation worth taking into account.

\begin{figure}[!htb]
\centering
  \includegraphics[width=0.49\linewidth]{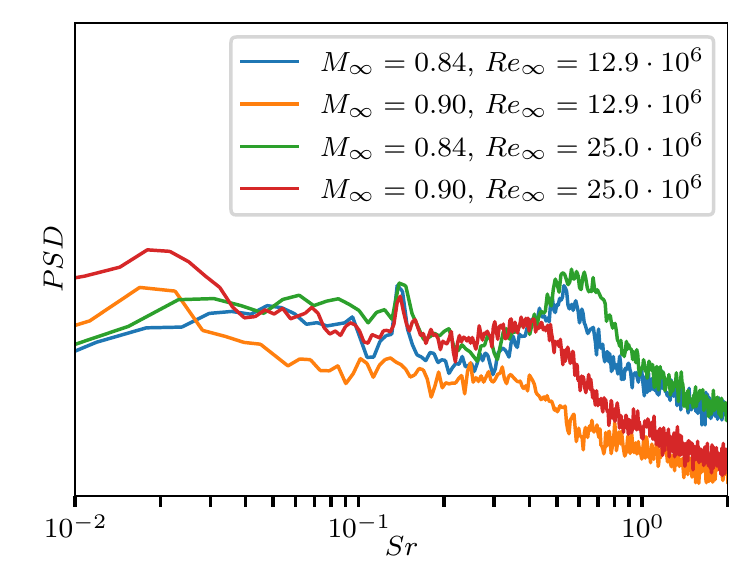}
  \caption{Unsteady pressure transducer spectra downstream from the shock ($\eta = 47\%$, $x/c= 0.7$) at $\alpha = \SI{4}{\degree}$, $q/E = 0.4\cdot 10^{-6}$ and different Mach and Reynolds numbers.}% $Sr$ for all spectra calculated using $u_{\infty}$ at $M_{\infty} = 0.84$ and $Re_{\infty} = 12.9 \cdot 10^6$.}
  \label{f:Re12_Re25_AoA4_KUP11802_spectra}
\end{figure}

As mentioned above, all of the discussed spectra show significant peaks around $Sr = 0.14$ across the board, coinciding with the largest peak visible at all conditions the accelerometer spectra in Fig.~\ref{f:ACC_Port_Re12_Re25_PSD_AccZ}, which is likely to stem from a structural eigenmode being excited by acoustic effects in the wind tunnel. Fig.~\ref{f:Re12_Re25_AoA4_KUP11802_spectra} shows this for a constant angle of attack $\alpha = \SI{4}{\degree}$ at different conditions. The shown spectra display a prominent buffet bump at all conditions, showcasing the effect of Mach and Reynolds number. The bump's center is reduced due to decreasing Reynolds number at constant Mach, both at $M_{\infty} = 0.84$ and $M_{\infty} = 0.90$. This is consistent with the earlier onset of buffet at low Reynolds number, and therefore a constant angle of attack $\alpha$ can be considered deeper within the buffet regime than the same angle of attack at higher Reynolds number. This is similar to the decrease of the local buffet bump frequency at increasing angle of attack. The same reasoning can be applied to an increase of Mach number at constant $Re_{\infty}$. Buffet onset occurs much earlier at $M_{\infty} = 0.90$, causing the shown incidence $\alpha = \SI{4}{\degree}$ to be in a deeper buffet regime than the same $\alpha$ at lower Mach number. Therefore the buffet bump is broader, less prominent, and occurs at lower Strouhal number.

\subsection{Buffet Cells and Spanwise Propagation}

As suggested by the recent literature above, the origin of the buffet bumps in the pressure transducer spectra is expected to be the propagation of buffet cells. The high temporal resolution of the unsteady pressure sensors is counterbalanced by their low number and therefore low spatial resolution. In addition, the upstream sensors alternate between being upstream and downstream from the shock at several flow conditions, which makes it difficult to distinguish between the occurring aerodynamic phenomena. The time-resolved PSP data, however, allow a detailed investigation of the unsteady shock motion that goes beyond the data available at these discrete points. The two data sources enable a complementary analysis.

Analysis of buffet cell is shown for two selected conditions at $M_{\infty} = 0.84$ and $M_{\infty} = 0.90$ at $Re_{\infty} = 12.9 \cdot 10^6$. The low Reynolds number is used due to lower noise floor at higher temperature, and due to the high sampling rate of 2000 frames per second which enables a better temporal resolution of the phenomena. The preceding sections have shown that no fundamental change occurs between the two Reynolds number settings, with the changes mostly confided to an altered chordwise shock position. The angle of attack of $\alpha = \SI{4}{\degree}$ is sufficiently high to detect significant shock motion, but not excessive as to avoid entering deep buffet and possibly loss of motion coherence.

\begin{figure}[!htb]
	\centering
	\begin{subfigure}{.49\textwidth}
		\centering
		\includegraphics[width=1\linewidth]{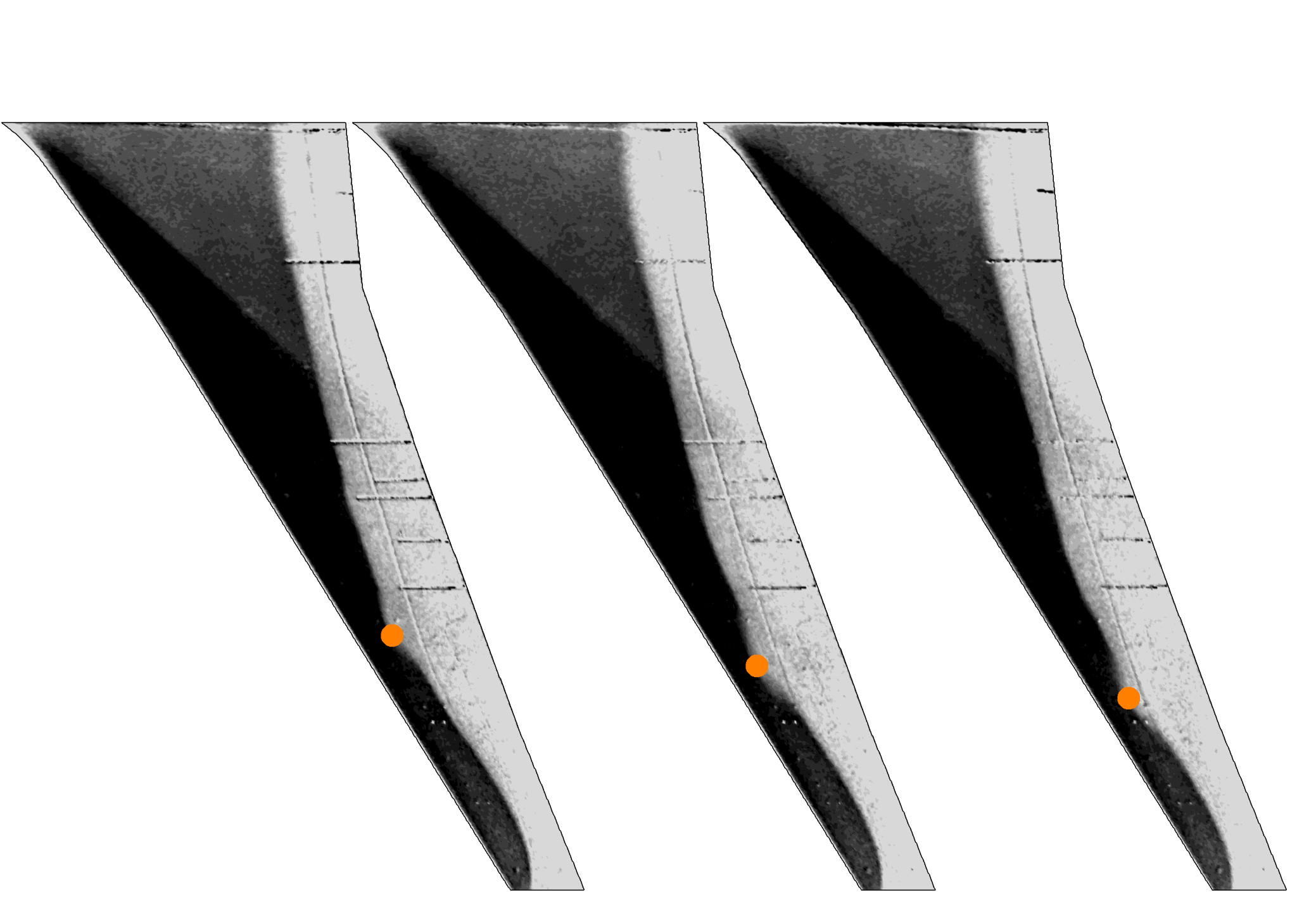}
		\caption{$M_{\infty} = 0.84$}
		\label{f:buffet_timeseries_M084}
	\end{subfigure}%
	\hfill
	\begin{subfigure}{.49\textwidth}
	\centering
	\includegraphics[width=1\linewidth]{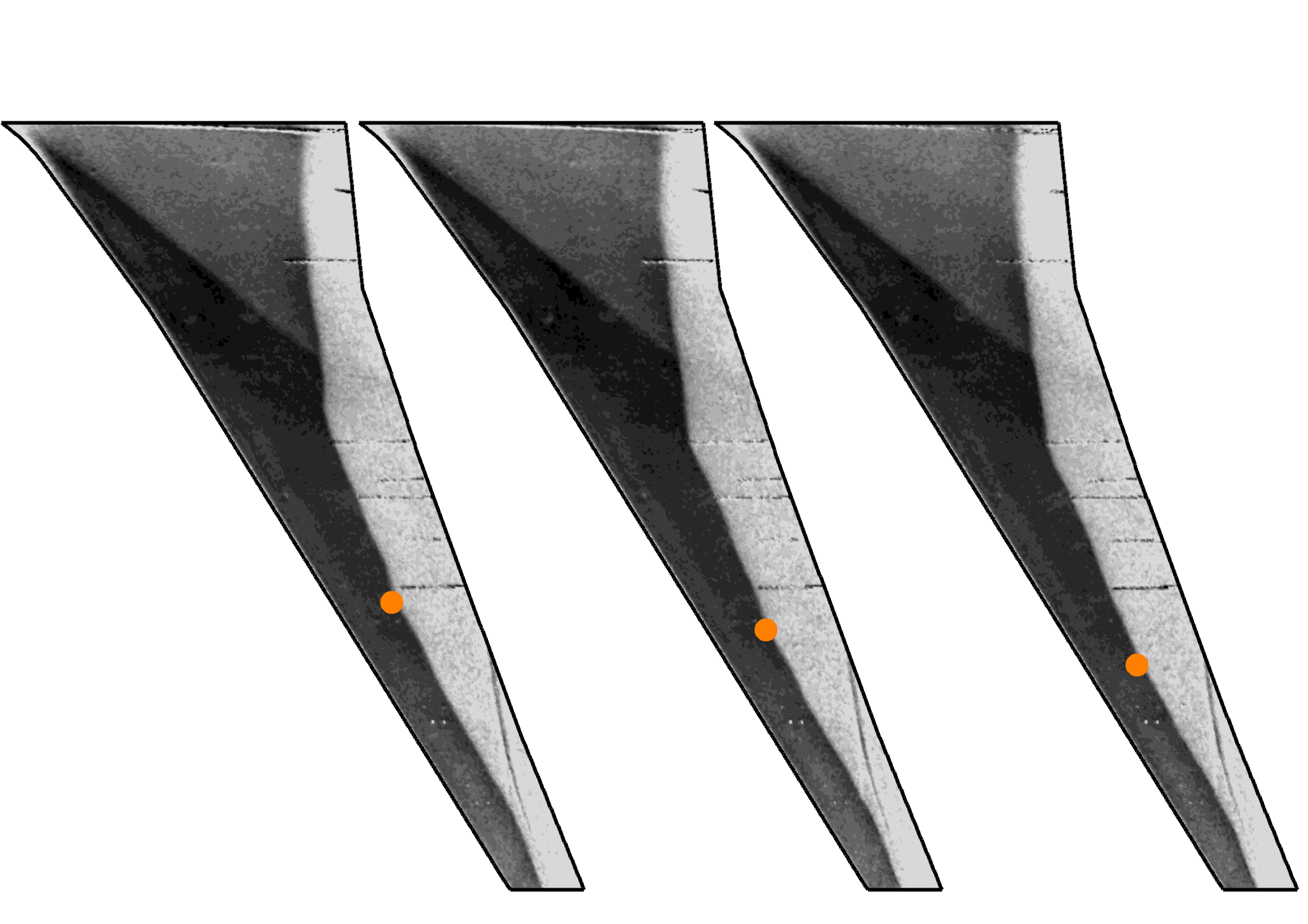}
	\caption{$M_{\infty} = 0.90$}
	\label{f:buffet_timeseries_M09}
\end{subfigure}%
	\caption{$c_p$ contours acquired using unsteady PSP at three consecutive timesteps at \SI{2000}{\hertz} at $Re_{\infty} = 12.9 \cdot 10^6$ and $\alpha=\SI{4}{\degree}$}
	\label{f:buffet_timeseries_overall}
\end{figure}

Fig.~\ref{f:buffet_timeseries_overall} shows three consecutive temporal snapshots of the pressure distribution on the wing suction side for $M_{\infty} = 0.84$ and $M_{\infty} = 0.90$, respectively. As already indicated by the pressure RMS data in Fig.~\ref{f:PSP_cprms_wing_upper_M090_allRe_aL2_aB}, the wavy motion of the shock front is more pronounced at the lower Mach number.

For this analysis method, the time averaged shock position is extracted from the unsteady PSP dataset by computing the maximum chordwise pressure gradient at each spanwise position between $\eta = 45 \%$ and $\eta = 90 \%$. A Savitzky-Golay filter is subsequently applied to smooth the resulting shock line prior to analysing the spanwise propagation. The shown snapshots visualize the spanwise motion of one buffet cell in the region between 60 and 80\% half span, whose crest is highlighted by orange dots in Fig.\ref{f:buffet_timeseries_M084}. The shock is comparatively steady in the inboard region, which agrees with the RMS values in Fig.~\ref{f:PSP_cprms_wing_upper_M090_allRe_aL2_aB}. Further outboard, near the wing tip, the undulating motion of the shock front transitions into a shock motion parallel to the leading edge. 
%The undulating motion of the shock is consistent with observations by \cite{sugioka:2018} and others. 
The undulation is pronounced and easily detectible with the naked eye. A wavy shape of the shock front is also present at $M_{\infty} = 0.90$ and $\alpha=4^\circ$, as shown in Fig.~\ref{f:buffet_timeseries_M09}. However, the chordwise amplitude of the shock motion is significantly smaller. Nevertheless, a spanwise propagation can be discerned at these inflow conditions as well.

The spatiotemporal motion characteristics of the shock front captured via unsteady PSP can be analyzed to obtain a propagation velocity of the disturbance. \cite{paladini:2019} employ spatial tracking of the phase shift of multiple signals at multiple fixed frequencies. This is of particular interest when a frequency range with high coherence between the signals is well known and a spatial distribution of multiple data points is available, as it is the case for PSP data. The propagation velocity can then be calculated by $u_s=\frac{2 \pi f}{\Delta \Phi/ \Delta \eta}$, where $f$ denotes the frequency and $\Delta \Phi/ \Delta \eta$ the gradient of the phase angle over the spanwise coordinate.

In the present work, the method is implemented via peak detection of the instantaneous shock line relative to the time-averaged shock line. For this method, the time averaged shock position is extracted from the PSP dataset and the propagation of buffet cells along this line is calculated. The phase angle $\Phi$ is extracted from cross-spectra along the shock line with a reference point at $\eta = 60\%$. 

\begin{figure}[!htb]
	\begin{subfigure}{0.48\textwidth}
		\centering
		\includegraphics[width=1\linewidth]{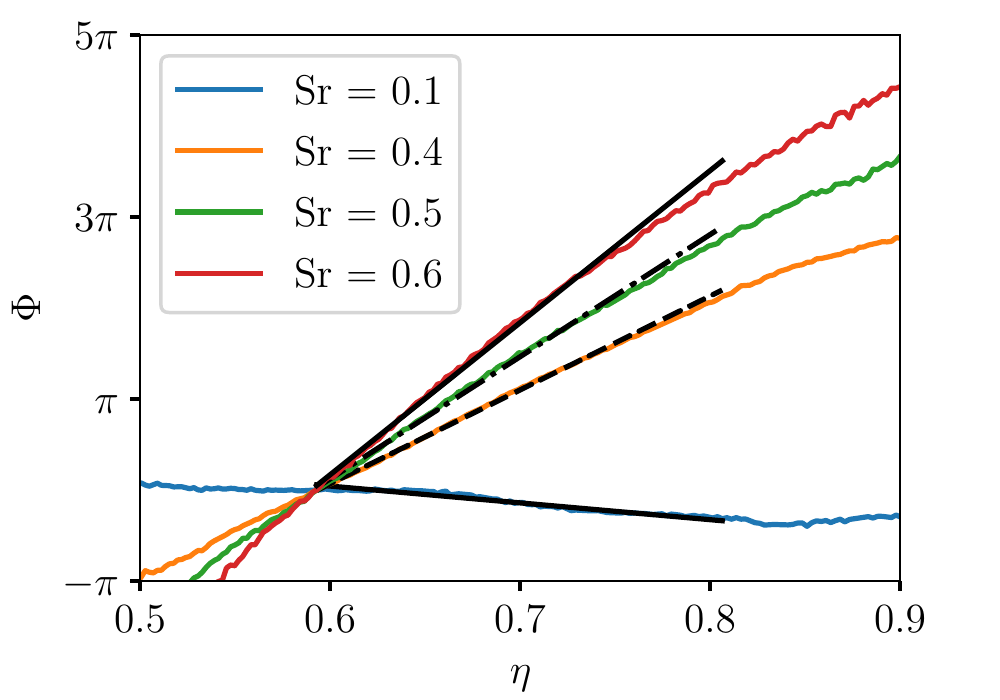}
		\caption{$M_{\infty} = 0.84$, $Re_{\infty} = 12.9 \cdot 10^6$, $\alpha=\SI{4}{\degree}$}
		\label{f:buffetcell_speed0225}
	\end{subfigure}
	\hfill
	\begin{subfigure}{.48\textwidth}
		\centering
		\includegraphics[width=1\linewidth]{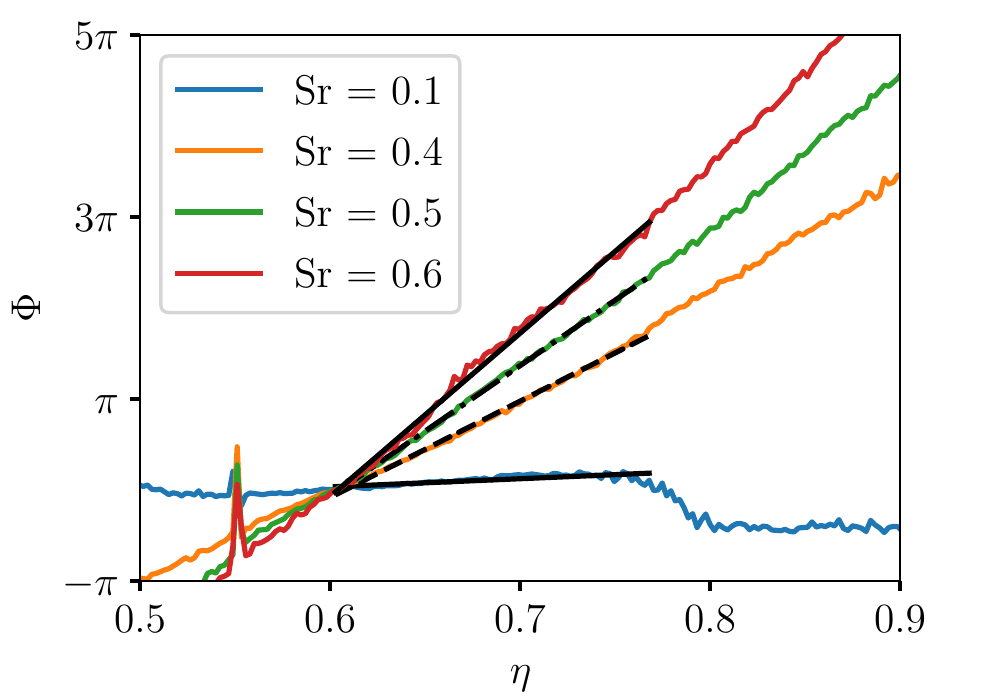}
		\caption{$M_{\infty} = 0.90$, $Re_{\infty} = 12.9 \cdot 10^6$, $\alpha=\SI{4}{\degree}$}
		\label{f:buffetcell_speed0227}
	\end{subfigure}
	\caption{Spanwise evolution of phase angle $\Phi$ at different frequencies.}
	\label{f:buffetcell_speed}
\end{figure}

The phase angle is extracted at frequencies with high levels of coherence, which occurs at very low frequencies around $Sr \approx 0.1$ and in the range of the buffet bump near $Sr = 0.4$. The phase angle variation of these frequencies between $\eta=0.5$ and $\eta=0.9$ is shown in Fig.~\ref{f:buffetcell_speed} for the two Mach numbers under investigation. The estimation of propagation velocity involves a linear fit of the phase angle in the vicinity of the reference point. The fitted slopes are indicated in Fig.~\ref{f:buffetcell_speed} and are carried out between 60\% and 80\% of half-span for consistency purposes.

At $M_{\infty} = 0.84$, the phase shift of $Sr=0.4$ and $0.5$ reveals a linear increase in this spanwise domain. The propagation velocity for both frequencies results in a value of $u_s/u_\infty=0.3$. Further outboard, the slopes decrease which can be attributed to the transition of the shock motion toward a chordwise oscillation of the shock front. The propagation velocity computation at $M_{\infty} = 0.90$ results in a value of $u_s/u_\infty=0.28$. At $Sr=0.1$, the phase angle remains approximately constant over span. This hints at a bulk shock motion in chordwise direction throughout the shown spanwise domain and confirms the observations of \cite{sugioka:2018}. In fact, the slightly negative slope at $M_{\infty} = 0.84$ indicates a possible inward propagation. ~\cite{masini:2020} described different propagation directions at different spanwise locations and at different frequencies. Future work is required to ascertain whether similar phenomena occur in the present dataset.

\begin{figure}[htb]
	\centering
	\begin{subfigure}{.49\textwidth}
		\centering
		\includegraphics[width=1\linewidth]{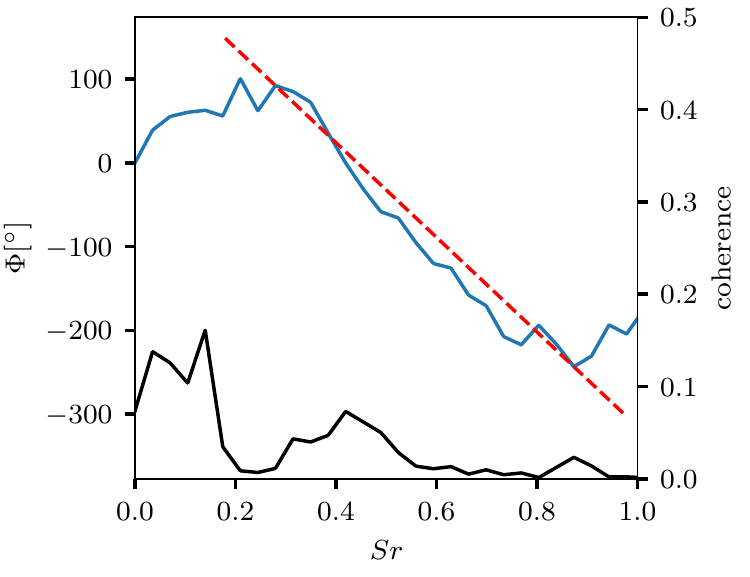}
		\caption{$M_{\infty} = 0.84$}
		\label{f:phase_angle_M084}
	\end{subfigure}
	\begin{subfigure}{.49\textwidth}
		\centering
		\includegraphics[width=1\linewidth]{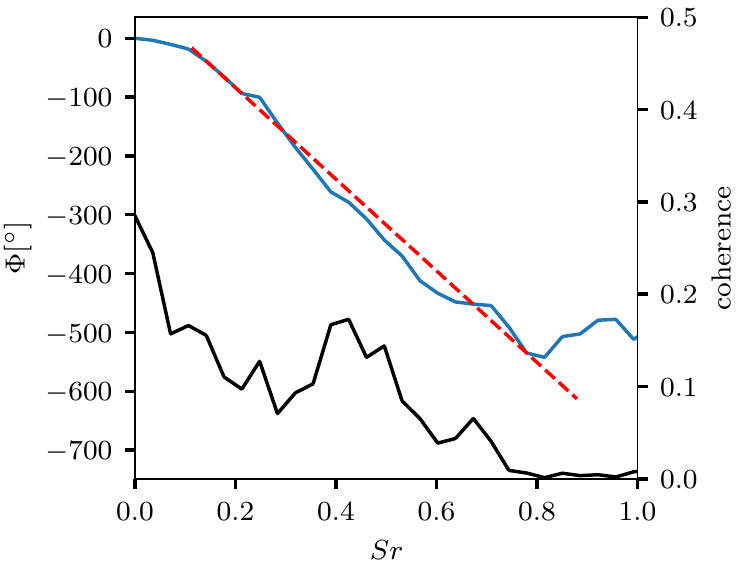}
		\caption{$M_{\infty} = 0.90$}
		\label{f:phase_angle_M090}
	\end{subfigure}
	\caption{Coherence and phase angle at $Re_{\infty} = 12.9 \cdot 10^6$ and $\alpha = \SI{4}{\degree}$ between a pair of pressure transducers ($\eta = 51\%$, $x/c = 0.7$ and $\eta = 63\%$, $x/c = 0.55$). Linear slope fitted in the region denoted by dashed red line in regions of high coherence associated with buffet.}
	\label{f:kulite_phase_angle}
\end{figure}

\cite{paladini:2019} employ a second method that can be used to derive the propagation velocity from time resolved pressure data at discrete points. A phase shift can also be obtained from the cross spectrum of two signals at two discrete points. The range of frequencies where convective phenomena occur can be narrowed down to regions of high coherence. In the buffet frequency range, high coherence goes along with a linear evolution of the phase angle in the frequency domain, which implies a constant propagation speed. This speed can finally be calculated by $u_s=\frac{2\pi s}{d\Phi/df}$, where $d\Phi /df$ is the slope of the phase angle over the frequency and $s$ is the distance between the two points. A pair of sensors at 51\% and 63\% span were selected, at $x/c = 0.7$ and $x/c = 0.55$, respectively. The selection occurred based on the presence of a distinct frequency region with high coherence. Fig.~\ref{f:kulite_phase_angle} shows the phase angle and coherence of between the two sensor signals at two Mach numbers. Despite computing the cross spectrum and the magnitude squared coherence using Welch's method and employing between 400 and 600 signal segments for averaging and variance reduction, it needs to be noted that the coherence never reaches high absolute values and remains on the order of 0.1 in the frequency regions associated with the buffet bumps. At $M_{\infty} = 0.84$, the linear fit results in a propagation speed of $u_s = \SI{70}{\meter\per\second} = 0.32 u_{\infty}$. In Fig.~\ref{f:phase_angle_M090}, a propagation velocity of $u_s = \SI{54}{\meter\per\second} = 0.24 u_{\infty}$ can be derived.

Overall, the results confirm the assumption that this range of values of $u_s$ has rather universal character for tapered and moderately swept wings. The propagation velocity agrees well with recent literature. Among others, \cite{paladini:2019} obtained values in the range of $u_s/u_\infty=0.245\pm0.015$ for different wings with moderate sweep. \cite{masini:2020} obtained a similar value of $u_s/u_\infty=0.26$ on a civil aircraft wing. \cite{sugioka:2018} derived a convection speed of $u_s/u_\infty=0.53$ on an 80\% scaled CRM. Moreover, \cite{timme:2020} observed a non-dimensional phase speed of \SIrange{0.26}{0.32}{} on the CRM and \cite{ehrle:2020} showed a spanwise propagation velocity of the buffet cells between $u_s/u_\infty =$ \SIrange{0.24}{0.28}{} at the same aircraft configuration. Considering this spread of results for the convection speed, the present investigation with values around $u_s/u_\infty=0.3$ aligns well with those findings.

\section{Conclusion}\label{sec13}

The present work provides an of the first cryogenic wind tunnel measurement campaign carried out in the ETW facility, focusing on establishing a baseline dataset for the XRF-1 aircraft model and study a variety of off-design conditions at extreme angles of attack.

The measurements in the European Transonic Windtunnel employed conventional experimental techniques including a model-internal force balance, pneumatic pressure measurements, accelerometers and dynamic surface pressure transducers. These were complemented by the use of steady-state and time-resolved pressure sensitive paint (PSP) acquisition setup involving multiple cameras in the bottom wind tunnel wall. The campaign was designed to first obtain polar data using continuous angle of attack traversals over a wide incidence range for several combinations of Mach and Reynolds numbers and dynamic pressure levels. The Mach numbers of $0.84$, $0.87$ and $0.90$ were selected to represent conditions near typical cruise and high speed flight. Three Reynolds number levels $Re_{\infty}$ = $\left\lbrace 3.3 \cdot 10^6, 12.9\cdot 10^6, 25.0\cdot 10^6 \right\rbrace$ span a wide range between typical ground testing conditions and realistic flight conditions at altitude. A set of such incidences was identified for the highest and the lowest Mach number and subsequently measured using all the above mentioned techniques. Additional incidences in the linear range of the lift polar and at predefined negative $\alpha$ values were recorded using steady-state PSP in order to obtain a richer dataset and a reference for the planned future entries.

The results showed that the distinction between linear and nonlinear ranges deteriorates at increasing Mach number. This is due to large areas of separated flow downstream from the shock on the wing occurring at high Mach numbers even at moderate incidences. The influence of Reynolds number at constant $M_{\infty}$ is comparatively small, with downstream displacement of the shock observed at all conditions.

The onset of buffet was found to be very different at the Mach numbers of $0.84$ and $0.90$. The onset occurs at much lower incidences at high Mach numbers and is gradual in terms of $c_{p,RMS}$ increase. In contrast, significant increase of pressure sensor unsteadiness at $M_{\infty} = 0.84$ was detected, which takes places at similar angles of attack as the occurrence of nonlinearity in the lift polar. Analysis of trailing edge $c_p$ divergence confirmed these general characteristics, with the divergence occurring at mid-wing and beginning at much lower angles of attack for high Mach numbers.

Multiple distinct oscillation frequencies were detected which persist over wide ranges of inflow conditions, which can be ascribed to structural and wind tunnel effects. Most important of these is the spectral peak at $Sr = 0.14$. Aerodynamic oscillations associated with shock buffet motion were detected in the Strouhal number range between 0.2 and 0.8, which includes higher frequencies than the typically described range between 0.2 and 0.6. The chordwise and spanwise variation of buffet frequency is similarly consistent with recent publication, with little detectable chordwise frequency shift. The present work focused on the unsteady pressure transducer data, with a more thorough modal analysis of the unsteady PSP dataset planned in the future.

The occurrence of buffet changes dramatically with Mach number. While the shock deforms at $M_{\infty} = 0.84$ in a manner forming a localized bulge with extreme upstream displacement around $\eta = 70\%$, the shock displacement is much more uniform at high Mach number. In fact, at high angles of attack approaching $\alpha = \SI{6}{\degree}$ it reaches the inboard wing and causes fluctuation at the buffet frequency near the trailing edge. The selected unsteady PSP measurement points span a regime from moderate buffet to deep buffet. While the former is associated with clearly visible buffet bumps in the unsteady pressure transducer spectra, the latter region at very high $\alpha$ is characterized by a breakdown of clear spectral features and the occurrence of high fluctuations across the board.

Analysis of unsteady $c_p$ data from PSP and the unsteady pressure transducers provided a means to compute the spanwise propagation velocity $u_s$ of the shock undulation in moderate buffet conditions. The resulting values of $u_s / u_{\infty} \approx 0.3 .. 0.32$ for $M_{\infty} = 0.84$ and $u_s / u_{\infty} \approx 0.24 .. 0.28$ for $M_{\infty} = 0.90$ at the typical buffet frequencies are in line with the survey provided by \cite{paladini:2019} and with the analysis by \cite{masini:2020}. This confirms the universal nature of the phenomenon. A low frequency motion around $Sr = 0.1$ and below showed little phase shift, indicating bulk chordwise motion of the shock.

The present study showed that the buffet frequency behavior described in the context of similar congurations by \cite{paladini:2019} and other authors remains consistent up to high and flight-like Reynolds numbers of $25 \cdot 10^6$. A frequency shift due to Reynolds number was observed at mid-wind unsteady pressure sensors, which was more pronounced at low Mach number. This is consistent with the significant $Re_{\infty}$-induced displacement of the shock, which is similarly more pronounced at lower Mach.

\backmatter

\bmhead{Acknowledgments}

The authors gratefully acknowledge the Deutsche Forschungsgemeinschaft DFG (German Research Foundation) for funding this work in the framework of the research unit FOR 2895. The authors would like to thank the Helmholtz Gemeinschaft HGF (Helmholtz Association), Deutsches Zentrum für Luft- und Raumfahrt DLR (German Aerospace Center) for financing the wind tunnel measurements and Airbus for providing the wind tunnel model. The authors would also like to thank all the scientists involved in the research initiative for the kind cooperation, as well as the colleagues at Airbus and ETW, whose valuable support has made the project possible.

\begin{appendices}

\section{Experiment Summary}\label{secA1}

The experimental campaign was laid out in a manner to make the most efficient use of available nitrogen. Operation of the iPSP cameras was the pacing item for the experiment schedule. The camera internal memory was able to hold 21840 images at full resolution, with read-out a lengthy process that cannot be feasibly done while the wind tunnel runs at full Mach number consuming large amounts of nitrogen. Given a maximum overall time for measurements, each measured condition is therefore necessarily the result of a trade-off between the number of acquired flow conditions and the number of images acquired for each.

The ETW sting is capable of rotating the model about the longitudinal axis inside the measurement section without requiring access to the model. The measurement plan made use of this, enabling recording of polars in both fin-up and fin-down configurations. The data was collected in several blocks over the course of several days, with successively decreasing temperature and increasing Reynolds number. After the acquisition of continuous polars at each Reynolds number, the resulting data was analyzed in-situ and used to fine-tune the flow conditions for the steady and unsteady pressure-sensitive paint runs.

\subsection{Flow Conditions}

The data points were set up to permit analysis of isolated effects of Mach number, Reynolds number and dynamic pressure wherever possible. The main runs were conducted at Mach numbers of $0.84$, $0.87$ and $0.90$. The Reynolds numbers were varied between $3.3\cdot 10^6$ at ambient temperature and $12.9\cdot 10^6$ and $25.0 \cdot 10^6$ at cryogenic conditions. Two levels of dynamic pressure $q/E$ were used, with the data at the intermediate Reynolds number of $12.9\cdot 10^6$ acquired at both $q/E = 0.2 \cdot 10^-6$ and $q/E = 0.4 \cdot 10^-6$. The resulting data for each combination is shown in Table~\ref{tab:exp_matrix}. Since unsteady PSP requires steady-state PSP data for reference, each data point at which iPSP measurements were carried out included a prior steady-state PSP measurement point.

\begin{table}[htb]
\caption{Experimental flow conditions and acquired data. }
\label{tab:exp_matrix}
\centering
\begin{tabular}{|ccccc|}
\hline
 & \thead{$Re_{\infty} = 3.3 \cdot 10^6$ \\ $q/E = 0.2 \cdot 10^{-6}$} & \thead{$Re_{\infty} = 12.9 \cdot 10^6$ \\ $q/E = 0.2 \cdot 10^{-6}$} & \thead{$Re_{\infty} = 12.9 \cdot 10^6$ \\ $q/E = 0.4 \cdot 10^{-6}$} & \thead{$Re_{\infty} = 25.0 \cdot 10^6$ \\ $q/E = 0.4 \cdot 10^{-6}$} \\
$M_{\infty} = 0.84$ & \cellcolor{yellow} polars only & \cellcolor{yellow} polars only & \cellcolor{olive} PSP/iPSP & \cellcolor{olive} PSP/iPSP\\
$M_{\infty} = 0.87$ & \cellcolor{yellow} polars only & \cellcolor{orange} no data & \cellcolor{brown} PSP & \cellcolor{brown} PSP\\
$M_{\infty} = 0.90$ & \cellcolor{yellow} polars only & \cellcolor{red} upper side PSP/iPSP  & \cellcolor{olive} PSP/iPSP  & \cellcolor{olive} PSP/iPSP  \\

\hline
\end{tabular}
\end{table}

\subsection{Selection of Fixed-Incidence Measurement Points}

As follows from Table~\ref{tab:exp_matrix}, the steady and unsteady PSP measurements were conducted at high Reynolds numbers, with lower $Re_{\infty}$ data acquired only using conventional measurement techniques. The goal of the PSP measurements was to characterize the flow phenomena from buffet onset until incidences beyond buffet. This required a priori determination of the angles of attack associated with such conditions. At each of the shown Mach and Reynolds number combinations involving PSP, the polar and $c_p$ data was used to identify a linear region and an incidence range where buffet and large scale unsteadiness is likely. This was carried out after recording continuous pitch traversal polars at each combination of $M_{\infty}$ and $Re_{\infty}$ shown in Table.~\ref{tab:exp_matrix}.

Based on the $\Delta C_L$ method by \cite{lawson}, between three and five incidences were defined in the vicinity of the lift curve break at $M_{\infty} = 0.84$ and $M_{\infty} = 0.90$. Higher Reynolds number data at $Re_{\infty} = 25.0\cdot 10^6$ received higher priority in this selection, as high $Re$ experiments constitute one of the main research priorities of the initiative. Therefore the $Re_{\infty} = 25.0\cdot 10^6$ datasets encompass a larger $\alpha$ range than those at $Re_{\infty} = 12.9\cdot 10^6$.

The resulting set of stabilized points at fixed incidence can be assigned as follows to the four scenarios defined by Lutz et al.~\cite{lutz:2022}:
\begin{enumerate}
\item Steady state and unsteady PSP measurements around buffet conditions at \SIrange{3}{5}{\degree}($M_{\infty} = 0.84$) and \SIrange{2.5}{6}{\degree} ($M_{\infty} = 0.90$ $q/E = 0.4$).
\item High angle of attack conditions at $M_{\infty} = 0.90$, $Re_{\infty} = 12.9 \cdot 10^6$ and $q/E = 0.2$ with steady and unsteady PSP measurements on the upper surfaces only. 
\item Acquisition of steady state PSP data at negative angles of attack \SI{-4}{\degree} ($M_{\infty} = 0.84$), \SI{-2}{\degree} ($M_{\infty} = 0.90$)
\item Acquisition of steady state PSP data in the linear range at $\alpha = \SI{0}{\degree}$ and $\alpha = \SI{1.5}{\degree}$ for all three Mach numbers
\end{enumerate}

While $Re_{\infty} = 25.0\cdot 10^6$ data was prioritized by acquiring more data incidences, measurements at the lowest temperatures required for this Reynolds number were also the most challenging for optical systems. The lower Reynolds number measurements at $Re_{\infty} = 12.9\cdot 10^6$ were conducted at higher temperatures of T=\SI{180}{\kelvin}, creating conditions enabling better PSP camera image quality and signal to noise ratio. This allowed the unsteady PSP measurements at these datasets to be carried out at a sampling rate of \SI{2000}{\hertz} as opposed to \SI{1000}{\hertz} at the other conditions. 

In addition, camera window icing during $Re_{\infty} = 12.9\cdot 10^6$ $q/E = 0.2 \cdot 10^{-6}$ runs specifically aimed at achieving the highest possible angles of attack further deteriorated image quality. Nevertheless, all intended incidences were successfully measured.

\end{appendices}

\section*{Declarations}

\paragraph{Funding} The experiments leading to these results were funded by the Deutsche Forschungsgemeinschaft research unit FOR 2895.
\paragraph{Ethics approval}
Not applicable
\paragraph{Consent to participate}
Not applicable
\paragraph{Consent for publication}
Not applicable
\paragraph{Availability of data and materials} The data confidentiality is covered by a framework non disclosure agreement between all parties participating in the DFG FOR 2895 initiative.
\paragraph{Code availability}
The analysis was conducted using in-house code written in the Python language
\paragraph{Author contributions}
All authors contributed to the study conception and design. D. Yorita was responsible for the pressure sensitive paint measurements. A. Waldmann participated in the experimental campaign and created the original draft. M. Ehrle, J. Kleinert and T. Lutz contributed data analysis. All authors commented on previous versions of the manuscript. All authors read and approved the final manuscript.
\paragraph{Competing interests} The authors declare that they have no conflicts of interest.

\bibliography{bibtex_database}

\begin{thebibliography}{33}
\providecommand{\natexlab}[1]{#1}
\providecommand{\url}[1]{{#1}}
\providecommand{\urlprefix}{URL }
\providecommand{\doi}[1]{\url{https://doi.org/#1}}
\providecommand{\eprint}[2][]{\url{#2}}
 \bibcommenthead

\bibitem[{{Abbas-Bayoumi} and {Becker}(2011)}]{abbasbayoumi:2011}
{Abbas-Bayoumi} A, {Becker} K (2011) An industrial view on numerical simulation
  for aircraft aerodynamic design. Journal of Mathematics in Industry 1(1):10.
  \doi{10.1186/2190-5983-1-10},
  \urlprefix\url{https://doi.org/10.1186/2190-5983-1-10}

\bibitem[{Abdrashitov(1939)}]{abdrashitov:1939}
Abdrashitov G (1939) Tail buffeting. Tech. Rep. Report No. 395, Central
  Aero-Hydrodynamical Institute, Moscow

\bibitem[{Crouch et~al(2009)Crouch, Garbaruk, Magidov, and
  Travin}]{crouch:2009}
Crouch JD, Garbaruk A, Magidov D, et~al (2009) Origin of transonic buffet on
  aerofoils. Journal of Fluid Mechanics 628:357--369

\bibitem[{D'Aguanno et~al(2022)D'Aguanno, {Camps Pons}, Schrijer, and {van
  Oudheusden}}]{daguanno:2022}
D'Aguanno A, {Camps Pons} C, Schrijer F, et~al (2022) {Experimental study of
  the effect of wing sweep on transonic buffet}. In: AIAA SciTech Forum

\bibitem[{Dandois(2016)}]{dandois:2016}
Dandois J (2016) Experimental study of transonic buffet phenomenon on a 3d
  swept wing. Physics of Fluids 28(1):116

\bibitem[{{Ehrle} et~al(2020){Ehrle}, {Waldmann}, {Lutz}, and
  {Kr\"amer}}]{ehrle:2020}
{Ehrle} MC, {Waldmann} A, {Lutz} T, et~al (2020) Simulation of transonic buffet
  with an automated zonal des approach. CEAS Aeronautical Journal
  11:1025--1036. \doi{10.1007/s13272-020-00466-7},
  \urlprefix\url{https://doi.org/10.1007/s13272-020-00466-7}

\bibitem[{Elsenaar(1988)}]{agard303}
Elsenaar A (1988) Reynolds number effects in transonic flow. AGARDograph No.
  303 AGARD-AG-303, North Atlantic Treaty Organization (NATO)

\bibitem[{Garnier and Deck(2010)}]{garnier:2010}
Garnier E, Deck S (2010) Large-eddy simulation of transonic buffet over a
  supercritical airfoil. In: Vincenzo~Armenio JFBernard~Geurts (ed) Direct and
  Large-Eddy Simulation VII. Springer, Berlin, Heidelberg, p 549--554

\bibitem[{Giannelis et~al(2017)Giannelis, Vio, and Levinski}]{giannelis:2017}
Giannelis NF, Vio GA, Levinski O (2017) A review of recent developments in the
  understanding of transonic shock buffet. Progress in Aerospace Sciences
  92:39--84

\bibitem[{G{\"o}rtz et~al(2020)G{\"o}rtz, Abu-Zurayk, Ilic, Wunderlich, Keye,
  Schulze, Klimmek, Kaiser, S{\"u}el{\"o}zgen, Kier, Schuster, D{\"a}hne,
  Petsch, Kohlgr{\"u}ber, H{\"a}{\ss}y, Mischke, Weinert, Knechtges, Gottfried,
  Hartmann, and Fr{\"o}hler}]{goertz:2020}
G{\"o}rtz S, Abu-Zurayk M, Ilic C, et~al (2020) Overview of collaborative
  multi-fidelity multidisciplinary design optimization activities in the dlr
  project victoria. In: 2020 AAF (ed) AIAA Aviation Forum 2020,
  \urlprefix\url{https://elib.dlr.de/135334/}

\bibitem[{Hartmann et~al(2013)Hartmann, Feldhusen, and
  Schr\"oder}]{hartmann:2013}
Hartmann A, Feldhusen A, Schr\"oder W (2013) On the interaction of shock waves
  and sound waves in transonic buffet flow. Physics of Fluids 25(2):025

\bibitem[{Iovnovich and Raveh(2014)}]{iovnovich:2014}
Iovnovich M, Raveh DE (2014) Numerical study of shock buffet on
  three-dimensional wings. AIAA Journal 53(2):449--463

\bibitem[{Jacquin et~al(2009)Jacquin, Molton, Deck, Maury, and
  Soulevant}]{jacquin:2009}
Jacquin L, Molton P, Deck S, et~al (2009) Experimental study of shock
  oscillation over a transonic supercritical profile. AIAA Journal
  47(9):1985--1994

\bibitem[{Klein(2022)}]{klein:2022}
Klein C (2022) Time resolved pressure measurements by means of PSP in cryogenic
  conditions. \doi{10.2514/6.2022-1938},
  \urlprefix\url{https://arc.aiaa.org/doi/abs/10.2514/6.2022-1938},
  \eprint{https://arc.aiaa.org/doi/pdf/10.2514/6.2022-1938}

\bibitem[{Koike et~al(2016)Koike, Ueno, Nakakita, and Hashimoto}]{koike:2016}
Koike S, Ueno M, Nakakita K, et~al (2016) {Unsteady pressure measurement of
  transonic buffet on NASA common research model}. In: 34th AIAA Applied
  Aerodynamics Conference, p 4044

\bibitem[{Lawson et~al(2016)Lawson, Greenwell, and Quinn}]{lawson}
Lawson S, Greenwell D, Quinn MK (2016) Characterisation of Buffet on a Civil
  Aircraft Wing. \doi{10.2514/6.2016-1309},
  \urlprefix\url{https://arc.aiaa.org/doi/abs/10.2514/6.2016-1309},
  \eprint{https://arc.aiaa.org/doi/pdf/10.2514/6.2016-1309}

\bibitem[{Lee(2000)}]{lee:2000}
Lee B (2000) Vertical tail buffeting of fighter aircraft. Progress in Aerospace
  Sciences pp 193--279

\bibitem[{Lee and Brown(1992)}]{lee:1992}
Lee BHK, Brown D (1992) Wind-tunnel studies of f/a-18 tail buffet. Journal of
  Aircraft 29(1):146--152

\bibitem[{Lutz et~al(2016)Lutz, Gansel, Waldmann, Zimmermann, and {Schulte am
  H\"ulse}}]{lutz:2016}
Lutz T, Gansel PP, Waldmann A, et~al (2016) Time-resolved prediction and
  measurement of the wake past the crm at high reynolds number stall
  conditions. Journal of Aircraft 53(2):501--514. \doi{10.2514/1.C033351}

\bibitem[{Lutz et~al(2022)Lutz, Kleinert, Waldmann, Koop, Yorita, Dietz, and
  Schulz}]{lutz:2022}
Lutz T, Kleinert J, Waldmann A, et~al (2022) Research initiative for numerical
  and experimental studies on high speed stall of civil aircraft. Under Review
  in Journal of Aircraft 0(0)

\bibitem[{Mabey(1999)}]{mabey:1999}
Mabey D (1999) Unsteady aerodynamics: retrospect and prospect. The Aeronautical
  Journal 103(1019):1--18. \doi{10.1017/S0001924000065064}

\bibitem[{Mann et~al(2019)Mann, Thompson, and White}]{mann:2019}
Mann A, Thompson G, White P (2019) Civil aircraft wind tunnel feature rich
  testing at the edge of the envelope. FP68-AERO2019-mann

\bibitem[{Masini et~al(2020)Masini, Timme, and Peace}]{masini:2020}
Masini L, Timme S, Peace AJ (2020) Analysis of a civil aircraft wing transonic
  shock buffet experiment. Journal of Fluid Mechanics 884:A1.
  \doi{10.1017/jfm.2019.906}

\bibitem[{Paladini et~al(2019)Paladini, Dandois, Sipp, and
  Robinet}]{paladini:2019}
Paladini E, Dandois J, Sipp D, et~al (2019) Analysis and {Comparison} of
  {Transonic} {Buffet} {Phenomenon} over {Several} {Three}-{Dimensional}
  {Wings}. AIAA Journal 57(1):379--396. \doi{10.2514/1.J056473},
  \urlprefix\url{https://arc.aiaa.org/doi/10.2514/1.J056473}

\bibitem[{Rogers and Hall(1960)}]{rogers:1960}
Rogers EWE, Hall IM (1960) An introduction to the flow about plane swept-back
  wings at transonic speeds. The Journal of the Royal Aeronautical Society
  64:449--464. \doi{10.1017/S0368393100073259}

\bibitem[{Sugioka et~al(2018)Sugioka, Koike, Nakakita, Numata, Nonomura, and
  Asai}]{sugioka:2018}
Sugioka Y, Koike S, Nakakita K, et~al (2018) Experimental analysis of transonic
  buffet on a {3D} swept wing using fast-response pressure-sensitive paint.
  Experiments in Fluids 59(6):108. \doi{10.1007/s00348-018-2565-5},
  \urlprefix\url{http://link.springer.com/10.1007/s00348-018-2565-5}

\bibitem[{Sugioka et~al(2021)Sugioka, Nakakita, Koike, Nakajima, Nonomura, and
  Asai}]{sugioka:2021}
Sugioka Y, Nakakita K, Koike S, et~al (2021) Characteristic unsteady pressure
  field on a civil aircraft wing related to the onset of transonic buffet.
  Experiments in Fluids 62(1):20. \doi{10.1007/s00348-020-03118-y},
  \urlprefix\url{https://link.springer.com/10.1007/s00348-020-03118-y}

\bibitem[{Timme(2020)}]{timme:2020}
Timme S (2020) Global instability of wing shock-buffet onset. Journal of Fluid
  Mechanics 885:A37. \doi{10.1017/jfm.2019.1001}

\bibitem[{Uchida et~al(2021)Uchida, Sugioka, Kasai, Saito, Nonomura, Asai,
  Nakakita, Nishizaki, Shibata, and Sonoda}]{uchida:2021}
Uchida K, Sugioka Y, Kasai M, et~al (2021) Analysis of transonic buffet on
  {ONERA}-{M4} model with unsteady pressure-sensitive paint. Experiments in
  Fluids 62(6):134. \doi{10.1007/s00348-021-03228-1},
  \urlprefix\url{https://link.springer.com/10.1007/s00348-021-03228-1}

\bibitem[{Walter(2004)}]{etw_user_guide}
Walter U (2004) Etw user guide revision a. Tech. Rep. ETW/D/95001/A, European
  Transonic Windtunnel

\bibitem[{Welch(1967)}]{welch:1967}
Welch PD (1967) The use of the fast fourier transform for the estimation of
  power spectra: A method based on time averaging over short, modified
  periodograms. IEEE Transactions on Audio Electroacoustics 15(2):70--73

\bibitem[{Yorita et~al(2017)Yorita, Klein, Henne, Ondruss, Beifuss, Hensch,
  Guntermann, and Quest}]{yorita:2017}
Yorita D, Klein C, Henne U, et~al (2017) Investigation of a Pressure Sensitive
  Paint Technique for ETW (Invited). \doi{10.2514/6.2017-0335},
  \urlprefix\url{https://arc.aiaa.org/doi/abs/10.2514/6.2017-0335},
  \eprint{https://arc.aiaa.org/doi/pdf/10.2514/6.2017-0335}

\bibitem[{Yorita et~al(2018)Yorita, Klein, Henne, Ondrus, Beifuss, Hensch,
  Longo, Guntermann, and Quest}]{yorita:2018}
Yorita D, Klein C, Henne U, et~al (2018) Successful Application of Cryogenic
  Pressure Sensitive Paint Technique at ETW. \doi{10.2514/6.2018-1136},
  \urlprefix\url{https://arc.aiaa.org/doi/abs/10.2514/6.2018-1136},
  \eprint{https://arc.aiaa.org/doi/pdf/10.2514/6.2018-1136}

\end{thebibliography}

\end{document}